\documentclass[fleqn,a4paper,usenatbib,usegraphicx,useAMS]{mn2e}
\usepackage{natbib,graphicx}
\usepackage{amsmath}
\usepackage{times}

\catcode`\@=11
\def\gta{\ifmmode{\,\mathrel{\mathpalette\@versim>\,}}
    \else{$\,\mathrel{\mathpalette\@versim>}\,$}\fi}
\def\lta{\ifmmode{\,\mathrel{\mathpalette\@versim<\,}}
    \else{$\,\mathrel{\mathpalette\@versim<}\,$}\fi}
\def\@versim#1#2{\lower 2.9truept \vbox{\baselineskip 0pt \lineskip
    0.5truept \ialign{$\m@th#1\hfil##\hfil$\crcr#2\crcr\sim\crcr}}}
\catcode`\@=12  

\def\figref#1{Fig.~\ref{#1}}
\renewcommand{\eqref}[1]{equation~(\ref{#1})}

\newcommand{\eqsref}[2]{equations~(\ref{#1}) and (\ref{#2})}
\newcommand{\blankeqref}[1]{(\ref{#1})}
\newcommand{\Eqref}[1]{Equation~(\ref{#1})}
\newcommand{\tabref}[1]{Table~\ref{#1}}
\def\secref#1{Section~\ref{#1}}

\def\percent{\,\,\text{per cent}}

\def\near{\sim\!}
\newcommand{\ttp}[1]{\times 10^{#1}}
\newcommand{\vect}[1]{{\bf #1}}
\newcommand{\vecthat}[1]{\hat{\vect{#1}}}

\def\msun{M_\odot}

\newcommand{\rsun}{R_0}

\def\kms{\,{\rm km}\,{\rm s}^{-1}}
\def\Myr{\,{\rm Myr}}

\def\Gyr{\,{\rm Gyr}}
 
\def\pc{\,{\rm pc}}
\def\kpc{\,{\rm kpc}}

\def\d{{\rm d}}

\def\a{{\rm a}}
\def\g{{\rm g}}
\def\fvfps{{\sc fvfps}}
\def\stackel{St\"{a}ckel}
\def\vJ{\vect{J}}
\def\DvJ{\delta\vJ}
\def\vT{\mbox{\boldmath$\theta$}}
\def\vO{\vect{\Omega}}

\def\eigen{\vecthat{e}}
\def\hessian{\vect{D}}

\def\jr{J_r}
\def\rlim{r'_\text{lim}}
\def\tdyn{t_\text{dyn}}

\def\eff{\text{eff}}
\def\vw{\vect{w}}
\def\tide{\text{tide}}
\def\peri{{\rm p}}
\def\apo{{\rm a}}

\def\twidr{\tilde{r}}

\def\squarefigshrink{0.6}
\def\doubsquarefigshrink{0.3}

\title[The mechanics of tidal streams]
{The mechanics of tidal streams}

\author[A. Eyre and J. Binney]{Andy Eyre\thanks{Email: eyre@thphys.ox.ac.uk} and James  Binney\\
Rudolf Peierls Centre for Theoretical Physics, 1 Keble Road, Oxford OX1 3NP, UK\\}

\begin{document}

\label{firstpage}

\date{\today}

\pagerange{\pageref{firstpage}--\pageref{lastpage}}
\pubyear{2010}

\maketitle

\begin{abstract}
  We present an analysis of the mechanics of thin streams, which are
  formed following the tidal disruption of cold, low-mass clusters in
  the potential of a massive host galaxy. The analysis makes extensive
  use of action-angle variables, in which the physics of stream
  formation and evolution is expressed in a particularly simple
  form. We demonstrate the formation of streams by considering
  examples in both spherical and flattened potentials, and we find
  that the action-space structures formed in each take on a consistent
  and characteristic shape.  We demonstrate that tidal streams
  formed in realistic galaxy potentials are poorly represented by single
  orbits, contrary to what is often assumed.  We further demonstrate
  that attempting to constrain the parameters of the Galactic
  potential by fitting orbits to such streams can lead to significant
  systematic error. However, we show that it is possible to predict
  accurately the track of streams from simple models of the action-space
  distribution of the disrupted cluster.
\end{abstract}


\section{Introduction}

Recent years have seen a tremendous advance in the quality and
quantity of observational data for substructure in the halo of our
Galaxy. Of particular note is the outstanding success of the Sloan
Digital Sky Survey \citep[SDSS,][]{sdss}, which has uncovered large
numbers of streams in the halo of the Milky Way
\citep{odenkirchen-delineate,majewski-sag,yanny-stream,field-of-streams,orphan-discovery,
  grillmair-orphan,gd1-discovery,ngc5466,
  grillmair-2009,newberg-streams-2009}. These streams appear to be the
relics of tidally stripped globular clusters and dwarf galaxies that
have fallen into the Galaxy, although it is often
the case that no obvious progenitor object can be associated with a
particular stream.

It has been empirically noted that such tidal streams appear to delineate the
orbit of the progenitor object from which they formed
\citep{mcglynn-streams-are-orbits, johnston-delineate}. A consequence of this
assumption is that measurements of separate segments of a single stream can
be equated to sampling different phases of a single orbit.  If either
complete phase-space information \citep{willett}, or a certain subset of it
\citep{jin-reconstruction,binney08,eb09b}, is known for a sequence of phases
along a single orbit, it is possible to diagnose the Galactic potential with
exquisite precision \citep{binney08}.  Thus the assumption that such streams
delineate orbits promises tremendous diagnostic power and had been frequently
invoked \citep{binney08,eb09b,oden-2009,willett,koposov,galplx}.

Such techniques are hampered by the considerable difficulty in obtaining
observational data of adequate precision for stars in distant streams.  More
worrisome, however, is the mounting evidence that streams do not precisely
delineate orbits
\citep{dehnen-pal5,choi-etal,montuori-nbody-streams,eb09a}---in these
circumstances, which we will demonstrate to be generic---any conclusions
drawn from the fitting of orbits to streams could be systematically in error.

It is therefore of vital importance for the continued application
of such techniques that we obtain an understanding of the
relationship between tidal streams and the orbits of the stars that
comprise them. In particular, we must determine under what
circumstances tidal streams delineate orbits, by what measure they are
in error when they do not, and what can be done to correct this error.
There has not to date been an exposition of stream formation that
fundamentally addresses these issues.

Most studies have either focused on N-body simulations
\citep{choi-etal,montuori-nbody-streams}, the confusion of which makes the
predominant physics hard to isolate, or have attempted to describe the
problem in terms of conventional phase-space coordinates and classical
integrals \citep{dehnen-pal5,choi-etal}, which makes the problem intractably
hard. The work of \cite{choi-etal} made some progress towards understanding
the dynamical structure of clusters at the point of disruption, and they
provide a qualitative picture of the evolution of tidal tails, understood in
terms of classical integrals. However, they are unable to make predictions
for stream tracks on the basis of this picture alone, and they are ultimately
forced to rely on N-body simulation for quantitative results.  In this paper,
we approach the problem using action-angle variables, in which the physics of
stream formation finds a natural and simple expression
\citep{HelmiW99,tremaine99}.

We confine our investigation to the formation of long, cold streams,
such as may form from tidally disrupted globular clusters. We do so
for two reasons. First, a low mass for the progenitor cluster
simplifies the description of its orbital mechanics, because of the
lack of dynamical friction and other feedback effects in its interaction
with the host galaxy.  Second, thin, long streams provide the
strongest constraints upon the Galactic potential, because any orbit
delineated by them can be observationally identified with less
ambiguity \citep{binney08,eb09a}.  Hence, it is long, cold streams
that are of primary interest for use as probes of the potential.

We study the mechanics of stream formation immediately following the
tidal disruption of a progenitor cluster. In much of the work that
follows, the assumption is made that stream stars feel only the
potential of the Galaxy, i.e.~the stream stars do not
self-gravitate. This assumption is generally a fair one: the stars in
streams are generally spaced too widely for their self-gravity to be
of consequence \citep{dehnen-pal5}.  Indeed, we will demonstrate below
that self-gravity becomes negligible shortly after stars are stripped
from our model clusters.

The remainder of the paper is arranged as follows: Section \ref{sec:aavars}
discusses the action-angle variables in which we perform our analysis.
\secref{sec:tremaine} discusses the basic mechanics of stream formation and
propagation. \secref{sec:d-examples} discusses the detail of stream formation
in spherical systems, with the aid of some examples. \secref{sec:mapping}
relates the action-angle structure of streams to the corresponding real-space
manifestation.  \secref{sec:actions} examines the action-space distribution
of disrupted clusters using N-body simulation.  \secref{sec:fitting}
describes the consequences of optimising potential parameters under the
faulty assumption that streams follow orbits.  \secref{sec:nonsph} discusses
stream formation in flattened systems, using an axisymmetric \stackel\
potential as an example. Finally, \secref{sec:conclusions} presents our
concluding remarks.

\section{Action-angle variables}
\label{sec:aavars}

We use action-angle variables to analyse stream formation and propagation. The
usefulness and theoretical basis of action-angle variables is extensively
discussed in \S3.5 of Binney \& Tremaine (2008; hereafter BT08).

Action-angle variables are a set of
canonical coordinates, like Cartesian phase-space coordinates, that
can be used to describe systems in Hamiltonian mechanics.
Actions $\vJ$ are useful because they constants of motion. Moreover, 
the angle variables $\vT$ that are conjugate to them evolve linearly in time
 \begin{equation}
\vT(t) = \vT(0) + \vO\,t,
\end{equation}
where we have introduced the frequencies
\begin{equation}
\Omega_i(\vJ) \equiv \dot{\theta_i} = {\partial H \over \partial J_i}.
\label{eq:eqnsmotion}
\end{equation}
Hence, the motion of a system described by action-angle variables
is very easy to predict.

\S3.5 of \citealt{bt08} discusses at length the calculation of
action-angle coordinates in various systems. Here, we note that
standard methods for calculating action-angle coordinates require the
Hamilton-Jacobi equation  to
separate (\S3.5.1 of BT08). This condition is met by all spherically symmetric
potentials, but excludes any non-spherical potential that is not of
\stackel\ form (see \secref{sec:stackel} below). Given
separability, the action corresponding to a coordinate $q$ is given by
 \begin{equation}
J_q = {1 \over 2\pi} \oint p_q \, dq,
\label{eq:jact}
\end{equation}
where the integral is over the closed path that encloses
a single oscillation of the coordinate $q$ along an orbit.

We note that just as spherical systems are naturally described by
spherical polar coordinates $(r,\vartheta,\phi)$, the natural actions
for such systems are $(J_r, L)$, where $J_r$ is the radial action, and
$L = J_\vartheta + |L_z|$ is the angular momentum.  Throughout this
paper, and without loss of generality, when spherical systems are
under study we set $L = L_z$ and confine the motion to the $(x,y)$
plane.

In the sections that follow, we utilize the equations
from \S3.5.2 of BT08 to transform between the
action-angle coordinates $(J_r,L,\theta_r,\theta_\phi)$ 
and conventional phase-space coordinates, when investigating
spherical potentials. The action-angle coordinates
used when investigating non-spherical potentials 
are discussed in \secref{sec:stackel} below.

\section{The formation of streams in action-angle space}
\label{sec:tremaine}

Consider a low-mass cluster that has recently been tidally disrupted,
with just enough time having passed so that the stripped stars no
longer feel the effects of one another's gravity.  The cluster is on a
regular orbit, identified by its actions $\vJ_0$, in a fixed
background potential, which has a Hamiltonian $H(\vJ)$. Suppose
further that in the locality of $\vJ_0$, the Hamiltonian is well
described by the Taylor expansion,
\begin{equation}
H(\vJ) = H_0 + \vO_0\cdot\DvJ + {1 \over 2}\DvJ^T\cdot\hessian\cdot\DvJ,
\label{eq:taylor-H}
\end{equation}
where $\delta\vJ = \vJ - \vJ_0$, $\vect{D}$ is the Hessian of $H$
\begin{equation}
D_{ij} = \left.{\partial^2 H \over \partial J_i \, \partial J_j}\right|_{\vJ_0},
  \label{eq:hessian}
\end{equation}
and $\vO_0$ is the frequency vector of the cluster's orbit
\begin{equation}
\Omega_{0,i} = \left.{\partial H \over \partial J_i}\right|_{\vJ_0}.
\end{equation}
The frequency vector $\vO$ of a nearby orbit $\vJ$ is then
\begin{equation}
\vO(\vJ) = \vO_0 + \hessian\cdot\DvJ.\label{eq:omega}
\end{equation}
If the disrupted cluster has some spread in actions $\Delta \vJ$ and
angles $\Delta \vT_0$, corresponding to the spread in position and
velocity of its constituent stars, then the spread in angle-space after
some time $t$ is given by (\citealt{HelmiW99,tremaine99}; BT08 \S8.3.1),
\begin{equation}
  \Delta \vT(t) = t \Delta \vO + \Delta\vT_0 \simeq t\Delta\vO, \label{eq:angle_t}
\end{equation}
where the near equality is valid when $t \, \Delta\vO \gg \Delta\vT_{0}$,
and where we have introduced the spread in frequencies, $\Delta \vO$,
which are related to the spread in actions via
\begin{equation}
  \Delta \vO = \hessian \cdot \Delta \vJ.
  \label{eq:d-dot-j}
\end{equation}
In the absence of self-gravity, the action-space distribution $\Delta
\vJ$ is frozen for all time.  $\hessian$ and $\vO$ are functions of
$\vJ$ only, and so are similarly frozen.  The secular evolution of a
disrupted cluster is therefore to spread out in angle-space, with its
eventual shape determined by $\Delta \vO = \hessian \cdot \Delta \vJ$,
and its size growing linearly with $t$.

For a given $\Delta \vJ$, what does $\Delta \vO$ look like?
\Eqref{eq:angle_t} and \eqref{eq:d-dot-j} show that $\hessian$ acts as a
linear map between a star's position in action-space and its position in
angle-space.  $\vect{D}$ is a symmetric matrix, so it is
characterised by its mutually orthogonal eigenvectors $\eigen_n$ ($n=1,3$)
and its real eigenvalues $\lambda_n$. 

Consider a cluster whose stars are distributed isotropically 
within a unit sphere in action-space  centred on $\vJ_0$.
The corresponding angle-space structure, resulting from the
mapping of this sphere by $\hessian$, will be an ellipsoid
with semi-axes of length $t\lambda_n$ and direction $\eigen_n$.

If the $\lambda_n$ are finite and approximately equal, such an
isotropic cluster will spread out in angle-space with no
preferred direction. Eventually, the cluster will uniformly populate the whole of
angle-space. In real-space the cluster will uniformly fill the entire
volume occupied by the orbit $\vJ_0$: such a cluster would not
form a stream.

If one of the $\lambda_n$ is much smaller than the others, then
$\hessian$ will map the cluster into a highly flattened
ellipsoid in angle-space.  In real-space, the cluster will occupy a
2-dimensional subspace of the orbital volume of $\vJ_0$. The precise
configuration of this subspace is likely to be complex, but it
would not form a stream.

If two of the $\lambda_n$ are small, then $\hessian$ maps
the cluster into a line in angle-space. In this case, the resulting
real-space structure will be a filament. The density of this
filament will fall linearly with $t$. In a real galaxy, such
a structure may therefore persist with a significant overdensity
for some time. It is this case that describes the formation of
tidal streams (\S8.3.1 of BT08) investigated in detail in this paper.
Finally, if all the $\lambda_n$ are zero, there will be no spread,
and even an unbound cluster will remain extant indefinitely.

We note that there is no {\em a priori} reason for any of the
$\lambda_n$ to be small.
The existence of $\vect{D}$ imposes no conditions on $H$ in
general, save that it must be twice differentiable near to $\vJ_0$.
We can write a Hamiltonian for which, for a particular $\vJ_0$ at least,
the $\lambda_n$ and the $\eigen_n$ take any specified values.
It must therefore be a peculiar property of realistic galactic
potentials that causes disrupted clusters to form streams.

\subsection{Validity}

Let us consider briefly the validity of the preceding analysis.
The key assumption is that the Taylor expansion \blankeqref{eq:taylor-H}
holds, which it does if
\begin{equation}
D_{ij} \gg {1 \over 3}{\partial D_{ij} \over \partial J_k} \delta J_k =
{1 \over 3}{\partial^3 H \over \partial J_i \partial J_j \partial J_k} \delta J_k.
\label{eq:condition2}
\end{equation}
If $H$ is dominated by some low power of $J$, this condition becomes
\begin{equation}
\delta J_i \ll J_i.
\label{eq:condition}
\end{equation}
 In general then, we expect
our analysis to be valid if the spread in action of the stars
in the cluster is small compared to the actions themselves, which is likely
to be true for cold clusters on high-energy orbits around massive hosts.
The condition (\ref{eq:condition2}) is met in detail
for all the examples considered in this paper.

\subsection{The problem}

We have seen that the condition for a stream to form is
that one of the eigenvalues $\lambda_n$ of the linear map, $\hessian$,
must be much larger than the other two. Herein, we will number the
$\lambda_n$ and their corresponding $\eigen_n$ such that
\begin{equation}
\lambda_1 > \lambda_2 \ge \lambda_3.
\end{equation}
Under what conditions does the direction of this stream point along
the progenitor's orbit?

Taken over a short period of time---say, the time taken for an individual
star to travel along the length of a stream---every star in the stream will
follow an almost identical trajectory. If phase is disregarded and the
trajectories superimposed, they would overlay almost perfectly. Streams form
because {\em slight} differences between the trajectories of individual stars
grow secularly over the lifetime of a stream, which may be many orbital
periods of the individual stars. There is no fundamental dynamical reason why
the track of the resulting stream should delineate the progenitor's orbit,
nor the orbit of any one of the stars, nor any orbit in the governing
potential.

In action-angle coordinates, the trajectory in angle-space of an orbit
$\vJ$ is given by the frequency vector, $\vO(\vJ)$.  A necessary
condition for a stream to delineate precisely the progenitor orbit is
that the long axis of the angle-space distribution $\Delta\vT$ is
aligned with the progenitor frequency $\vO_0$. However, we shall see in
Section \ref{sec:trajectory-j} that because
real-space position $\vect{x}(\vJ,\vT)$ is a function of $\vJ$ as well
as angle $\vT$, this condition alone is {\em not} sufficient to ensure
real-space alignment between streams and orbits.

We note that the angle-space distribution, $\Delta\vT$, depends on both
the potential, $\hessian$, and the action-space distribution, via
$\Delta \vJ$, and in practical cases it is necessary to consider both
when determining its gross alignment.

The problem of stream formation and evolution is patently a
complicated one. To further our analysis, we will examine separately
the role of the linear map $\hessian$, the real-space map
$\vect{x}(\vJ,\vT)$, and the action-space distribution $\Delta
\vJ$. Having done so, we will proceed to use this understanding to
examine the formation of streams in practical cases.

\section{The linear map $\hessian$ in spherical cases}
\label{sec:d-examples}

In order to analyze the behaviour of the linear map, let us restrict
ourselves to a case in which our cluster is both isotropic and small
in $\Delta\vJ$.  The angle-space distribution $\Delta\vT$ is now an
ellipsoid with semi-axes of length $\lambda_n$, and with the
semi-major axis of the distribution aligned with the principal
direction of the linear map, $\eigen_1$. In this scenario, the stream
will be delineated by a single orbit if the condition
\begin{equation}
\eigen_1 = \hat{\vO}_0,
\end{equation}
is met, since $\Delta\vJ$ is small and so $\vect{x}(\vJ,\vT)
\simeq \vect{x}(\vJ_0,\vT)$.

We will proceed by examining the forms of $\vO$ and $\hessian$, and
hence the angle-space geometry of streams, for some example spherical
potentials. We remind ourselves that such systems can always be completely
described by two actions.

\subsection{Kepler potential}
\label{sec:kepler}

There are remarkably few potentials
of interest for which the Hamiltonian can be written as a
function of $\vJ$ in closed form. One such potential is the
Kepler potential
\begin{equation}
  \Phi(r) = -{GM \over r},
\end{equation}
for which the Hamiltonian is,
\begin{equation}
  H(\vect{J}) =-  {(GM)^2 \over 2 (J_r + L)},
\end{equation}
where $J_r$ is the radial action, and $L$ is the angular momentum.
The frequencies can be worked out by direct differentiation, giving
\begin{equation}
\vO =
\begin{pmatrix}
\Omega_r\\
\Omega_\phi
\end{pmatrix}=
{ (GM)^2 \over 2\left(J_r + L\right)}
\begin{pmatrix}
  1 \\
  1
\end{pmatrix}.
\end{equation}
The linear map $\hessian$ is calculated via a further round of
differentiation, and has eigenvalue/vector pairs
\begin{align}
&\left[\lambda_1,\eigen_1\right] =\left[\lambda_1,
\begin{pmatrix}
e_{1,r}\\
e_{1,\phi}
\end{pmatrix}\right]=
\left[{-2 (GM)^2 \over \left(J_r + L\right)^3},\begin{pmatrix}
1\\
1
\end{pmatrix}\right],\\
&\left[\lambda_2,\eigen_2\right] =\left[\lambda_2,
\begin{pmatrix}
e_{2,r}\\
e_{2,\phi}
\end{pmatrix}\right]
= \left[0,
\begin{pmatrix}
-1\\
1
\end{pmatrix}\right].
\end{align}
The linear map $\hessian$ has only one finite eigenvalue, and so the
Kepler potential will always form streams from disrupted
clusters. Further, we observe that since $\eigen_1 \propto \vO$ in all
circumstances, streams will always perfectly delineate progenitor
orbits in Kepler potentials, regardless of the action-space
distribution of the cluster from which they form. In
this potential streams exhibit secular spread strictly along the orbit, and do
not grow wider with time.

\subsubsection{Numerical tests}

We have confirmed the above prediction numerically. We use a Kepler
potential with mass $M = 1.75 \ttp {11} \msun$, which reproduces a
fiducial circular velocity $v_c = 240\kms$ at the approximate solar
radius of $\rsun=8\kpc$.  A cluster of 50 test particles was created,
sampled from a Gaussian distribution in action-angle space, defined by
$\sigma_J = 1 \kpc\kms$ and $\sigma_\theta = 5\ttp{-3}$ radians. This
cluster was placed on the orbit K1, given in \tabref{tab:orbits}, in
the aforementioned Kepler potential. The orbit is highly eccentric,
with a pericentre radius of about $1.5\kpc$ and an apocentre of about
$13\kpc$.

\begin{table}
  \centering
  \caption
  {Actions and apses for selected orbits in the spherical potentials
    used in this paper. We generally choose $J_\phi = L$ so all orbits
    remain in the $(x,y)$ plane.}
  \begin{tabular}{l|ll|ll}
    \hline
    & $J_r / \kpc\kms$ & $L / \kpc\kms$ & $r_{\rm p} / \kpc$ & $r_\a / \kpc$ \\  
    \hline
    K1 & $780.$ & $1016.$ & $1.5$ & $13$ \\ 
    I1 & $313.$ & $1693.$ & $5$ & $13$ \\ 
    I2 & $0.$ & $1920.$ & $8$ & $8$ \\
    I3 & $207.$ & $1920.$ & $6$ & $12.5$ \\ 
    I4 & $215.4$ & $3127.$ & $11$ & $20$ \\ 
    I5 & $571.7$ & $2536.$ & $7$ & $20$  \\
    \hline
  \end{tabular}
  \label{tab:orbits}
\end{table}

\begin{figure}
  \centerline{
     \includegraphics[width=\squarefigshrink\hsize]{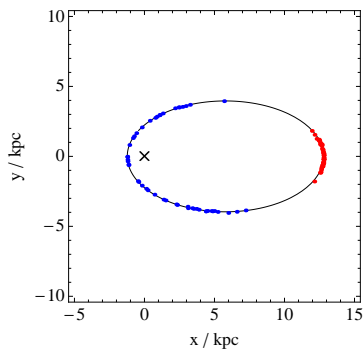}
     }
     \caption{The solid line shows the orbit K1 (\tabref{tab:orbits})
       in a Kepler potential, on which a cluster of 50 test particles
       has been evolved. The particles were released at apocentre. The
       red dots show the positions of the test particles near
       apocentre, after 24 complete orbits, at $t=4.02\Gyr$.  The blue
       dots show the same test particles near pericentre,
       approximately half an orbit later. In both cases, the dots
       delineate the progenitor orbit precisely.  }
  \label{fig:kepler}
\end{figure}

\begin{figure}
  \centerline{
    \includegraphics[width=\squarefigshrink\hsize]{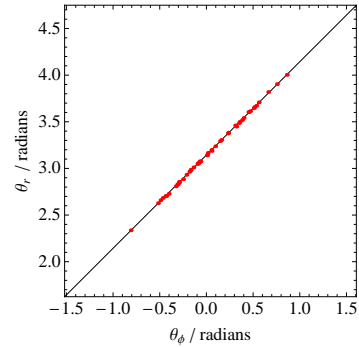}
  }
  \caption {Angle-space configuration for the particles shown at
    apocentre in \figref{fig:kepler}.  The solid line shows the
    frequency vector $\vO_0$, with which the stream particles are
    perfectly aligned.  Notably, there is no secular spread in the
    direction perpendicular to the stream motion, as predicted in
    \secref{sec:kepler}.  }
  \label{fig:kepler-angles}
\end{figure}

The cluster was released at apocentre, and evolved for $4.02\Gyr$,
equal to 24 complete orbits, by integrating the equations of motion
for each particle. \figref{fig:kepler} shows the
real-space configuration of the particles at the end of this time, and
\figref{fig:kepler-angles} shows the configuration of
the same particles in angle-space.

The cluster has elongated to form a stream that \figref{fig:kepler}
shows to cover approximately half the orbit when the centroid is at
pericentre. The stream does not spread in width in either real-space
or angle-space. In both figures, the stream delineates the cluster's
orbit perfectly. \figref{fig:kepler} shows that this is true
irrespective of the real-space location of the centroid, so the
prediction of the preceding section is validated.

\subsection{Spherical harmonic oscillator}
\label{sec:sho}

The spherical harmonic oscillator potential
\begin{equation}
\Phi(r) = {1\over2}\Omega^2 r^2,
\end{equation}
where $\Omega$ is a constant, applies for motion within
a sphere of uniform density (\S3.1a of BT08), and is therefore
of relevance to galaxy cores in the absence of a black hole.
The Hamiltonian is
\begin{equation}
H(J_r,L) = \Omega \left(L + 2J_r\right),
\label{eq:2dH-sho}
\end{equation}
 and the frequencies are
\begin{equation}
\vO =
\begin{pmatrix}
\Omega_r\\
\Omega_\phi
\end{pmatrix}=
\Omega
\begin{pmatrix}
2 \\
1
\end{pmatrix},
\end{equation}
while $\hessian$ is a null matrix, implying that the
eigenvalues are identically zero.

From this latter fact, we conclude that clusters in harmonic
potentials will always remain in the same configuration in
angle-space, and will not spread out. Hence, streams cannot form in
harmonic potentials. We further note that, in any case, it would be
difficult to tidally strip a cluster in a harmonic potential, since
the tidal force $dF_{\rm tide}$ across a cluster
\begin{equation}
dF_{\rm tide} \simeq {\partial^2 \Phi \over \partial r^2} \d r
= \Omega^2 \d r,
\end{equation}
is independent of galactocentric radius $r$. Thus, a cluster that is bound
at apocentre in such a potential will remain bound elsewhere along its orbit.
\subsection{Isochrone potential}
\label{sec:isochrone}

The isochrone potential (\S2.2.2d of BT08) is a simple potential
which has several useful properties. It behaves as a harmonic
oscillator in the limit of small radius, and as a Kepler potential
at large radius, thus providing a reasonable model for a spherical
galaxy across all radii.
The form of the potential is
\begin{equation}
  \Phi(r) = -{GM \over b + \sqrt{b^2 + r^2}},
\end{equation}
where $b$ is a constant scale. The Hamiltonian is
\begin{equation}
  H(\vect{J}) =- {(GM)^2 \over 2[J_r + {1 \over 2}
    (L + \sqrt{L^2 + 4GMb})]^2}.
\label{eq:isochrone-H}
\end{equation}
To proceed, we require the frequencies, which are obtained by partial
differentiation of \eqref{eq:isochrone-H}
\begin{align}
  \Omega_r &= { (GM)^2 \over [J_r + {1 \over 2}
    (L + \sqrt{L^2 + 4GMb})]^3},\label{eq:iso-freqs-r}\\
  \Omega_\phi &= {1 \over 2}
  \left( 1 + {L \over \sqrt{L^2 + 4GMb}} \right) \Omega_r. \label{eq:iso-freqs-p}
\end{align}
To construct $\hessian$, we also require the derivatives of the
frequencies with respect to the actions, which are obtained directly
from equations (\ref{eq:iso-freqs-r}) and (\ref{eq:iso-freqs-p}).

\begin{figure}
  \centering{
    \includegraphics[width=\squarefigshrink\hsize]{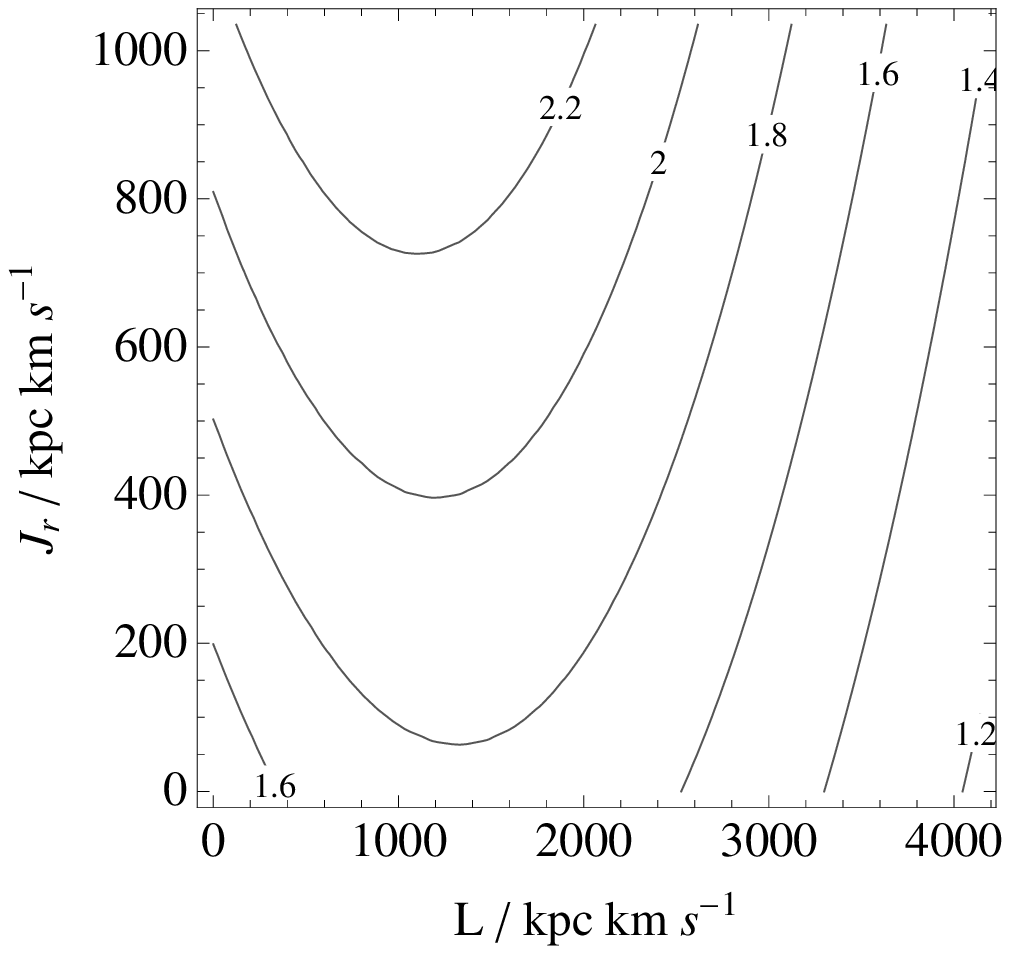}\\
    \includegraphics[width=\squarefigshrink\hsize]{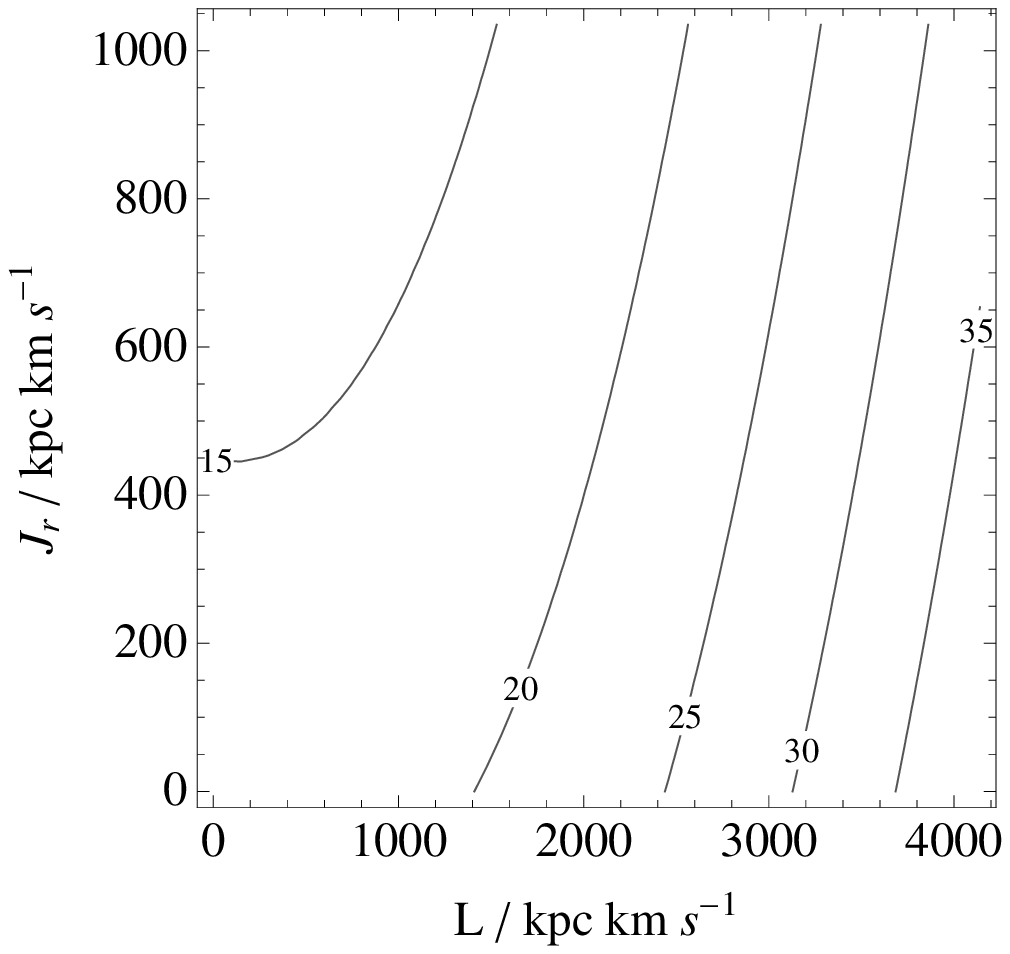}
  }
\caption{Details of the stream geometry in the isochrone potential of
    \secref{sec:isochrone}. Upper panel: the misalignment angle
    $\vartheta$, in degrees, between the principal direction of
    $\hessian$ and $\vO_0$, shown as a contour plot against the
    actions of the progenitor orbit.  In all cases, $\vartheta$ is
    between $1.2\deg$ and $2.2\deg$ in angle-space. Lower panel: The
    ratio of the eigenvalues $\lambda_1/\lambda_2$. The
    ratio is $>10$ everywhere and rises sharply with increasing
    $L$.  The actions shown in both plots cover a range of
    interesting orbits, which are described in
    \tabref{tab:isochrone-extrema}.  }
  \label{fig:isochrone-hessian}
\end{figure}

\begin{table}
  \caption
  {
    The coordinate extrema of selected orbits from \figref{fig:isochrone-hessian},
    illustrating the variety of orbits covered by that figure. The actions
    are expressed in $\kpc\kms$, while the apses are in \kpc.} 
\begin{center}
  \begin{tabular}{ll|ll}
    \hline
    $J_r$ & $L$ & $r_\peri$ & $r_\apo$ \\
    \hline
    $1000$ & $4000$ & $13$ & $46$\\
    $1000$ & $\sim 0$ & $\sim 0$ & $12$\\
    $0$ & $4000$ & $20.3$ & $20.3$\\
    \hline
  \end{tabular}
\end{center}
  \label{tab:isochrone-extrema}
\end{table}

The algebraic forms of both $\vO$ and $\hessian$ are untidy, and
little progress can be made by simply inspecting them. Instead, we
proceed by working numerically with a specific example.
We use the isochrone potential with the parameters
$M=2.852 \ttp {11} \msun$, $b = 3.64\kpc$, which is chosen to
have a rotation curve maximized with $v_c = 240\kms$ at the fiducial
solar radius of $\rsun = 8\kpc$.

What then is the geometry of streams formed in this potential?
\figref{fig:isochrone-hessian} shows the misalignment angle 
between $\hat{\vO}_0$ and $\eigen_1$, given by
\begin{equation}
\vartheta = \arccos \left(\hat{\vO}_0 \cdot \eigen_1\right),
\label{eq:misangle}
\end{equation}
as well as the ratio of the eigenvalues $\lambda_1/\lambda_2$, both as
functions of $\vJ$. The range of $\vJ$ shown in
\figref{fig:isochrone-hessian} covers a variety of interesting orbits,
described in \tabref{tab:isochrone-extrema}.

The upper panel shows that the principal direction of $\hessian$ is
misaligned with the progenitor orbit by $1$--$2\deg$ for all values of
$\vJ$. The misalignment is at a minimum for both low energy and high
energy circular orbits, and at a maximum for eccentric orbits with a
guiding centre close to $r=b$.  The lower panel shows that the ratio
of the eigenvalues $\lambda_1/\lambda_2$ varies from 15 to 35 across
the range, with the ratio maximized for high energy circular orbits,
and minimized for high energy plunging orbits.

On this basis, we expect an isotropic cluster of test particles in this
potential to form a stream in angle-space that is misaligned with
$\vO_0$ by $1$--$2\deg$ and is $15$--$35$ times longer than it is
wide. Notably, such a stream is precluded from being delineated
by any single orbit, close to $\vJ_0$, in the correct isochrone potential.

\subsubsection{Numerical tests}
\label{sec:isochrone-tests}

We created a cluster of 150 test particles, randomly sampled from
a Gaussian distribution in action-angle space, defined by $\sigma_J =
0.2 \kpc\kms$ and $\sigma_\theta = 10^{-3}$ radians. This cluster was
placed on the orbit I1 from \tabref{tab:orbits}, which has
an apocentre radius of $\sim13\kpc$ and a pericentre radius of
$\sim 5\kpc$.

\begin{figure}
  \centerline{
    \includegraphics[width=\squarefigshrink\hsize]{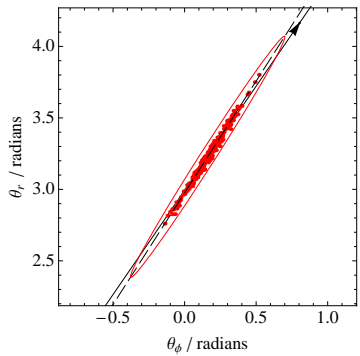}
  }
  \caption{The angle-space distribution of a cluster of test particles,
    evolved on orbit I1 in the isochrone potential of
    \secref{sec:isochrone-tests} for 94 complete azimuthal
    circulations. The particles are shown at apocentre. The frequency
    vector $\vO_0$ of the progenitor orbit is shown with an arrowed
    black line.  The stream is slightly misaligned with $\vO_0$; the
    dashed line is a straight line fit to the positions of the test
    particles, and clearly demonstrates this misalignment. Also
    plotted with a red solid line is the image in angle-space of a
    circle in action-space, mapped by $\hessian$. The shape and
    orientation of the image reflects the $\lambda_n$ and $\eigen_n$
    of $\hessian$ for this orbit. The ellipse is clearly misaligned
    with the underlying orbit, but is perfectly aligned with the
    stream particles.  }
  \label{fig:isochrone-angles}
\end{figure}

The cluster was released at apocentre, and evolved for 94 complete
azimuthal circulations, equal to a period of $t=22.75\Gyr$.
\figref{fig:isochrone-angles}
shows the angle-space configuration of the particles
after this time. The arrowed, black line shows the orbit of the underlying
cluster, $\vO_0$. The dashed line shows a straight line fit to
the distribution of particles, which is clearly misaligned
with the black line. We further note that, unlike
the stream in the Kepler potential shown in 
\figref{fig:kepler-angles}, this stream has clearly
increased in width since inception.

We can predict the shape of this distribution precisely. Plotted as a red
ellipse in \figref{fig:isochrone-angles} is the angle-space image under
$\hessian$ of a circle in action-space. After some time, the angle-space
distribution of an isotropic cluster of test particles should take the form
of a scaled version of this image. We see that the image and the particle
distribution are indeed comparable, and that the dashed line is perfectly
aligned with the principal axis of the image.

\begin{figure}
  \centerline{
    \includegraphics[width=\squarefigshrink\hsize]{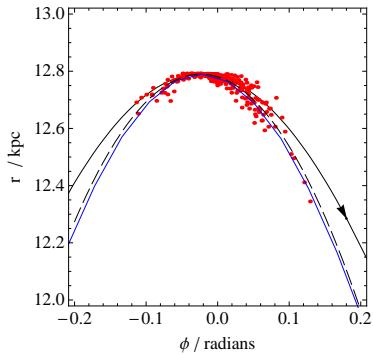}
  }
  \caption{ Real-space configuration for the
    stream of test particles shown in \figref{fig:isochrone-angles},
    plotted in polar coordinates.  The arrowed black line shows the
    trajectory of the progenitor orbit.  The stream formed by the dots
    falls away in radius faster than does the orbit, in both forwards
    and backwards directions: the stream clearly does not follow the
    progenitor orbit. The dashed line is a quadratic curve
    least-squares fitted to the stream, which shows that the stream
    has a substantially greater curvature than does the underlying
    orbit. The solid blue line is the track predicted by mapping the
    dashed line from \figref{fig:isochrone-angles} into real-space: it
    agrees perfectly with the stream.}
  \label{fig:isochrone}
\end{figure}

How does this misalignment manifest itself in real-space?
\figref{fig:isochrone} shows the real-space configuration of the
cluster at the end of the simulation. The orbit of the cluster is
drawn with a solid black line, while the dashed line is a quadratic
curve that has been least-squares fitted to the particle data.

The progenitor orbit is clearly a poor representation of the
stream. Although the orbit passes through the centroid of the stream,
as expected, the curvature of the orbit is too low to match the stream
adequately. Thus, the small misalignment in angle-space is manifest as
a significant change in stream curvature at apocentre.

Since we know the orientation of the long axis of the angle-space
distribution for this stream, we can predict the stream track in
real-space. The blue line in \figref{fig:isochrone} shows the mapping
into real-space of the dashed line from
\figref{fig:isochrone-angles}. The mapping is done by solving
numerically for the real-space roots of the equations that relate
action-angle variables to real-space coordinates: for spherical
potentials, the appropriate equations are given in \S3.5.2 of
BT08. \figref{fig:isochrone} shows that, unlike the progenitor
orbit, this mapped line delineates the stream perfectly.

Although we have only considered the unrealistic case of a stream
formed from an isotropic cluster, we have nonetheless produced
a stream which is misaligned from its progenitor orbit in both
angle-space and real-space. 

We conclude that, in the isochrone potential, the track of the stream
makes a poor proxy for the orbit of its stars, and that in general,
streams in this potential cannot be relied upon to delineate orbits.

\section{Mapping streams from action-angle space to real space}
\label{sec:mapping}

The example of \figref{fig:isochrone} shows that an angle-space misalignment
can alter the curvature of the real-space stream.  In this section, we
examine why this is so, and show the range of real-space effects that are a
consequence of angle-space misalignment. We will also examine the effects of
a finite action-space distribution on the real-space track of a stream.

\subsection{Non-isotropic clusters}
\label{sec:nonisotropic}

Up until this point, we have considered our streams to form from a
cluster of particles that is isotropic in $\vJ$, resulting in a stream
that is perfectly aligned in angle-space with the principal direction
of $\hessian$.

It is not obvious that this is a fair assumption. The structure in
angle-space, given by equations \blankeqref{eq:angle_t}
and \blankeqref{eq:d-dot-j}, is linearly dependent
upon the action-space distribution that generates it. By properly
choosing that distribution, we can create streams of arbitrary shape
in angle-space.
Nature does not create
clusters with arbitrary action-space distributions, so
arbitrary-shaped streams do not emerge. But what kind of action-space
distribution should be considered reasonable?

In \secref{sec:actions} below, we will investigate the action-space
distribution of real clusters, by means of N-body simulation. Here, we
only require a qualitative understanding, in order to know what kind
of gross action-angle structures we should learn to map. Since the action-space
distribution fundamentally arises from the random motion of stars within the cluster,
it is unlikely to be dominated by complex structure. We
will therefore assume that the gross structure is ellipsoidal. But what
should be the axis ratio of this ellipse?

Consider a cluster with isotropic velocity dispersion, $\sigma$, that
is on an orbit with apocentre $r_\apo$ and pericentre $r_\peri$, where it is
tidally disrupted. In our spherical system, the radial action is
given by the closed integral
\begin{equation}
J_r = {1 \over 2\pi} \oint p_r \, dr,
\label{eq:jr}
\end{equation} 
where the integration path is one complete radial oscillation.
Now consider a particle whose radial momentum $p_r$ differs from that of the
cluster average by $\delta p_r \sim \sigma$. 
We can take a finite difference over \eqref{eq:jr} and thus obtain an expression for the
the difference in radial action between the particle and the cluster
\begin{equation}
\delta J_r \sim {1 \over \pi} \delta p_r \Delta r \sim{1\over\pi} \sigma \Delta r,
\label{eq:deltajr}
\end{equation}
where $\Delta r = (r_\apo - r_\peri)$ is the amplitude of the radial oscillation.
Now consider another particle, whose azimuthal velocity differs from that of the
cluster by $\delta v_t \sim \sigma$. The difference in angular momentum
between this particle and the cluster is
\begin{equation}
\delta L \sim r_{\rm p} \, \delta v_t \sim r_{\rm p} \sigma,
\end{equation}
where we have performed our calculation at pericentre,
because that is where the cluster is stripped.
For the purposes of this section, we are interested in the relative size
of the spread in radial action $\Delta J_r$ and the spread in
angular momentum $\Delta L$ for a disrupting cluster. We see that their
quotient
\begin{equation}
{\Delta J_r \over \Delta L} \sim {\Delta r \over \pi r_{\rm p}}.
\label{eq:djr/dl}
\end{equation}
Although this ratio will take on every value between $(0, \infty)$ as
we move from a circular orbit to a plunging one, for the orbits likely
to be occupied by stream-forming clusters, it will typically be within
an order-of-magnitude of unity.

\begin{figure}
  \centering{
    \includegraphics[width=\squarefigshrink\hsize]{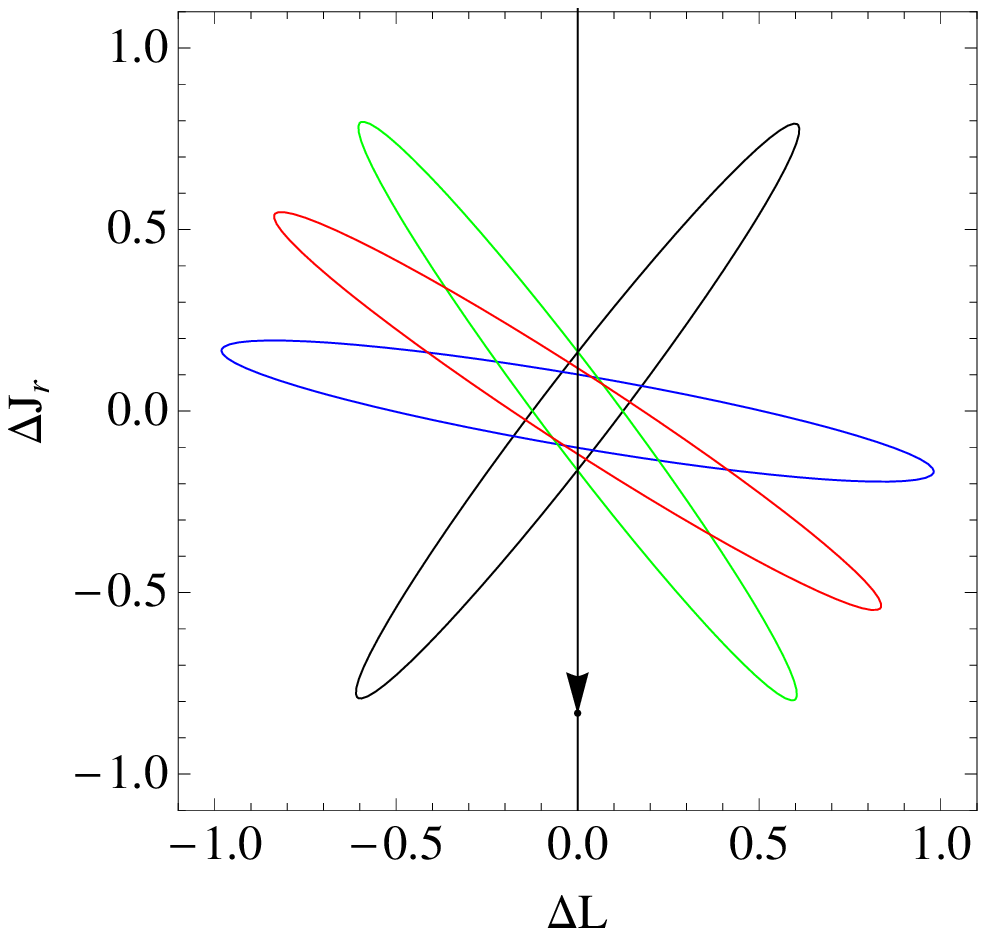}\\
    \includegraphics[width=\squarefigshrink\hsize]{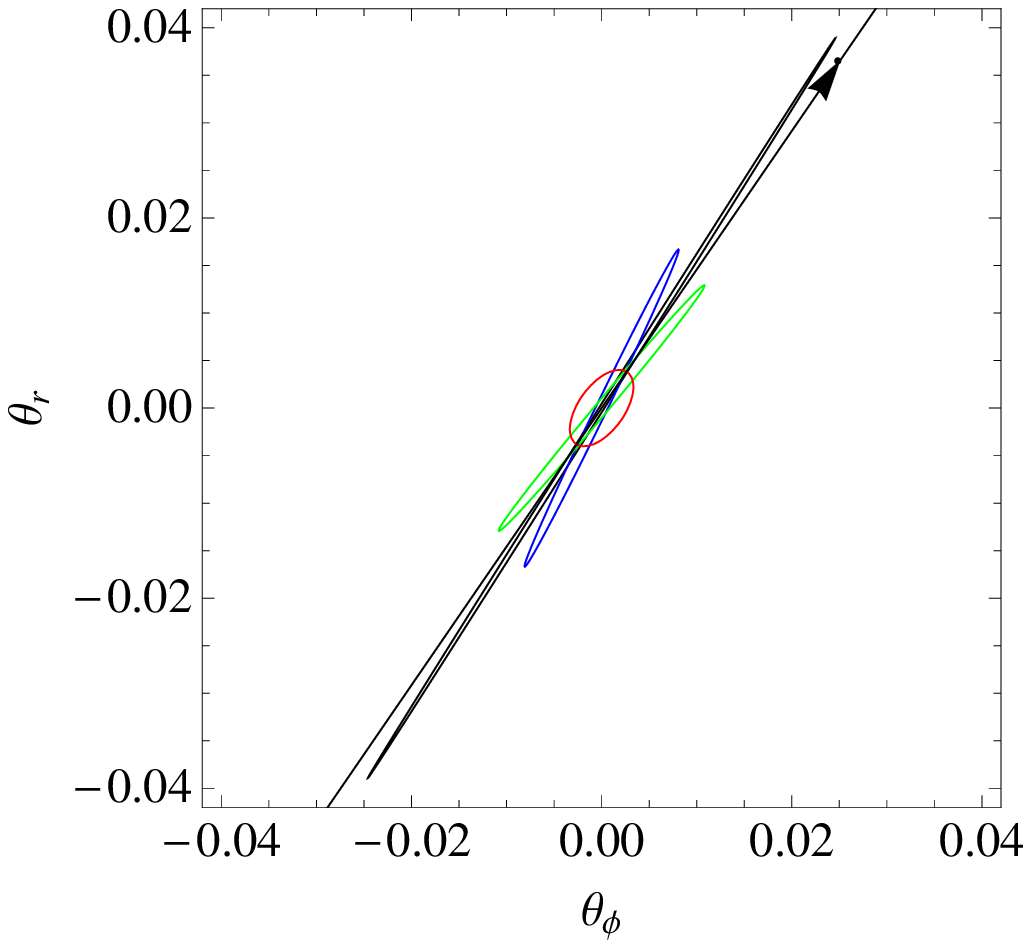}
  }
  \caption
{The upper panel shows a selection of ellipses in action-space,
    each of axis ratio 10, but oriented in different directions.
    The lower panel shows the image in angle-space that results from
    mapping each of the action-space ellipses with the linear map $\hessian$,
    calculated for the orbit I1 in the isochrone potential of 
    \secref{sec:isochrone-tests}. In the lower panel, the arrowed black
    line is the frequency vector $\vO_0$; in the upper panel, the arrowed
    black line is the inverse map of the frequency vector, $\hessian^{-1}
    \vO_0$.
    We see that regardless of the shape in action-space, the 
    mapped images are all elongated and roughly aligned with the
    orbit, although the alignment is generally not perfect.
    In this example, the misalignment of each of the red and green images
    is about $10\deg$.
  }
  \label{fig:isochrone-ellipses}
\end{figure}

The upper panel of \figref{fig:isochrone-ellipses} shows a set of
ellipses, with axis ratio 10, placed at various orientations in
action-space: an axis ratio of 10 would not be considered untypical by
the arguments of the previous paragraph. The lower panel of
\figref{fig:isochrone-ellipses} shows the images under $\hessian$ of these ellipses in
angle-space when evaluated on orbit
I1 in the isochrone potential of the previous section.   All
the images are both elongated and roughly oriented towards the
principal direction.  We conclude that the images of most action-space
ellipses under this map---and thus, most streams formed in this
potential---would be highly elongated and oriented to within a few
degrees of the principal direction, which is itself oriented to within
a few degrees of the frequency vector $\vO_0$.

Since the ratio of the eigenvalues for this orbit is $\sim 17$, it is
not possible to produce an image in angle-space that is not elongated
towards the principal direction, by mapping an action-space ellipse of
axis-ratio 10.  We note from \figref{fig:isochrone-hessian} that,
for this potential, the ratio of eigenvalues does not vary much, and
nor does the principal direction stray from $\vO_0$ by more than a few
degrees. We therefore conclude that reasonable action-space
distributions will always result in the formation of streams in this
potential, and that such streams will always be oriented in angle-space to
within a few degrees of $\vO_0$.

\subsection{The mapping of action-angle space to real space}
\label{sec:map}

\begin{figure*}
  \centerline{
    \includegraphics[width=\doubsquarefigshrink\hsize]{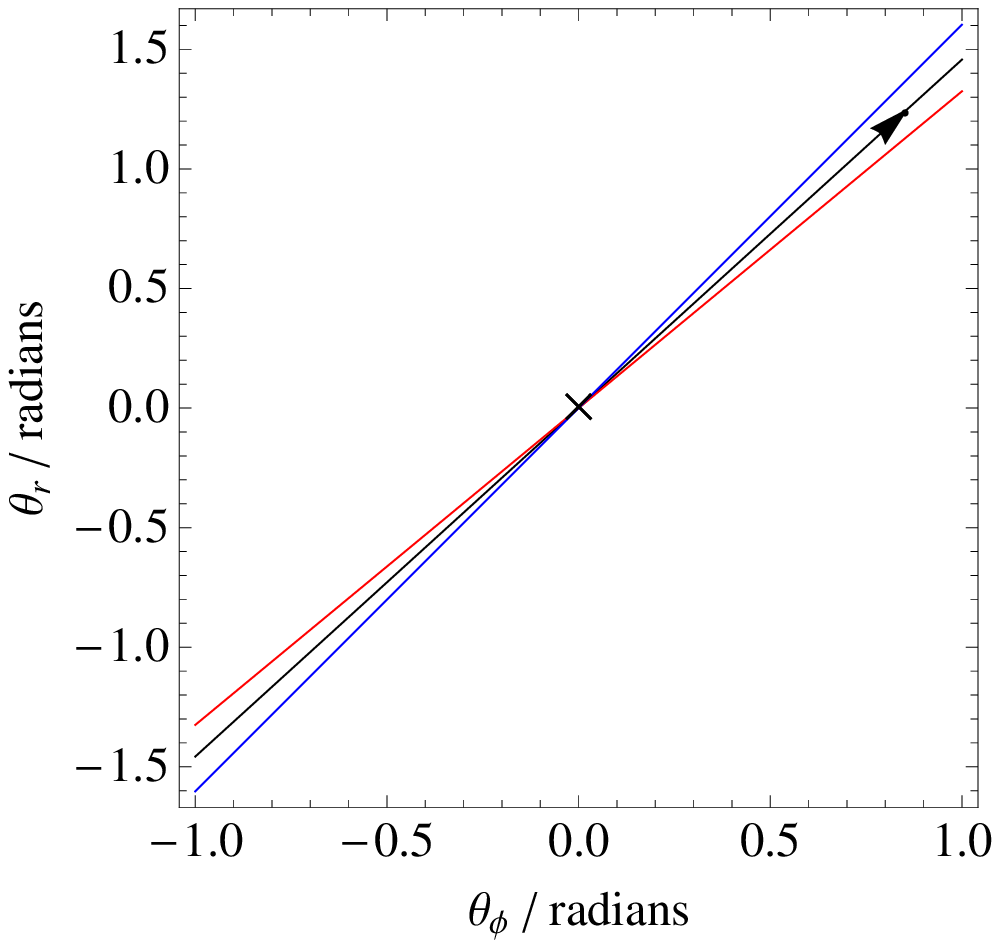}
    \includegraphics[width=\doubsquarefigshrink\hsize]{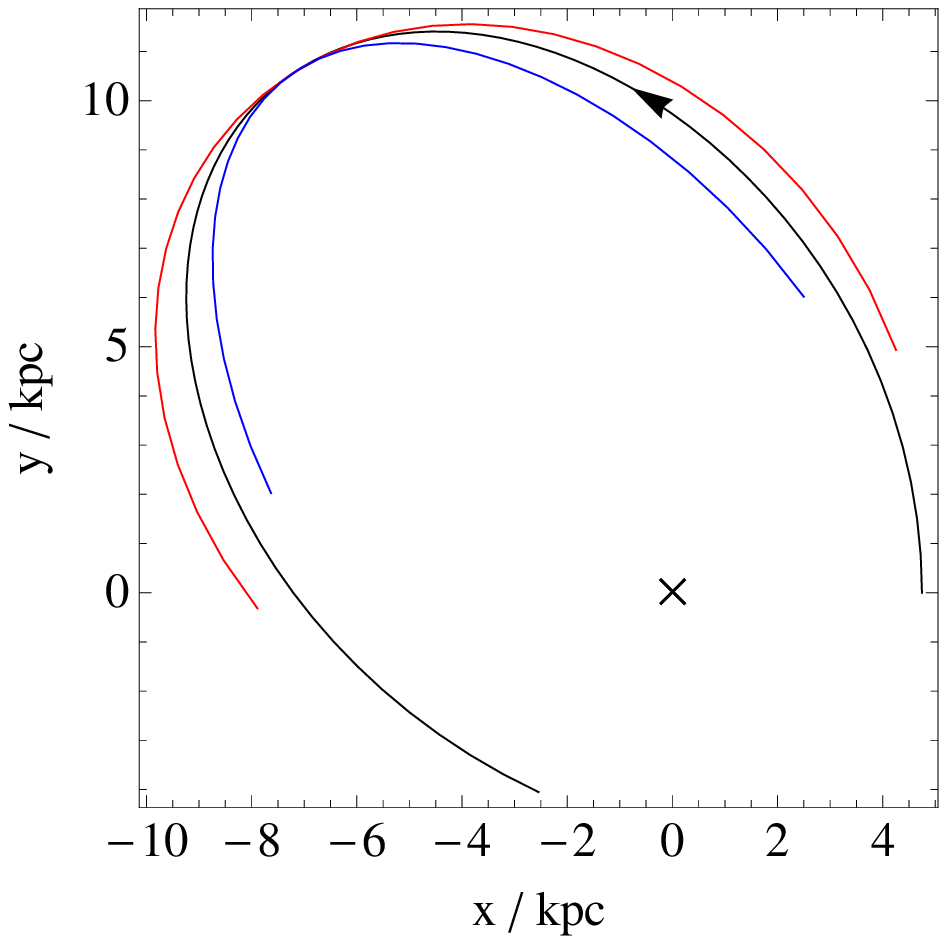}
  }
  \centerline{
    \includegraphics[width=\doubsquarefigshrink\hsize]{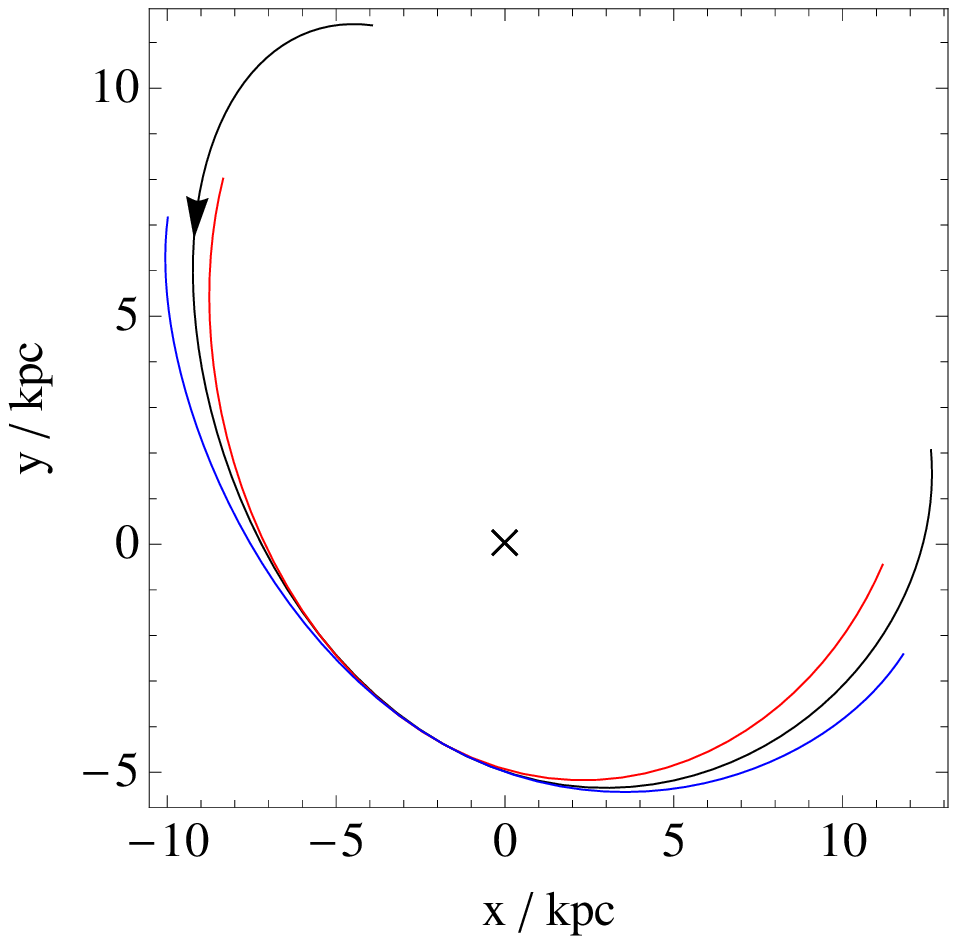}
    \includegraphics[width=\doubsquarefigshrink\hsize]{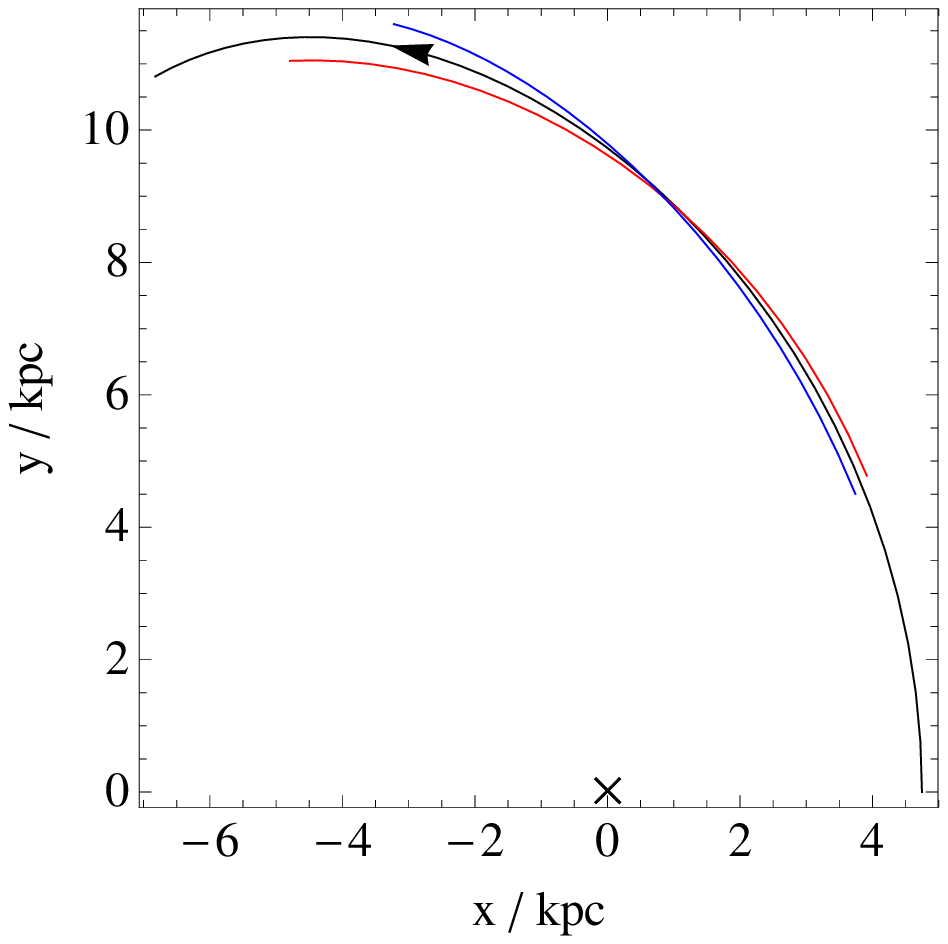}
  }
  \caption{ The upper-left panel shows three trajectories in angle
    space.  The black line is the trajectory of orbit I1 in the
    isochrone potential of \secref{sec:isochrone-tests}. The red line
    has a frequency ratio $\Omega_r / \Omega_\phi$ that is $10
    \percent$ lower than I1. The blue line has a frequency ratio that
    is $10 \percent$ higher than I1. The lines are phase-matched at
    their point of intersection. We note that the
    red line has retarded radial phase (relative to the black line) on
    the leading tail, and advanced radial phase on the trailing
    tail. Conversely, the blue line has advanced radial phase in the
    leading part, and retarded radial phase in the trailing part.  The
    other three panels show the real-space images of these three
    lines, when phase-matched: (upper-right) near apocentre,
    (lower-left) near pericentre, and (lower-right) at a point far
    from apsis. In each of these three panels, the centroid of the
    potential is marked with a a cross.}
  \label{fig:mapping}
\end{figure*}

The upper-left panel of \figref{fig:mapping} shows three trajectories
in angle space. The black line is the trajectory of I1 in the
isochrone potential given in \secref{sec:isochrone-tests}. The red
line has a frequency ratio $\Omega_r / \Omega_\phi$ that is $10
\percent$ lower than the black line, and is therefore rotated from it
by  $\sim2.9\deg$.  Conversely, the blue line has a frequency
ratio that is $10 \percent$ higher than the black line, and is
therefore rotated from it by $\sim2.5\deg$.

The red and blue lines were chosen to represent likely streams in angle-space
that could form in the isochrone potential, given the results of the
previous section. We note that the red line has retarded radial phase (relative to the
black line) on the leading tail, and advanced radial phase on the
trailing tail. Conversely, the blue line has advanced radial phase in
the leading part, and retarded radial phase in the trailing part.


How do these lines map into real-space? The upper-right panel
of \figref{fig:mapping}
shows the real-space curve obtained from the lines in
the upper-left panel of that figure, having chosen the point of
intersection such that the lines are phase-matched near apocentre.
The curves in the upper-right panel have been
drawn by assuming that all points along each line have the same $\vJ$.
Thus, this plot represents the real-space curves of streams oriented
in angle-space according to the upper-left panel, but
formed from clusters of vanishingly small $\Delta\vJ$.

In the upper-right panel, we see that the red line has
systematically lower curvature than the black line. Conversely, the
blue line has systematically greater curvature than the black line.
This is because the red curve has retarded radial phase
on the leading tail, and advanced radial phase on the trailing tail,
and thus is flattened with respect to the orbit. Similarly,
the blue line has advanced radial phase on the leading tail, and
retarded radial phase on the trailing tail, and thus appears
curved with respect to the orbit.

The lower-left panel of \figref{fig:mapping} shows the same lines, but now
phase-matched at pericentre. Similarly to the upper-right panel, the red line
again appears less curved with respect to the orbit, and the blue line
appears more curved with respect to the orbit. The lower-right panel of that
figure also shows the same lines, phase-matched at a point well away from
apsis. In this case, a misalignment between the stream and $\vO_0$ in
angle-space manifests itself as a real-space directional misalignment, rather
than a curvature error as at apsis.  This occurs because, unlike at apsis,
the mapping between angle-space and real-space plane polar coordinates is
relatively undistorted near this point.

In summary, realistic angle-space distributions in an isochrone potential,
which may be misaligned with the progenitor frequency by a few degrees,
will produce real-space streams that exhibit differing curvature
from the progenitor orbit when observed at apsis, and differing
directional alignment from the orbit when observed far away from apsis.

\subsection{Are trajectories insensitive to small changes in $\vJ$?}
\label{sec:trajectory-j}

To this point, all real-space tracks have been derived from streams in
angle-space under the assumption that at all points along that stream the
action is that of the progenitor, $\vJ_0$.  This assumption is only strictly
valid in the case of a vanishingly small action-space distribution, and for
asymptotically large time since disruption of the cluster. If a mapping into
real-space from a line in angle-space is made under this assumption, then a
stream generated by a sufficiently broad action-space distribution will not
be accurately represented, even though the representation in angle-space may
be exact.  This is because the small changes in action that give rise to the
small changes in frequency also cause small changes in real-space trajectory
as well.

When computing a stream track in real-space, it is possible to
correct for this effect. By inverting \eqref{eq:d-dot-j}
and eliminating $\Delta\vO$ using \eqref{eq:angle_t}, we find
that for a star separated from a fiducial point on the stream by angle
$\delta\vT$, the difference in action between the star and the fiducial
point, $\delta\vJ$ is given by
\begin{equation}
\delta\vJ = {1 \over t_\d} \hessian^{-1} \, \delta\vT,
\label{eq:correction}
\end{equation}
where $t_\d$ is the time since the star and the fiducial point were
coincident.
We may therefore guess the correction $\delta \vJ$ for a star's true action
$(\vJ_0 + \delta \vJ)$ from its position in the stream, provided we know $t_\d$.

We typically take the fiducial point to be the centroid of the stream,
following which we may assume $t_\d$ to be the time since the first
pericentre passage of the cluster on its present orbit. Although this
assumption neglects the possibility that the star could have been torn
away during a subsequent pericentre passage, we note that during tidal
disruption, it is the fastest moving stars which become unbound. The
cluster core that remains after a pericentre passage therefore has
lower velocity dispersion. Stars subsequently torn from that cluster
will therefore have a smaller distribution in action-space.
Consequently, the stars with the largest $\delta\vT$ from the
centroid---i.e.~those for which the $\delta\vJ$ correction will be
most important---must have been torn away at the earliest time, and so
the assumption that $t_\d$ equals the time since the first pericentre
passage remains good.

 Just how important is this effect? For small changes in $J_r$, the
trajectory changes we discuss are expressed as changes in the radial
amplitude $\Delta r$, while the guiding-centre radius $r_\g$, which
is purely a function of $L$, is held constant. We
can estimate the magnitude of the effect as follows. 

Consider a cluster on an
orbit that is close to circular, whose radial action is given by \eqref{eq:jr}.
Orbital energy $E$ is conserved, so close to apsis $r=r_0$, the
radial momentum $p_r$ is given by
\begin{align}
E &=  {1 \over 2} p_r^2(r) + \Phi_{\rm eff}(r)\nonumber\\
&\simeq {1 \over 2} p_r^2(r) + \Phi_{\rm eff}(r_0)
+ (r - r_0) \left.{\d \Phi_{\rm eff} \over \d r }\right|_{r_0},
\label{eq:p-near-apsis}
\end{align}
where we have defined the effective potential
\begin{equation}
\Phi_{\rm eff}(r)
= \Phi(r) + L^2/2r^2.
\end{equation}
Since at apsis $p_r = 0$,
\begin{equation}
E = \Phi_{\rm eff}(r_0).
\end{equation}
 Hence from \eqref{eq:p-near-apsis} we have
\begin{equation}
p_r(r) = \sqrt{-2(r - r_0)\left.{\d \Phi_{\rm eff} \over \d r }\right|_{r_0}}
= \sqrt{2(r-r_0)F_{\rm eff}(r_0)},
\end{equation}
where we have defined the effective force,
\begin{equation}
F_{\rm eff}(r') = -\partial\Phi_{\rm eff}/\partial r |_{r'}.
\end{equation} 
If $(r_\a,r_{\rm p})$ are apocentre and pericentre respectively, then
\begin{equation}
p_r(r) \simeq
\begin{cases}
\sqrt{2|F_\eff(r_\a)|(r_\a - r)} & \text{if } r \simeq r_\a, \\
\sqrt{2|F_\eff(r_{\rm p})|(r - r_{\rm p})} & \text{if } r \simeq r_{\rm p}. 
\end{cases}
\end{equation}
Hence, we define a global approximation to $p_r$
\begin{equation}
\tilde{p_r}(r) = { \sqrt{2\left|F_{\rm eff}\right|(r - r_{\rm p})(r_\a - r)} \over
\sqrt{\Delta r}},
\label{eq:prapprox}
\end{equation}
 where $\Delta r=r_\a - r_{\rm p}$ and $F_{\rm eff}$ a constant set equal to
the value of $F_{\rm eff}(r)$ at either apsis, since we assume it
takes approximately the same value at both. We note that
 \begin{equation}
\int_{r_{\rm p}}^{r_\a}\!\! \sqrt{(r - r_{\rm p})(r_\a - r)}\,\d r = {\pi \over 8} \Delta r^2,
\end{equation}
 so when combined with \eqref{eq:prapprox}, \eqref{eq:jr} for $J_r$ yields
 \begin{equation}
J_r = {1 \over 8}\sqrt{2\left|F_{\rm eff}\right| \Delta r^3}.
\label{eq:bodgejr}
\end{equation}
We can deduce the value of $F_{\rm eff}$ as follows. We note that
\begin{equation}
\Phi_\eff = \Phi + {L^2 \over 2r^2},
\end{equation}
and that its derivative is
\begin{equation}
{\d \Phi_\eff \over \d r} = {\d \Phi \over \d r} - {L^2 \over r^3}.
\end{equation}
If the rotation curve is relatively flat, then $F_\eff(r)$ evaluated
at $r_\a \simeq r_\g + \Delta r/2$ is
\begin{align}
F_\eff &= \left.{\d \Phi_\eff \over \d r}\right|_{r_\a}
= {v_c^2 \over r} - {L^2 \over (r_\g + {1 \over 2}\Delta r)^3}\nonumber\\
&\simeq {v_c^2 \over r_\g}\left(1 - {\Delta r \over r_\g}\right) - {L^2 \over r_\g^3}
\left(1 - {3 \over 2}{\Delta r \over r_\g}\right)\nonumber\\
& = -\Delta r \left({v_c^2 \over r_\g^2} + {3 L^2 \over 2 r_\g^4}\right)
= -{5 v_c^4 \Delta r \over 2 L^2},
\end{align}
and similarly when evaluated at $r_{\rm p}$, but with opposite sign.
\Eqref{eq:bodgejr} then becomes
\begin{equation}
J_r \simeq {\sqrt{5} v_c^2 \Delta r^2 \over 8 L}.
\label{eq:jrchange}
\end{equation}
Differentiating the above expression, we find
\begin{align}
{\d J_r \over \d \Delta r} &= 
{\sqrt{5} v_c^2 \Delta r \over 4L}\nonumber\\
&=\sqrt{\sqrt{5} J_r v_c^2 \over 2L}.
\label{eq:amplitude}
\end{align}
Hence, for a small change $\delta J_r$ we can estimate the corresponding change
in the radial amplitude, $\delta\Delta r$, which is likely to be a good estimate
for the positional error we would make in assuming that a star with
action $J_r + \delta J_r$ actually had action $J_r$.

\begin{figure*}
  \centering{
    \includegraphics[width=\doubsquarefigshrink\hsize]{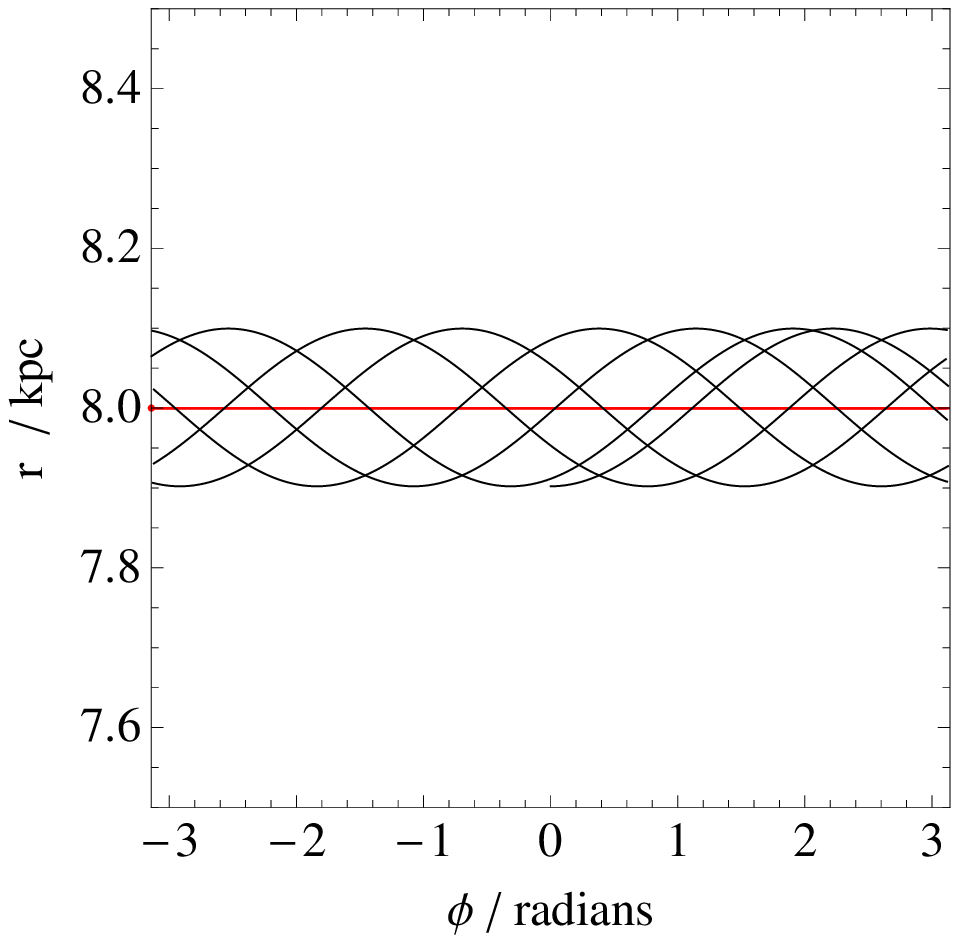}
    \includegraphics[width=\doubsquarefigshrink\hsize]{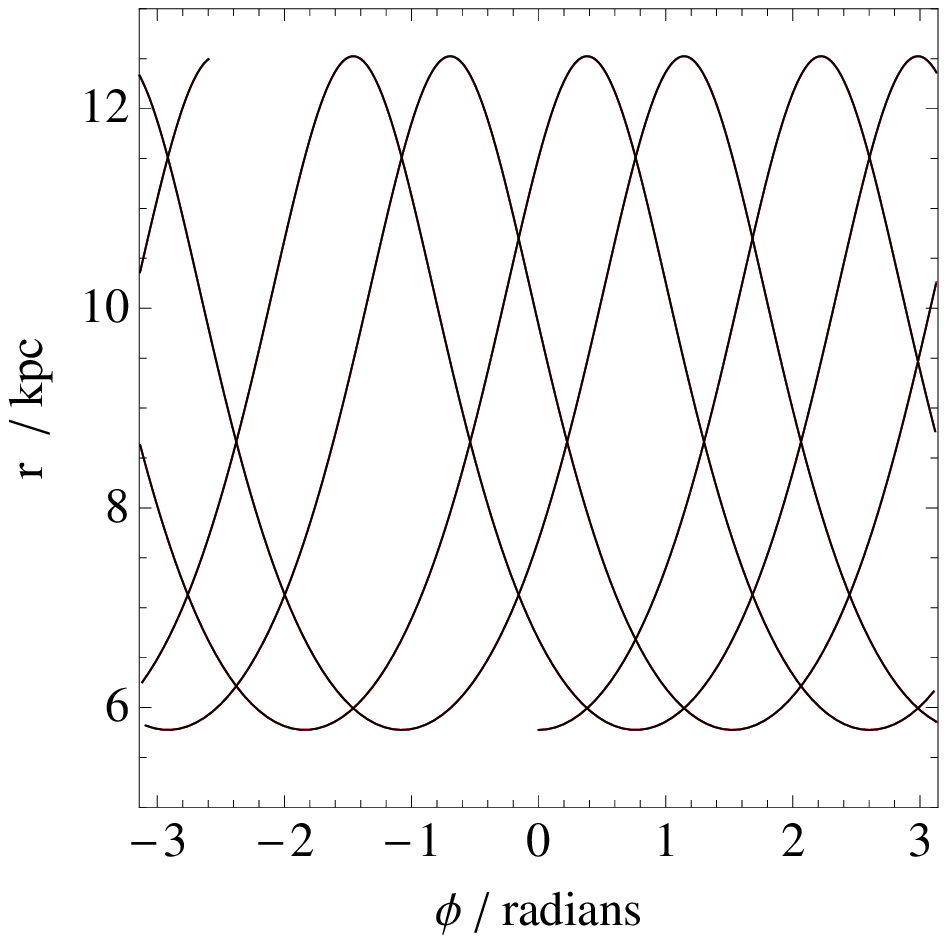}
    \includegraphics[width=\doubsquarefigshrink\hsize]{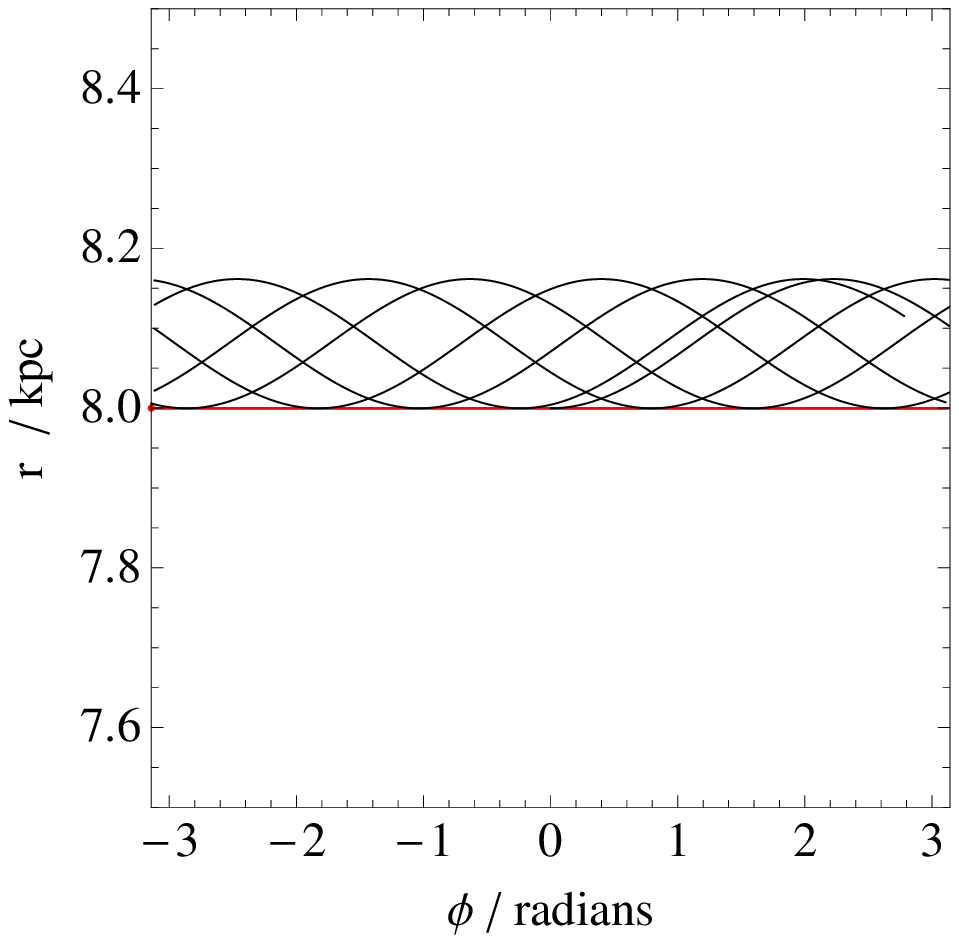}
  }
  \caption{The effect of small changes in actions
    on real-space orbital trajectories. The left panel shows that a
    small change in $\jr$ ($\delta \jr = 0.21 \kms\kpc$) to a circular
    orbit (I2 in \tabref{tab:orbits} shown in red) produces a change of
    $\delta\Delta r\sim 0.2\kpc$ in radial amplitude. The centre panel
    shows the effect of the same perturbation $\delta\jr$ on an
    eccentric orbit (I3 in \tabref{tab:orbits}) with the same angular
    momentum as the circular orbit, but with $\jr = 207\kms\kpc$. The
    effect on radial amplitude is so small as to be invisible in this
    plot.  The right panel shows the effect of an increment in angular
    momentum $L$ of $1\percent$ on the trajectory of the orbit I2. In
    this case, the radial amplitude changes by $\delta\Delta r\sim
    0.15\kpc$.  }
  \label{fig:j-trajectory}
\end{figure*}

We can confirm the predictions of the above equations numerically. The left panel of
\figref{fig:j-trajectory} shows the real-space trajectory
of the circular orbit I2, with radius $r=8\kpc$, in the isochrone potential of
\secref{sec:isochrone-tests}. Also plotted is the trajectory
of an orbit that has identical $L$ to I2, but $J_r = 0.21\kms\kpc$.

Clearly, \eqref{eq:amplitude} ceases to have meaning when faced with orbits
very close to circular, so we rely for our estimate on the integral form
\eqref{eq:jrchange} instead.  \Eqref{eq:jrchange} predicts $\Delta r =
0.16\kpc$ for this perturbation from circular, which appears from the left
panel of \figref{fig:j-trajectory} to be a reasonable estimate.

From \eqref{eq:amplitude} we see that the magnitude of the effect
diminishes with increasing $J_r$ as $\delta\Delta r \sim 1/\Delta r \sim 1/\sqrt{J_r}$. The
centre panel of \figref{fig:j-trajectory} shows this to be the
case. The panel shows two trajectories in the isochrone potential: one
for the orbit I3, and one for the same orbit with $J_r$ incremented by
$0.1\percent$. \Eqref{eq:amplitude} predicts a change $\delta\Delta r
\sim 3\pc$. Close inspection of the trajectories confirms an actual
$\delta\Delta r \sim 3\pc$, so the prediction is correct, but as is
clear from \figref{fig:j-trajectory}, corrections of such magnitude
are negligible.

What follows is an estimate for the positional error made by incorrectly
guessing $L$. Consider again a cluster on an orbit close
to circular. The angular momentum of the cluster is related to the
guiding-centre radius $r_\g$ and the circular velocity $v_c$ by
\begin{equation}
L = v_c r_\g.
\end{equation}
Consider now a star whose angular momentum is suddenly reduced by $\delta L$.
This star is now at apocentre, since its guiding-centre radius has been 
reduced by
\begin{equation}
\delta r_\g = {\delta L \over v_c},
\label{eq:dLdrg}
\end{equation}
where we have assumed that the rotation curve is flat. The pericentre
radius will have been reduced by of order twice the change in
guiding-centre radius, hence we may write
\begin{equation}
\delta \Delta r = {2 \delta L \over v_c}.
\label{eq:l-amp}
\end{equation}
Thus, we can predict the change in radial amplitude $\delta\Delta r$ for
a small change in angular momentum $\delta L$, which is likely to be a good estimate
for the positional discrepancy we would encounter in assuming that a star with
angular momentum $L + \delta L$ actually had angular momentum $L$.

Again, we can check the predictions of this expression numerically. 
The right panel of \figref{fig:j-trajectory} shows the trajectories of two orbits
in our isochrone potential. The red line is orbit I2, while
the black line is the same orbit, but with the angular momentum increased
by $1\percent$. \Eqref{eq:l-amp} predicts $\Delta r \simeq 0.16\kpc$
for this change. \figref{fig:j-trajectory} shows that this estimate
is close to exact.

The only example we have considered thus far for which this effect could be
of consequence is the isochrone-potential stream shown in
\figref{fig:isochrone}. For this cluster, \eqref{eq:amplitude} predicts that
a positional error of $\sim 2\pc$ would be accrued by assuming all stars have
the same radial action. Similarly, \eqref{eq:l-amp} predicts that a
positional error of $\sim 1.5\pc$ will be accrued by assuming that all stars
have the same angular momentum. These errors are insignificant, so no
corrections are required in this case.

However, the scale of the $\Delta \vJ$ distribution in the example of
\figref{fig:isochrone} was deliberately chosen to be very small, with
a velocity dispersion $\sigma \sim 4 \ttp {-2} \kms$. The
\eqsref{eq:amplitude}{eq:l-amp} show the magnitude of the trajectory
anomaly  rises linearly with the scale of $\Delta \vJ$. For a large
cluster with $\sigma \sim 20 \kms$ on the orbit I2, this would imply
trajectory errors of order $\sim 1\kpc$, which are not
negligible. Therefore, consideration of the effects of a finite
action-space scale on the real-space track of streams is necessary in
practical cases.

In the case of a large cluster, even a stream that is perfectly
aligned with its frequency vector in angle-space will not be
delineated by the progenitor orbit in real-space. The correction
described by \eqref{eq:correction}, to account for variation
in action down the stream will be required to correctly predict the
stream-track from its angle-space distribution. In general we would
not know the time since first pericentre $t_\d$ accurately, although
given $\vJ_0$ and a knowledge of the extent of the stream, we could
make a reasonable guess as to its value. However, even a poor, but
finite, guess for the value of $t_\d$ would likely produce a more
accurate real-space stream track than would assuming $\vJ=\vJ_0$
everywhere along the stream.

\section{The action-space distribution of disrupted clusters}
\label{sec:actions}

To this point, we have relied upon the qualitative estimate from
\secref{sec:nonisotropic} for what the action-space distribution
of a disrupted cluster might be. In this section, we investigate
the action-space distribution of various cluster models using N-body
simulation. We further utilize our N-body models to
confirm the misalignment between streams and orbits, and to demonstrate
that we can accurately predict the real-space track of the stream.

The action-angle coordinates, of the host-galaxy potential, have limited
usefulness when applied to the particles in a bound cluster, because the
actions are not constant with time. Nonetheless, they remain a valid set of
canonical coordinates and can be legitimately used to describe the
phase-space distribution of the cluster. Moreover, we shall see that a
disrupted cluster gives rise to a characteristic distribution in
action-space, from which the track of the stream can be predicted.

\begin{figure}
  \centering{
    \includegraphics[width=\squarefigshrink\hsize]{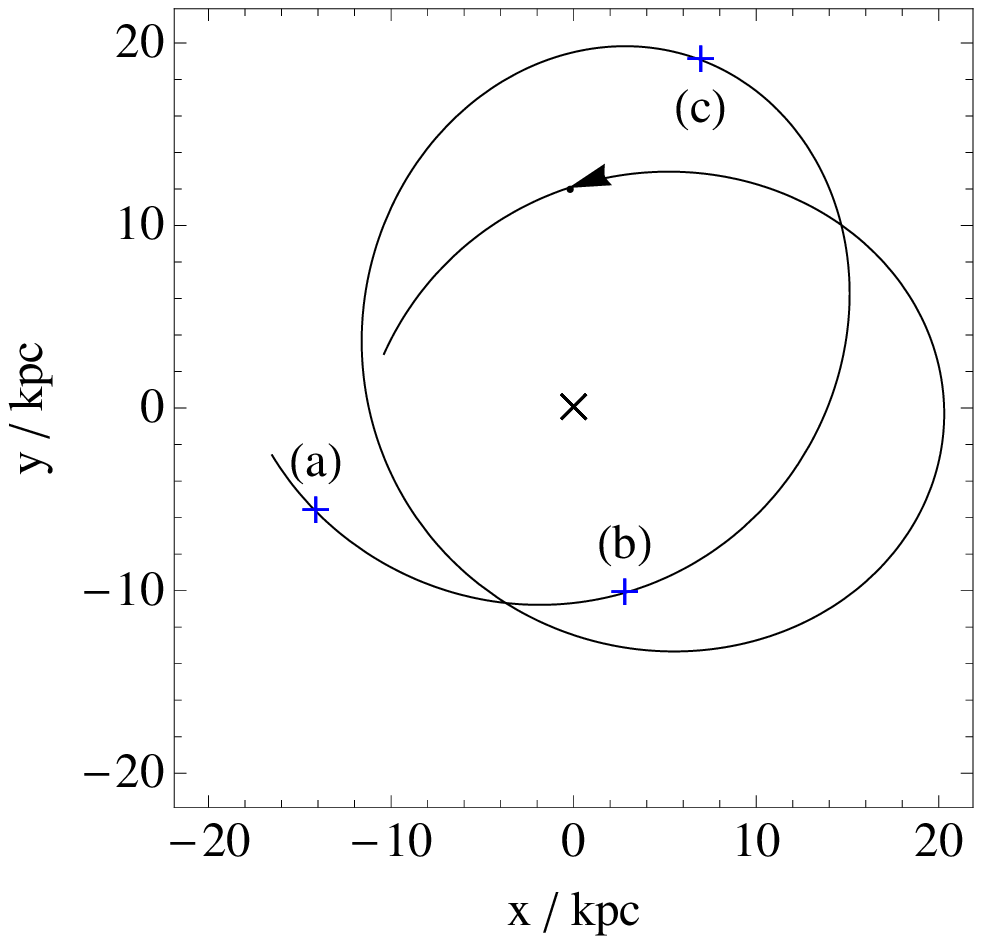}\\
    \includegraphics[width=\squarefigshrink\hsize]{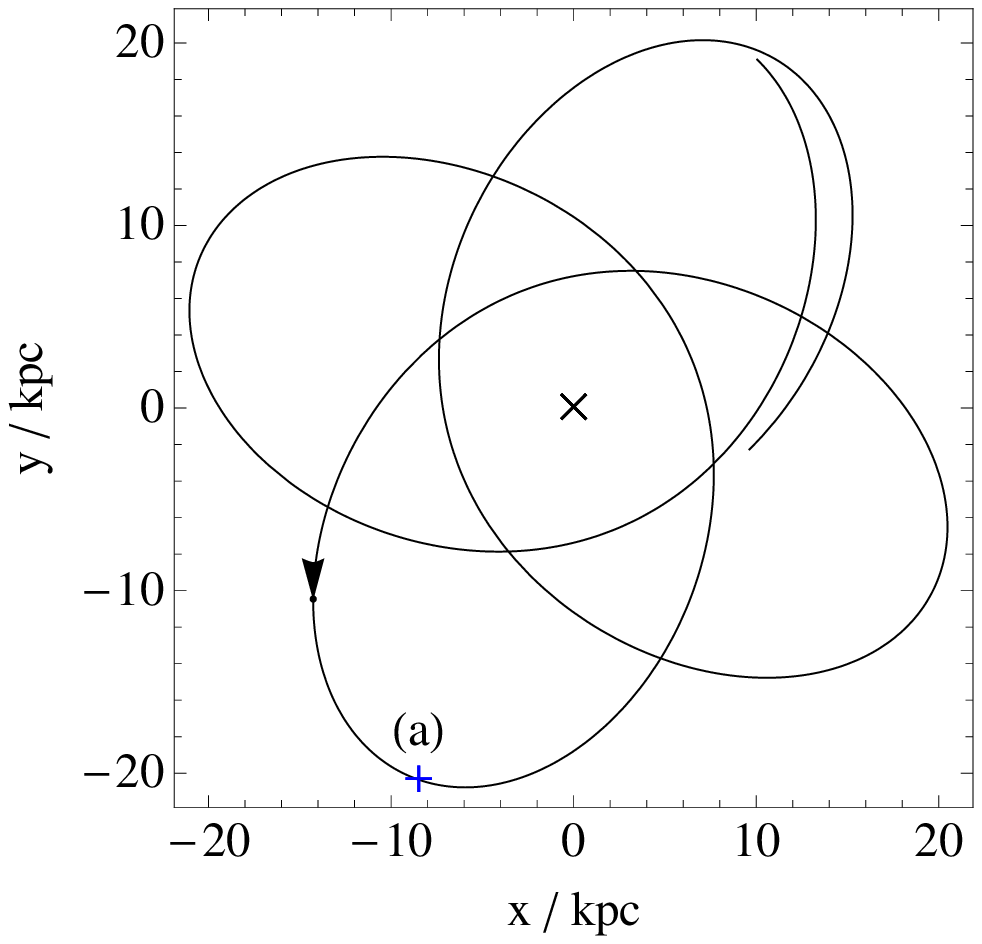}
  }
  \caption
{Plan views of the orbits used in this section. The top
    panel shows I4, with (a), (b) and (c) marking the positions of the
    cluster corresponding to the upper-left, upper-middle and upper-right
    panels of \figref{fig:nbody-run1} respectively. The bottom
    panel shows I5, with (a) marking the position of the cluster
    corresponding to the bottom-right panel of
    \figref{fig:nbody-runs234}. In both panels, the potential in
    use was the isochrone potential described in
    \secref{sec:isochrone-tests}.  }
\label{fig:nbody-orbit-plans}
\end{figure}

\figref{fig:nbody-orbit-plans} shows a segment of each of the
orbits I4 and I5 in the isochrone potential of
\secref{sec:isochrone-tests}. These orbits were chosen to be fairly
representative of those occupied by tidal streams in our Galaxy: they
have apocentre radius $\sim 20\kpc$ and are moderately eccentric to
allow for efficient tidal stripping. For
our investigation, we wish to launch model clusters on each of these
orbits, where the otherwise-stable cluster has been chosen such that
its outermost stars will be torn away by tidal forces when close to
pericentre. The process by which we choose our model clusters is
detailed in the next section.

\begin{figure*}
\centering{
  \includegraphics[width=\doubsquarefigshrink\hsize]{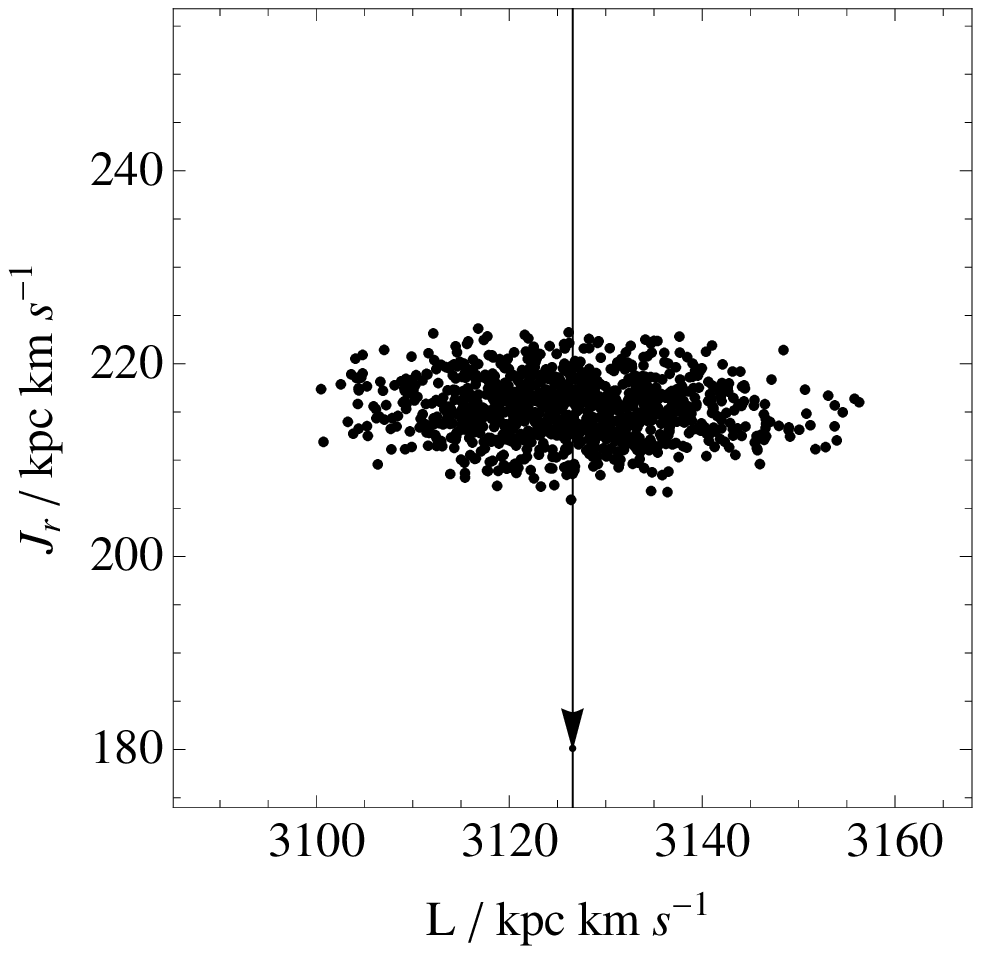}
  \includegraphics[width=\doubsquarefigshrink\hsize]{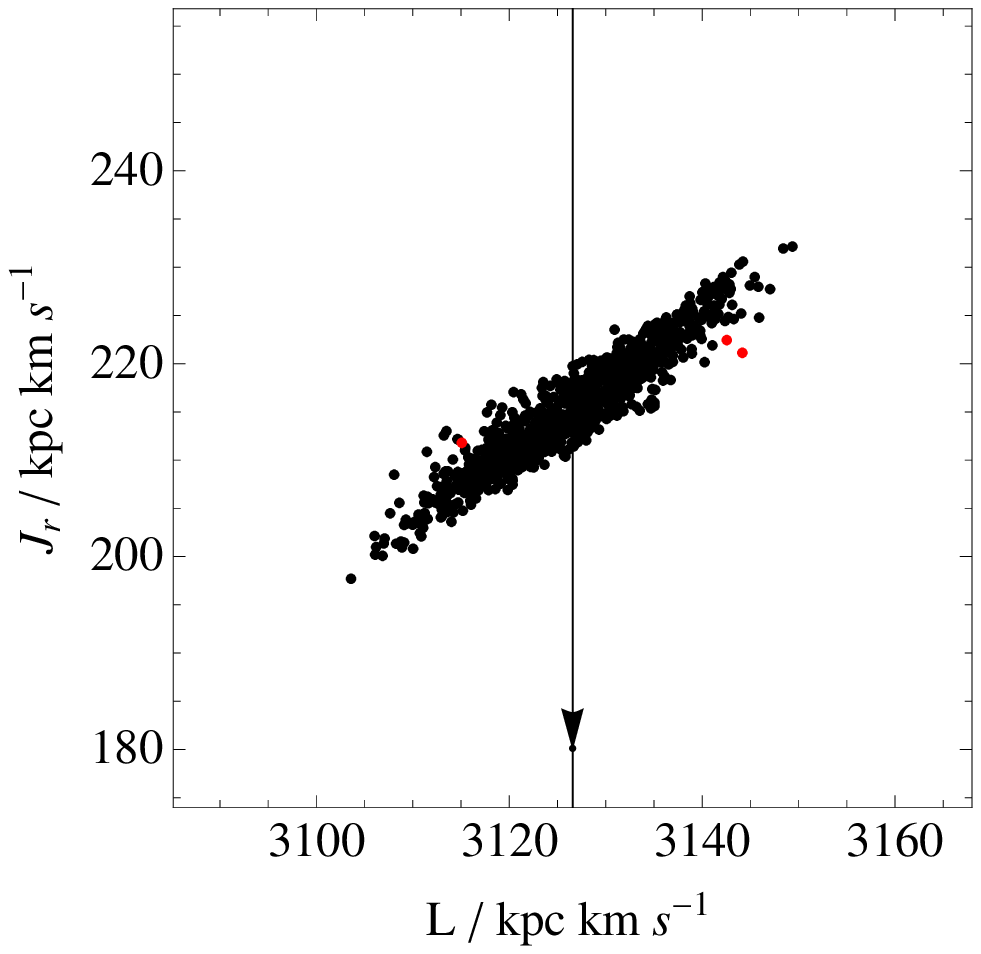}
  \includegraphics[width=\doubsquarefigshrink\hsize]{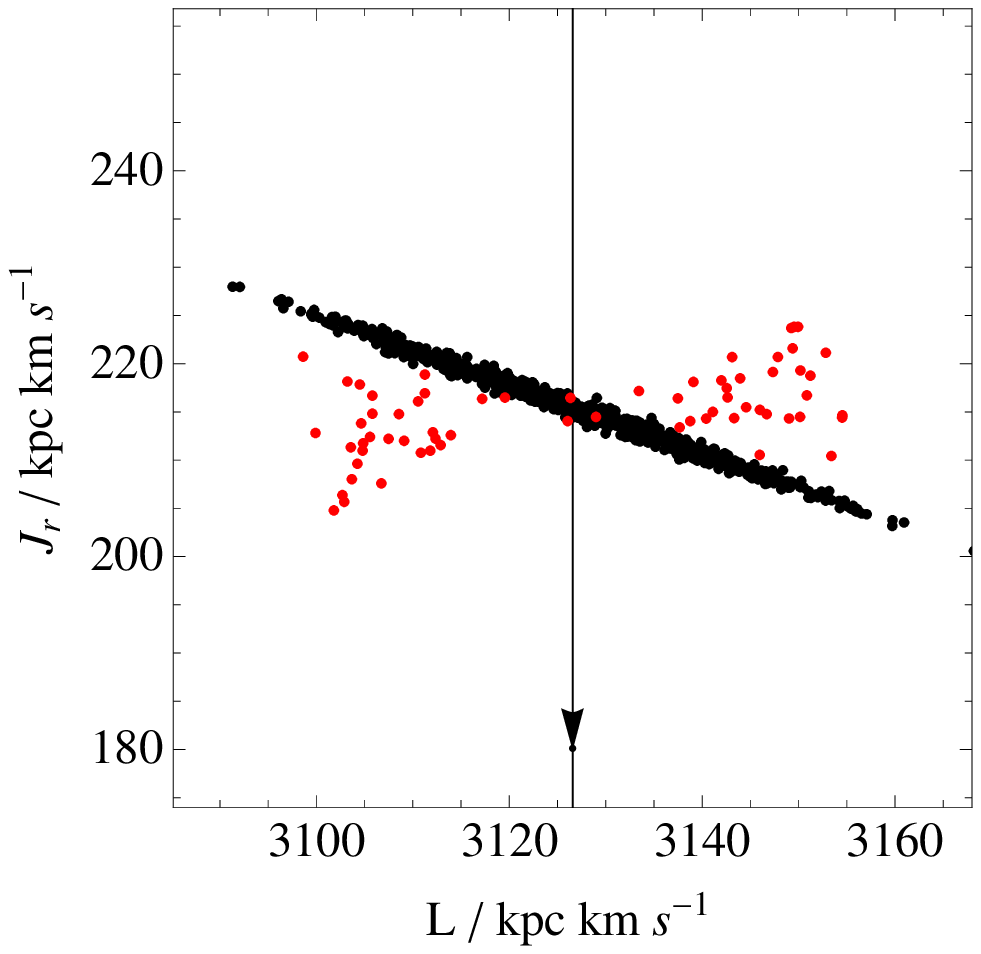}
}
\centering{
    \includegraphics[width=\doubsquarefigshrink\hsize]{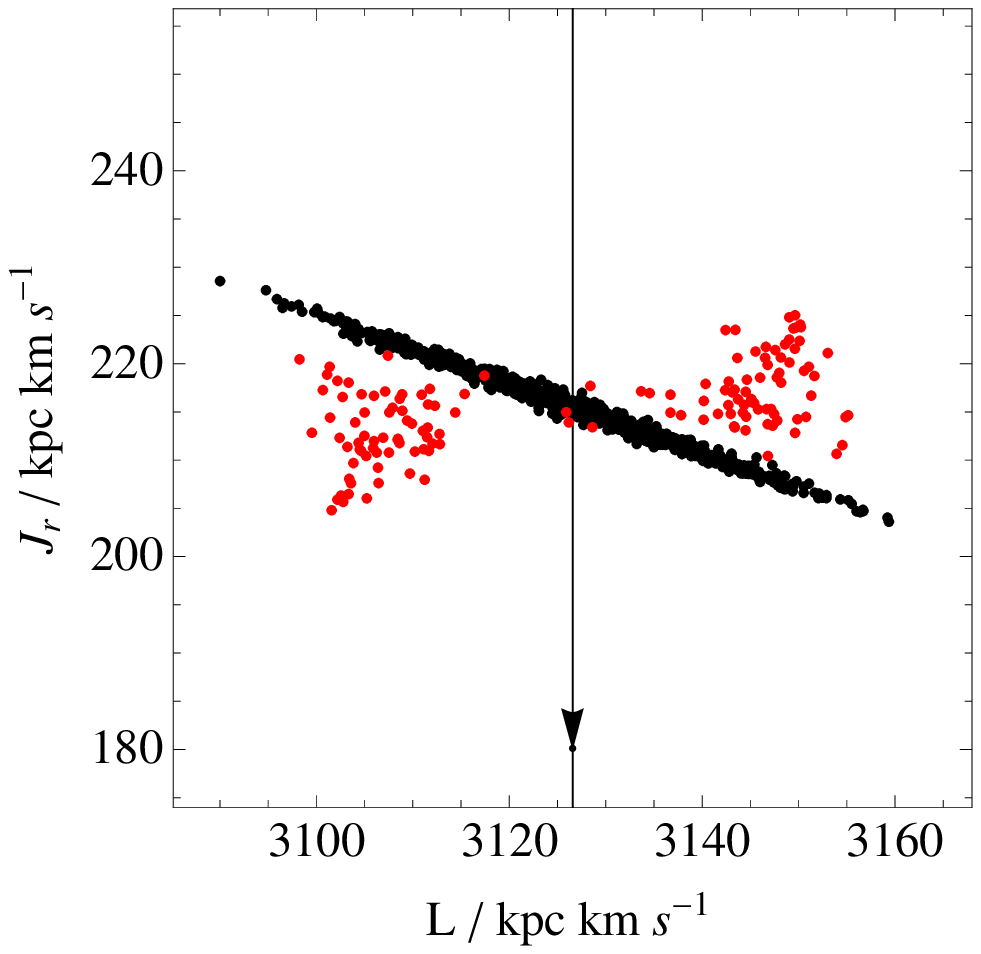}
    \includegraphics[width=\doubsquarefigshrink\hsize]{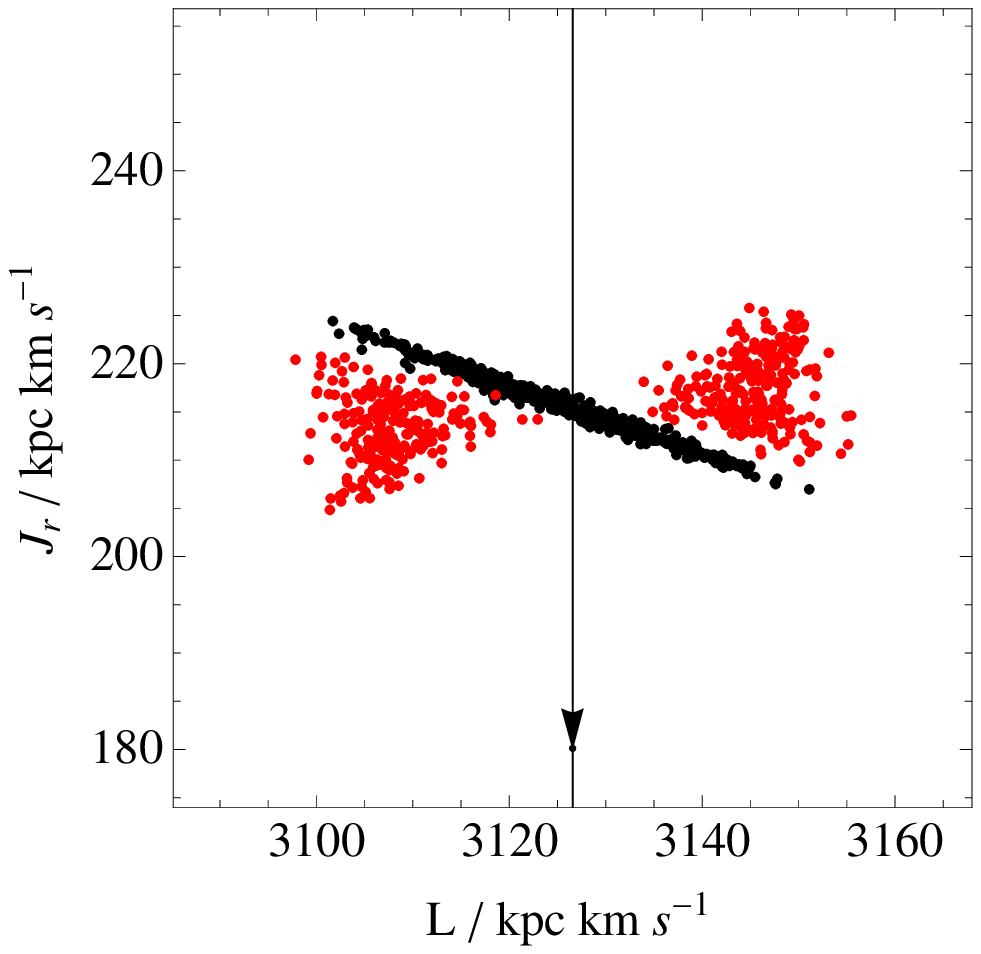}
    \includegraphics[width=\doubsquarefigshrink\hsize]{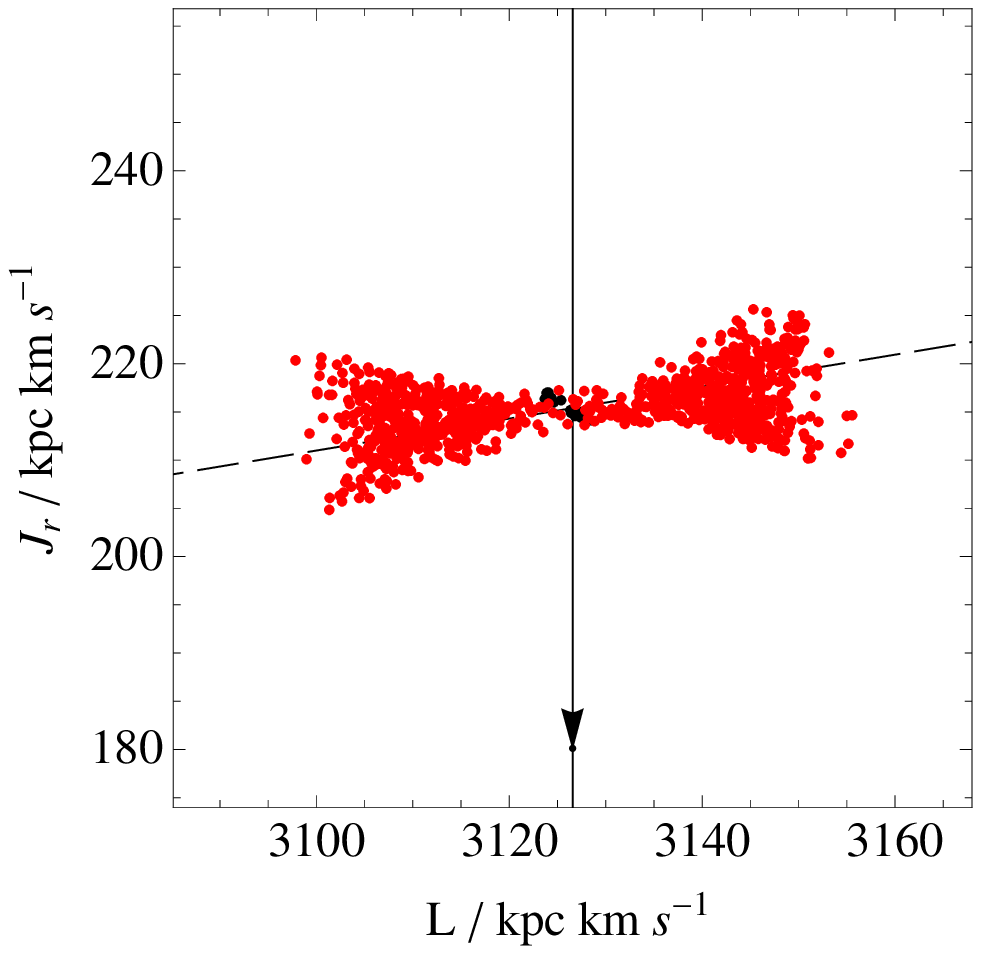}
  }
  \caption{ The action-space distribution of particles
    for the N-body cluster model C1, at different times along the
    orbit I4.  From left-to-right and top-to-bottom, these times are:
    shortly after release; first pericentre passage;
    first apocentre passage;
    second apocentre passage;
    7th apocentre passage;
    14th apocentre passage.
    The solid black line is the inverse map of the frequency vector,
    $\hessian^{-1}\vO_0$.  The dashed line in the bottom-right panel represents a
    least-squares linear fit to the particle distribution.  }
\label{fig:nbody-run1}
\end{figure*}

\subsection{Cluster models}
\label{sec:clusters}

We  work with King models \citep[][BT08]{kingmodel} for our
clusters, since these simple models are both easy to generate and
are fairly representative of some observed globular cluster profiles \citep[Fig.~6.18,]
[]{bm98}. The profile of a
King model can be defined
by $W \equiv \Phi_0/ \sigma^2$, the ratio of the central potential to
the squared-velocity parameter $\sigma^2$. For a given $W$, all
resulting models are similar in terms of
$\rho(\tilde{r}/\tilde{r}_0)/\rho_0$, where $\rho_0$ is the central
density and $\tilde{r}_0$ is the core radius. An exact model is
specified by choosing $\rho_0$ and $\tilde{r}_0$, in addition to $W$,
either directly or through a relation with another parameter.

For a given orbit, we specify our models as follows. Following the argument of
\cite{dehnen-pal5}, we note that a cluster of mass $M_c$ orbiting at a
galactocentric radius $r$ from the centre of a host galaxy with circular velocity
$v_c$, will be tidally pruned to the cluster radius $\tilde{r}_\tide$, where
\begin{equation}
\tilde{r}^3_\tide \simeq {GM_c \over v_c^2} r^2.
\label{eq:rtide}
\end{equation}
We freely choose a profile parameter $W$ and a cluster mass $M_c$, and we
also specify a galactocentric stripping radius $r_{\rm s} > r_{\rm p}$, where
$r_{\rm p}$ is the pericentre radius of the orbit concerned. We then
set $\tilde{r}_{\rm t}$, the cluster truncation radius,
equal to $\tilde{r}_\tide$ from \eqref{eq:rtide}, where 
$r\rightarrow r_{\rm s}$ and $v_c \rightarrow v_c(r_{\rm s})$. The resulting cluster
will remain intact while $r\gg r_{\rm s}$, but will have its outermost stars
tidally stripped when $r \sim r_{\rm p}$.

\begin{table*}
  \centering
  \caption
{Details of the cluster models used in this section. The defining parameters
  for each model are $(W,M_c,r_{\rm s})$, while $(\rlim,\sigma,\tdyn,\epsilon)$ are derived parameters.}
  \begin{tabular}{l|lll|llll}
    \hline
    & $W$ & $M_c$ & $r_{\rm s}$ & $\rlim$ & $\sigma$ & $\tdyn$ & $\epsilon$\\
    \hline
    C1 & 2 & $10^4\msun$ & $12\kpc$ & $48.6\pc$ & $1.18\kms$ & $12.6\Myr$ & $1.0\pc$ \\
    C2 & 2 & $10^5\msun$ & $12\kpc$ & $104.8\pc$ & $2.54\kms$ & $12.6\Myr$ & $2.2\pc$ \\
    C3 & 6 & $10^4\msun$ & $12\kpc$ & $48.6\pc$ & $1.14\kms$ & $2.36\Myr$ & $0.32\pc$ \\
    C4 & 2 & $10^4\msun$ & $11\kpc$ & $45.5\pc$ & $1.22\kms$ & $11.79\Myr$ & $0.94\pc$ \\
    C5 & 2 & $10^4\msun$ & $11\kpc$ & $45.7\pc$ & $1.22\kms$ & $11.79\Myr$ & $0.94\pc$ \\
    \hline
  \end{tabular}
  \label{tab:clusters}
\end{table*}

\subsection{The disruption of a cluster}
\label{sec:disruption}

The low-mass cluster model C1 (\tabref{tab:clusters}) was
specified for the orbit I4 (\tabref{tab:orbits}) according to the
schema in \secref{sec:clusters}. We chose a low value  $W=2$ for the
profile parameter of for our basic cluster model, in order to
ensure the presence of many particles near the cluster truncation
radius $\twidr_{\rm t}$ during successive stripping events.

A $10^4$ particle realization of the cluster model C1 was made by
random sampling of the King model distribution function (equation
4.110 of BT08). This cluster was placed at a point shortly after apocentre on the
orbit I4 in the isochrone potential of \secref{sec:isochrone-tests}.
The cluster was evolved forward in time in the aforementioned potential
by the \fvfps\ tree code of \citet{fvfps}, using a time step of
$\d t = \tdyn/100$ and a softening length $\epsilon$ as specified
in \tabref{tab:clusters}. The simulated time period was
$4.81\Gyr$, or almost 14 complete radial orbits.

\figref{fig:nbody-run1} shows the evolution of the action-space
distribution of the cluster model C1 as a function of time. In all the
panels of that figure, an arrowed black line shows the mapping of the
frequency vector from angle space into action space,
$\hessian^{-1}\vO_0$. This vector indicates the direction that maps onto
$\vO_0$ in angle-space, so any action-space distribution that is aligned with this
vector will be aligned with $\vO_0$ in angle-space.

In all cases involving the isochrone potential,
$\hessian^{-1}\vO_0$ is oriented exactly along the $J_r$ axis. We can understand
this from \eqref{eq:iso-freqs-p}, which shows that in the isochrone
potential, the frequency direction $\hat{\vO}_0$ is a function of $L$
only, and is independent of $J_r$. Thus, a line of constant
$\hat{\vO}_0$ must map into action-space as a line of constant $L$.
This is a peculiar feature of the isochrone
potential, and is not true for a general potential.

The upper-left panel of \figref{fig:nbody-run1} shows
the configuration of the cluster immediately after release, at
position (a) in the upper panel of \figref{fig:nbody-orbit-plans}.
The distribution is ellipsoidal without additional substructure, which we
expect since the distribution in action results entirely from the approximately
spheroidal density profile, and the approximately isotropic velocity dispersion,
of the cluster. We understand the ellipticity of the action-space distribution
from \secref{sec:nonisotropic}. Indeed, the prediction of \eqref{eq:djr/dl}
of $\Delta J_r/\Delta L \sim 0.3$ for this orbit can be seen to be approximately
correct.

In the upper-middle panel of \figref{fig:nbody-run1}, the cluster
has now moved from its point of release to its first pericentre
passage, marked as position (b) in
\figref{fig:nbody-orbit-plans}.  We see that the ellipse has
flattened somewhat, and has rotated anticlockwise. In the upper-right
panel, the cluster has now progressed to the subsequent apocentre passage,
marked as position (c) in \figref{fig:nbody-orbit-plans}.  Here, the
cluster is again flattened, but now it has rotated clockwise.

We can qualitatively understand this behaviour as follows. Consider
a particle in a cluster near apsis. What changes to the actions
will be made by perturbations to the velocity of this particle?
A perturbation $\delta v$ to the transverse velocity will cause a
change to the angular momentum
\begin{equation}
\delta L = r \delta v.
\label{eq:dL=dv}
\end{equation}
By means of the mechanism described by \eqref{eq:l-amp}, this $\delta L$
will cause a change in the guiding-centre radius $r_\g$, which will cause a
corresponding change in the radial action, according to
\eqref{eq:amplitude}. Conversely, a perturbation to the
radial velocity will cause negligible change to the radial action,
since
\begin{equation}
\delta E \simeq p_r \delta p_r = \dot{r} \delta v \sim 0,
\label{eq:de0}
\end{equation}
and $J_r(E,L)$ remains unchanged. Hence, the distribution in both $J_r$
and $L$ is governed primarily by the transverse velocity when the
cluster is at apsis, and the high degree of correlation observed
in \figref{fig:nbody-run1} reflects this. 

\begin{figure}
  \centering{
    \includegraphics[width=\squarefigshrink\hsize]{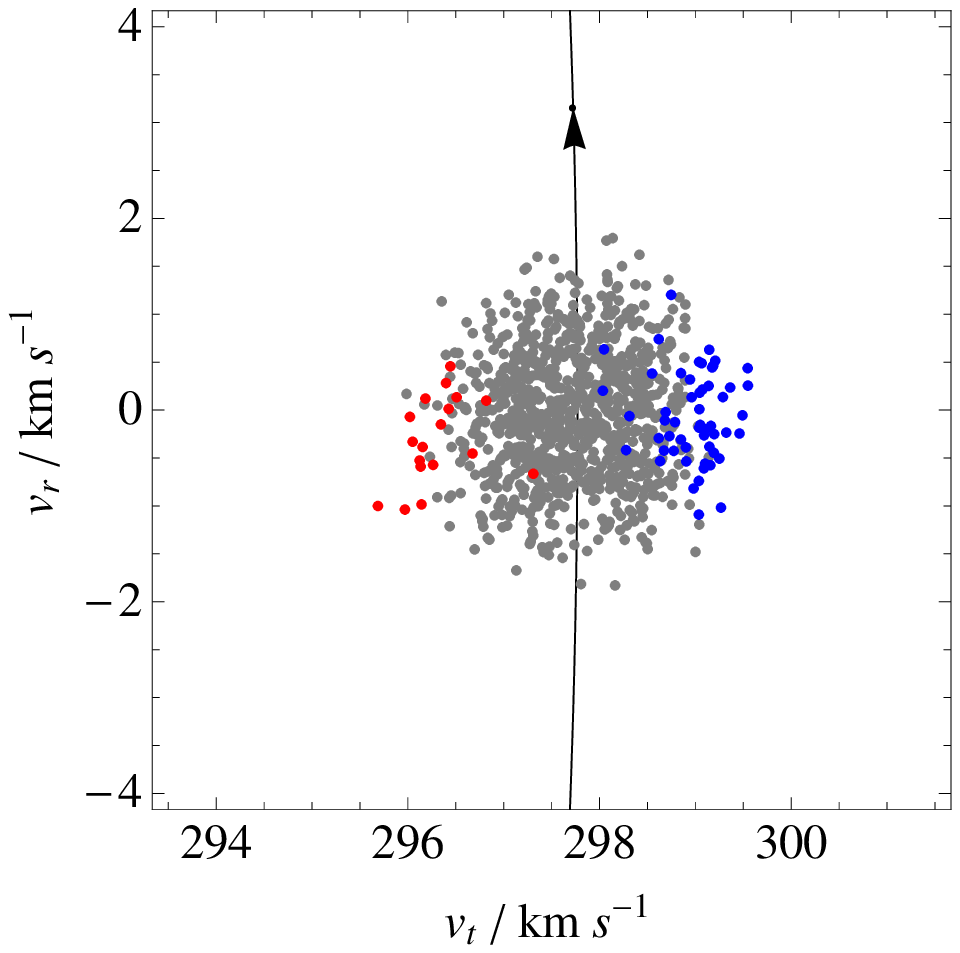}\\
    \includegraphics[width=\squarefigshrink\hsize]{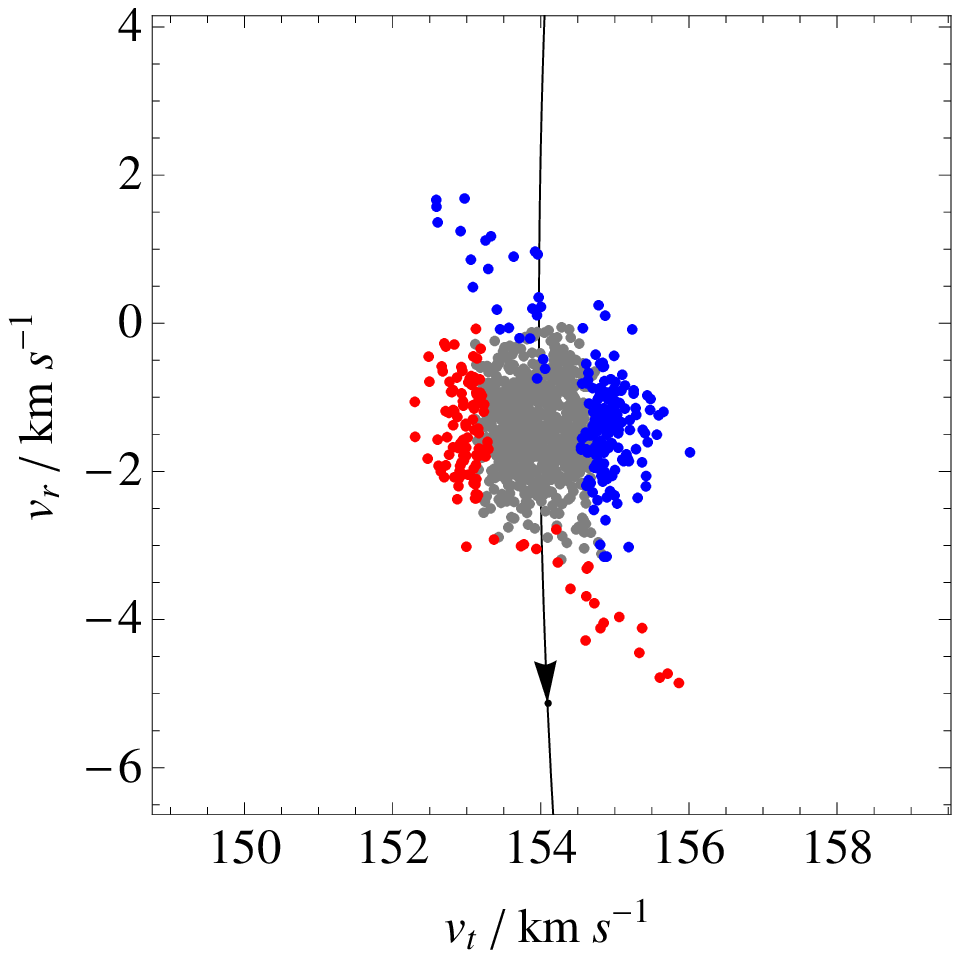}
  }
\caption{
 Radial velocity versus tangential velocity for the C1 cluster model, on
the I4 orbit, at (top panel) the first pericentre passage, and
(bottom panel) the first apocentre passage.
}
\label{fig:nbody-v}
\end{figure}

We can confirm this analysis by examining \figref{fig:nbody-v}. The upper
panel shows $(v_r,v_{\rm t})$ for the cluster near its first pericentre passage,
while the lower panel shows the same for the cluster near its subsequent
apocentre passage. The particles coloured red are those with $L < 3110\kpc
\kms$ and those coloured blue have $L > 3140\kpc\kms$. The boundary
between the colours is very sharp in the $v_{\rm t}$ direction, as one would
expect, since the particles have been coloured on the basis of $L$. However,
we note that particles with the full range of $v_r$ contribute to both the
blue and red regions equally, despite \figref{fig:nbody-run1}
showing that $J_r$ is very different for these two regions. Hence $J_r$
must be independent of $v_r$ near apsis.

We complete our explanation of the orientation of the action-space distribution
by taking note of the sign of the change in $J_r$ near apsis. Increasing
$L$ will always increase $r_\g$, and when the cluster is at pericentre,
this pushes $r_\g$ further away, and so increases the radial action.
Conversely, when the cluster is at apocentre, increasing $r_\g$ brings
it closer to the cluster, and so decreases the radial action. Thus, we see that
near pericentre, particles with high $L$ will have high $J_r$ and the distribution
will be rotated to have a positive gradient $\Delta J_r / \Delta L$.
Conversely, at apocentre, particles with high $L$
will have low $J_r$, and so the distribution will be rotated
to have a negative gradient $\Delta J_r / \Delta L$.

We can predict the value of the gradient of this distribution,
by combining \eqref{eq:amplitude} and \eqref{eq:l-amp}.
We find
\begin{equation}
{\d J_r \over \d L} \sim \pm
\sqrt{ 2\sqrt{5} J_r \over L},
\label{eq:gradient}
\end{equation}
where the sign of the radical depends on the apsis under
consideration, as detailed above. We can see from the upper-right
panel of \figref{fig:nbody-run1} that this equation predicts
approximately the correct gradient of $\d J_r / \d L \sim -0.5$, when
evaluated for the orbit I4.  We further note that this equation implies that
the gradient will be steeper for an orbit of greater eccentricity.

Let us again examine \figref{fig:nbody-run1}, and note that the black
particles are those still bound to the cluster, while the red
particles are those that are unbound, where we have defined `bound' to
mean all particles that are within $r_\tide$ of the cluster
barycentre. In the upper-right plot, which shows the cluster
configuration at the first apocentre passage, the few unbound
particles form an approximately horizontal distribution.  These are
particles that have been stripped from the cluster near to the
preceding pericentre passage, which is depicted in the upper-middle
panel of \figref{fig:nbody-run1}.

We note two things. First, the bulk of the stripped particles have
large $\Delta J$ from the cluster centroid; this is simply a
consequence of the high-speed stars being most likely to be stripped.
Second, we note that although the red particles in the upper-right
panel span approximately the same range in $L$ as the black particles
in the upper-middle panel, they span a range in $J_r$ that is only about
half that spanned by the black particles.

We explain this as a result of the cluster's self-gravity, as follows.
Particles that are stripped from the cluster mostly escape through the
Lagrange points $L_1$ and $L_2$. These two points lie near the cluster along
the line from the barycentre of the host to the cluster (Fig. 8.6 of BT08).
Thus, it is the particles' radial velocity which initially carries them away
from the cluster, and it is from this velocity component that the particles
pay most of the energetic penalty for escaping, with the consequence that the
radial velocity dispersion of the escaping stars is reduced. This reduction
in the radial velocity dispersion corresponds to a compression in $J_r$ of the
distribution of escaping particles. Once unbound, the particles are
carried further away from the cluster along a complex trajectory that ends up
with the now-free particle drifting away from the cluster according to the
mapping in \eqref{eq:angle_t}.  Thus, the final sum of the energetic penalty
is paid from the difference in action $\vJ - \vJ_0$ between the particle and
the cluster, with the net result that the unbound distribution uniformly
shrinks.  The latter effect is minor compared to the compression in $J_r$,
because much more work is done in becoming unbound than in escaping to
infinity once already unbound. The complete effect is to generate an unbound
distribution, that looks like the high $\Delta \vJ$ wings of the pericentre
distribution, but is compressed in $J_r$ and slightly shrunk.

Looking again at \figref{fig:nbody-run1}, we note that in
the lower-middle panel, which corresponds to the 7th apocentre
passage, many particles have escaped, and the size of the
bound distribution has visibly shrunk. We understand this as a
consequence of the most energetic particles having already
escaped the cluster, leaving behind a colder core. By the time
of the 14th apocentre passage, shown in the lower-right panel,
almost all the particles have escaped. We note that the positions
of many of the red particles have remained static between the lower-middle
and lower-right panels. The action-space distribution of the
unbound particles is therefore frozen in place, confirming that
self-gravity is unimportant after a stream has initially formed.

In conclusion, we have qualitatively understood the distortion of a cluster
in action-space as it passes through pericentre and apocentre along its
orbit. We have found that stars stripped at pericentre form a distribution
that is derived from the pericentre distribution of bound stars, but is
compressed in $J_r$.  We have found that the pericentre distribution exhibits
a high correlation between $J_r$ and $L$, and that the gradient of this
correlation in $(L,J_r)$ scales as $\near \sqrt{J_r / L}$.  We typically
assume that our cluster orbit will have large $L$ and comparatively smaller
$J_r$: this would only be untrue for extreme plunging orbits, which are not
likely to be relevant to the problem in hand.  Hence the gradient in
$(L,J_r)$ will typically be less than unity, and the compression will only
act to shrink it still further: the stripping mechanism produces an
action-space distribution that is both flattened and very roughly oriented
along $\hat{L}$.

\subsubsection{The effect of changing the cluster model or orbit}
\label{sec:changingorbit}

We now investigate the qualitative effects on the action-space distribution
of a disrupted cluster of changing the cluster model parameters, or the
cluster orbit.  The cluster models used in this section are C1 to C4,
detailed in \tabref{tab:clusters}. These clusters were created according to
the schema of \secref{sec:clusters}, taking the orbit to be I4 in the
isochrone potential of \secref{sec:isochrone-tests} for the models C1--C3,
and taking the orbit to be I5 in the same potential for the model C4.

The cluster model C1, the evolution of which along the orbit I4
was detailed in the previous section, is used as our baseline
to which we compare the distributions of the other clusters. The cluster
model C2 has the same profile parameter, $W=2$, as does C1, but is
10 times more massive. The result is a cluster that is both
heavier and proportionately larger while being stripped at the same
galactocentric radius, $r_{\rm s}$.

The cluster model C3 has the same mass as does C1, but is considerably more
concentrated, with a profile parameter $W=6$. The cluster has an identical
truncation radius $\twidr_{\rm t}$ and velocity scale $\sigma$, but has
significantly fewer particles near to $\twidr_{\rm t}$ than C1. The particles
of C3 are generally more tightly bound to the cluster than are those of C1.

The cluster model C4 has the same mass and profile parameter as C1, but
is specified for the orbit I5, which has lower $L$ than I4, and thus
a smaller pericentre radius. The resulting cluster is slightly more
compact, allowing it to survive to a closer galactocentric radius
than can C1.

A $10^4$ particle realization of each of the models C1--C3 was placed at a
point shortly after apocentre on the orbit I4, and evolved forward in
time by the \fvfps\ tree code, using a time step of $\d t = \tdyn/100$
and a softening length $\epsilon$ as specified in
\tabref{tab:clusters}. The total period of the simulation
was $2.36\Gyr$, or almost 7 complete radial orbits.
Additionally, a $10^4$ particle realization of C4 was placed at a
point shortly after apocentre on the orbit I5, and evolved forward in
time by the \fvfps\ tree code, for a total period of $2.21\Gyr$, or
almost 7 complete radial orbits.

\begin{figure*}
\centering{
  \includegraphics[width=\doubsquarefigshrink\hsize]{figs/nbody_j_run1_apo7.eps}
  \includegraphics[width=\doubsquarefigshrink\hsize]{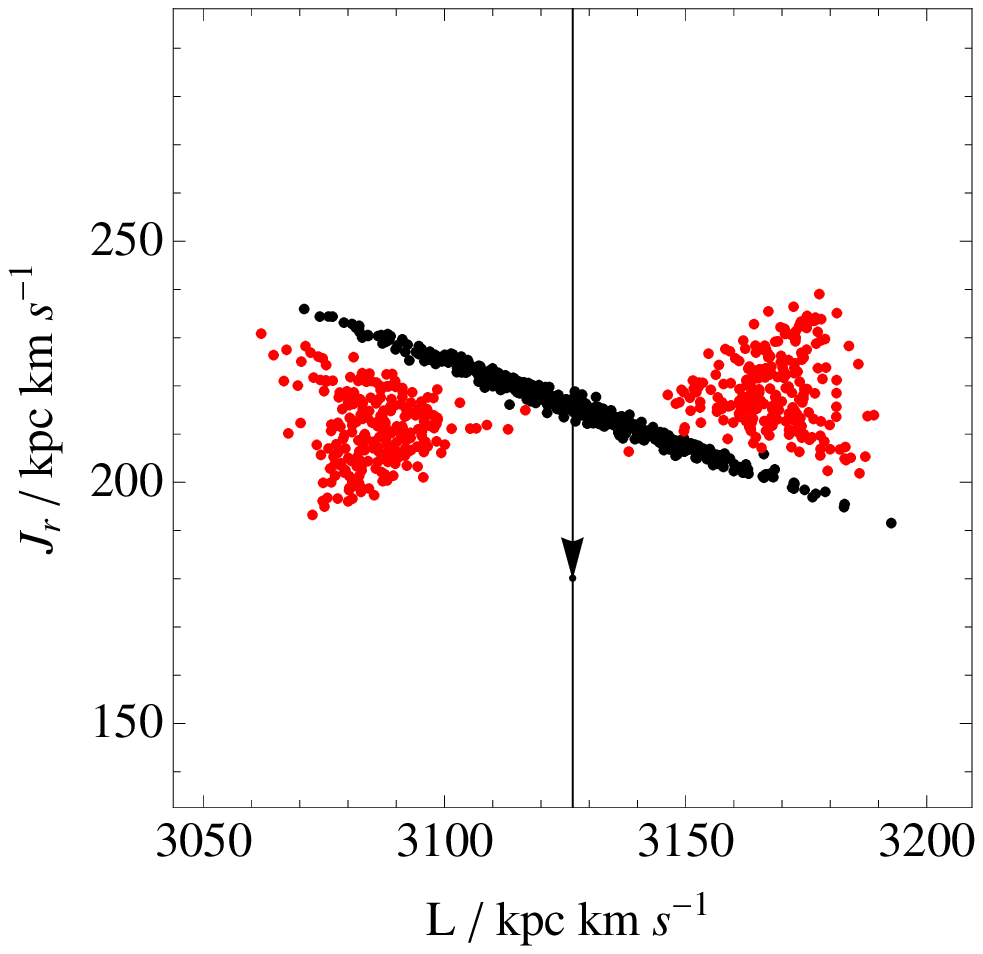}\\
  \includegraphics[width=\doubsquarefigshrink\hsize]{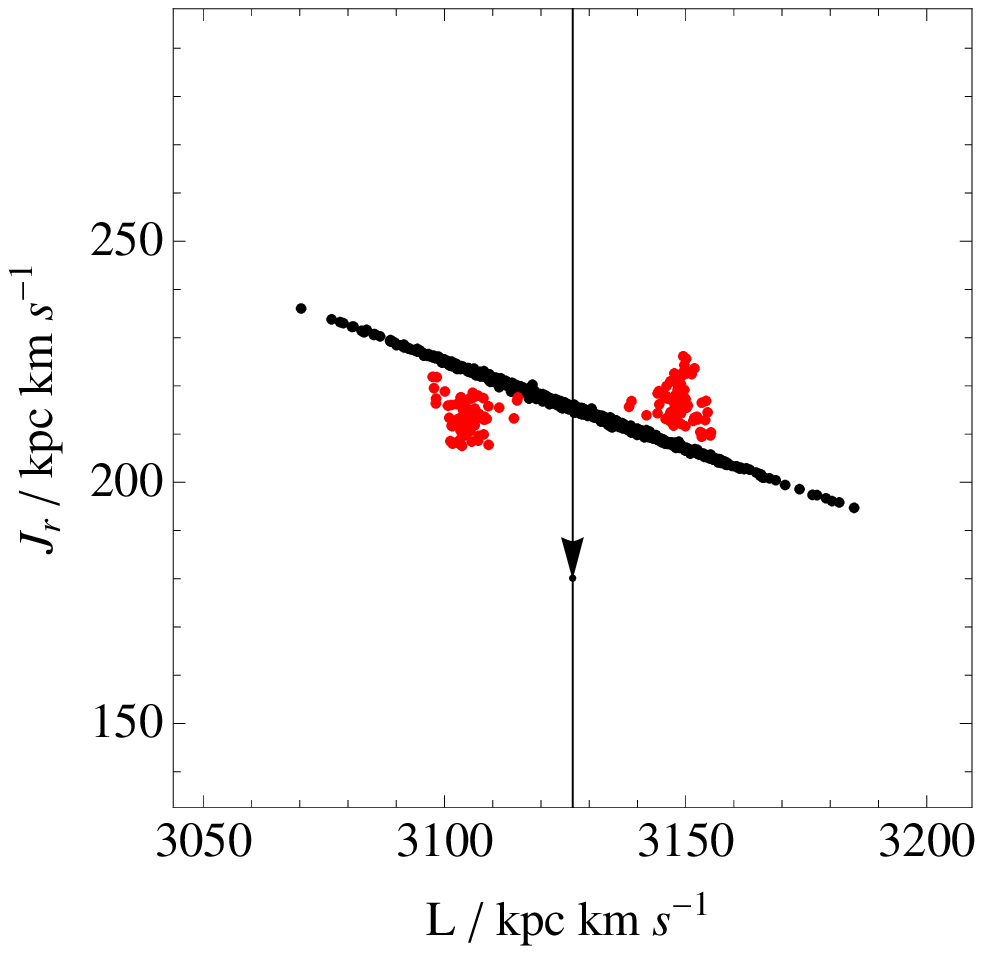}
  \includegraphics[width=\doubsquarefigshrink\hsize]{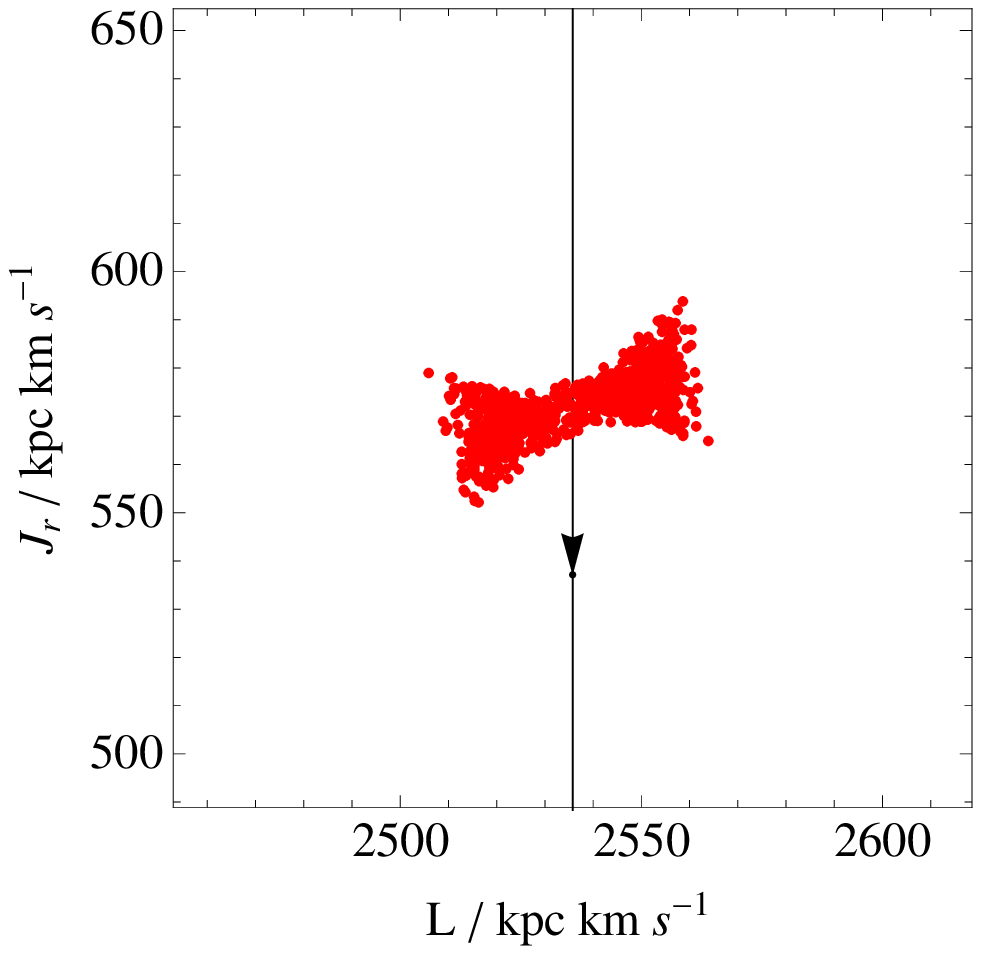}
}
\caption{The action-space distribution of particles for cluster models
on various orbits. Each plot shows the distribution at the 7th apocentre passage.
Black points are bound to the cluster, red points orbit free in the host
potential. From left-to-right and top-to-bottom, the panels show: the C1 cluster
on the orbit I4; the C2 cluster on the orbit I4; the C3 cluster
on the orbit I4; and the C4 cluster on the orbit I5. Note the change of scale
between successive plots.
}
\label{fig:nbody-runs234}
\end{figure*}

\figref{fig:nbody-runs234} shows the action-space distribution for
the cluster at the seventh apocentre passage, for each of these simulations.
The top-left panel shows C1 on the orbit I4, and is identical to the
lower-middle panel in \figref{fig:nbody-run1}.
The top-right panel shows the cluster C2 on the same orbit; the bottom-left
panel shows the cluster C3 on the same orbit; and the bottom-right panel
shows the cluster C1 on the orbit I5. 

In the top-right panel, we see that the action-space distribution of the
more massive cluster is qualitatively identical to that of C1, except
that the distribution has approximately twice the scale. We conclude
that cluster mass merely sets the scale of  distribution in
action-space of stripped stars.

The bottom-left panel shows the distribution from the more centrally
concentrated cluster C3. In this case, the region occupied by
bound particles is longer than for C1, reflecting the greater
internal random velocities of the more concentrated model.  However,
the gross structure of the unbound distribution is approximately the
same as for C1.  This is because the extent of this distribution is
defined by the least bound stars, stripped from the outer parts of
each cluster: the behaviour of these stars is determined only by the
mass of the cluster, and not by its internal configuration.  Thus we
conclude that cluster concentration does not alter the general shape
of the eventual action-space distribution.

The bottom-right panel shows the distribution from the cluster C4
on the orbit I5, which has the same apocentre radius as I4 but a
pericentre radius about $33\percent$ smaller. Unlike in the other panels,
 the cluster has become completely unbound by the 7th apocentre
passage on I5, which is likely to be a result of $(r_{\rm s} - r_{\rm p})$ being
slightly larger for C4 on I5, when compared to the other clusters on I4,
resulting in more efficient stripping at pericentre.

The distribution shown in this plot has approximately the same scale in
$\Delta L$ as does the distribution from I4, but has approximately twice the
scale in $\Delta J_r$. This is a consequence of the gradient $\d J_r / \d L$
given by \eqref{eq:gradient} being steeper for I5 than for I4. Further, since
we have already noted that this cluster was stripped faster than was C1, the
energetic penalty for escaping must be lower, and so we expect less
compression in $J_r$.  The resulting distribution is similar to that of the
top-left panel, but less compressed in the $\hat{J_r}$ direction. Thus we
conclude that changing the cluster orbit can distort the shape of the
action-space distribution of unbound particles, but does not affect its basic
structure.

\subsection{Predicting the stream from the action-space distribution}

\begin{figure}
  \centering{
    \includegraphics[width=.8\hsize]{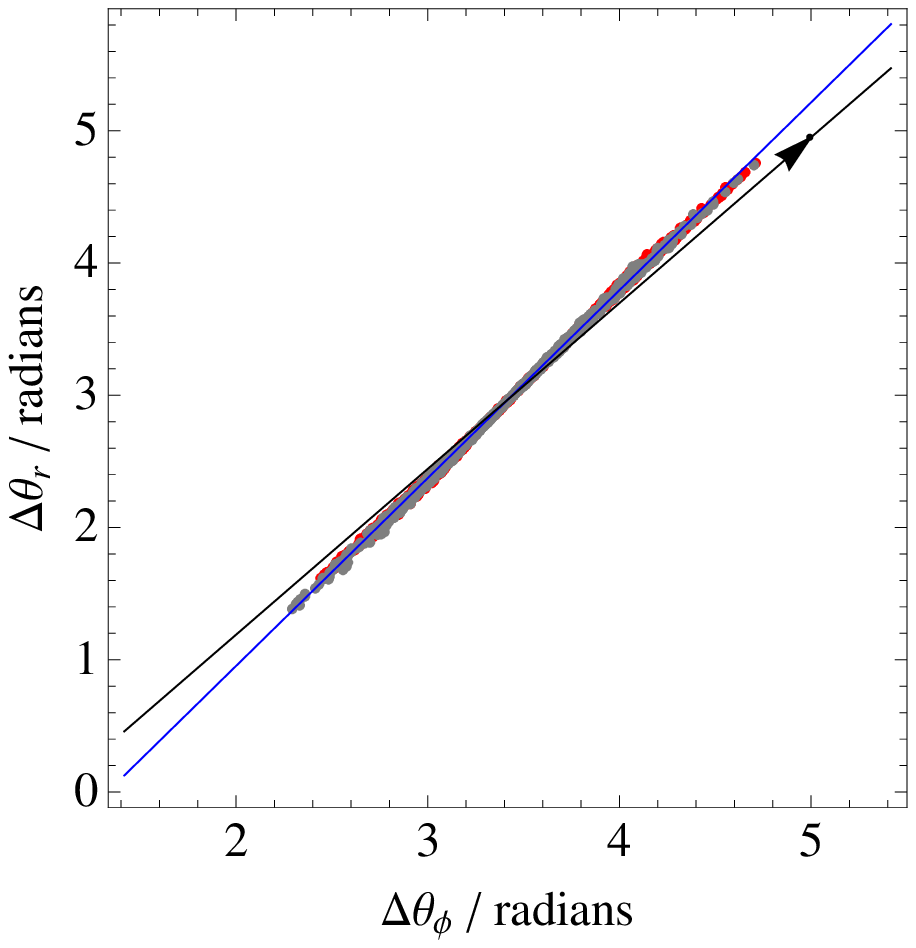}
    \includegraphics[width=.8\hsize]{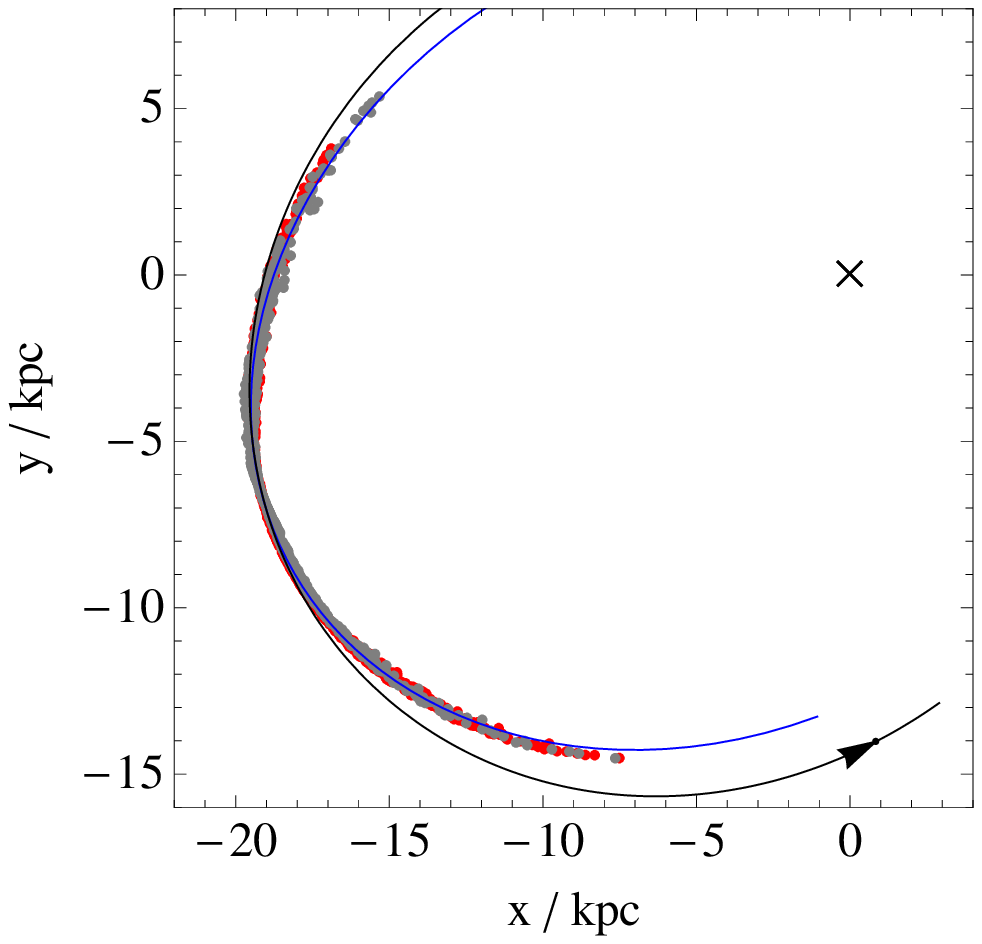}
  }
\caption
{The distribution of particles for cluster C1, near
    its 14th apocentre passage on orbit I4. The upper panel shows the
    angle-space distribution, while the lower panel shows the
    configuration in real space.  The grey particles show positions
    directly computed from the N-body simulation, while the red
    particles show those positions predicted from mapping the
    action-space distribution in the bottom-right panel of
    \figref{fig:nbody-run1}. The two distributions almost
    precisely overlap.  In both plots, the black arrowed curves show
    the trajectory of the progenitor orbit, while the blue curves show
    the mapping of the dashed line from \figref{fig:nbody-run1}.
    The blue curve is everywhere a much closer match to the stream particles than is the
    progenitor orbit.}
\label{fig:nbody:isoch}
\end{figure}

We have determined the action-space distribution for several disrupted cluster models
by means of N-body simulation. We now ask whether we can accurately predict the
real-space path of the stream, given the action-space distribution of
one of those models.

Suppose that we know the time $t_\d$ since a cluster's first
pericentre passage. The angle-space distribution is predicted by
\eqref{eq:angle_t}, where $\hessian$ is evaluated on the progenitor orbit
$\vJ_0$ and $t\rightarrow t_\d$. We will use as an example the action-space distribution
shown in the bottom-right panel of \figref{fig:nbody-run1}, which corresponds
to the 14th pericentre passage of the simulated cluster C1 on the orbit I4.
The upper panel of \figref{fig:nbody:isoch} shows the angle-space configuration
corresponding to this panel: the grey
particles are for angles directly computed from the results of the N-body simulation,
while the red particles are for those predicted by \eqref{eq:angle_t}, assuming
$t_\d$ is known. Also plotted is the frequency vector $\vO_0$, shown as a
black arrowed line.
The distributions of black and grey particles in this panel agree perfectly.
Furthermore, both distributions are obviously
misaligned with the progenitor orbit. 

The lower panel of \figref{fig:nbody:isoch} shows the real-space
configuration of particles for the same scenario. The grey particles
are again plotted directly from the results of the N-body simulation,
while the red particles result from the mapping into real-space
of the red particles from the upper panel.  As in angle-space, the two
distributions agree perfectly. Furthermore, the real-space
manifestation of the misalignment of the stream with the orbit can be
seen: the stream delineates a track that has substantially lower
curvature than the orbit.

Our attempt to predict the real-space stream configuration
from the action-space distribution has been completely successful.
However, any
complete model of the bottom-right panel of
\figref{fig:nbody-run1} must necessarily be rather
complicated. It might not be possible to guess the form of this
distribution without full N-body modelling. The dashed line in the
bottom-right panel of \figref{fig:nbody-run1} is a least-squares
fit of the action-space distribution to a line. This represents a  simpler
model of the action-space distribution, which it might well be
possible to guess {\em ab initio} for a cluster on a given orbit.

How good a prediction for the stream track can we get from this line? The blue
lines in \figref{fig:nbody:isoch} show the results of mapping this line
into both angle-space and then real-space. It is clearly an excellent fit to the
stream, in marked contrast to the orbit, which is a much poorer model
of the stream. Thus, even a very simple model of the
cluster in action-space---albeit one deduced from an accurate knowledge of the
distribution---allows
us to predict stream tracks accurately.

Finally, we note that during the mapping of this line into real-space,
we need to make a correction to the cluster's action, as described by
\eqref{eq:correction}, to account for the variation in action
down the stream.  Evaluating \eqref{eq:amplitude} and
\eqref{eq:l-amp} for the orbit I4, and taking $\d L \sim
25\kpc\kms$ and $\d J_r \sim 10\kpc\kms$ from the bottom-right panel
of \figref{fig:nbody-run1}, we predict errors in the real-space
of up to $\sim 0.2\kpc$ on account of the finite $L$ distribution,
and up to $\sim 0.15\kpc$ on account of the finite $J_r$ distribution.
These errors would be serious enough to be seen in
\figref{fig:nbody:isoch}, and thus the correction is required. We note,
however, that even these substantial errors are insignificant compared
to the several-kpc discrepancy between the stream and the orbit.

\section{The consequences of fitting orbits}
\label{sec:fitting}

We confirmed in \secref{sec:isochrone} that streams formed in the
isochrone potential are not well delineated by orbits, and we provided
a realistic example where this is the case. Many practical techniques
attempt to use tidal streams to constrain the parameters of
the Galactic potential, by erroneously assuming that such streams {\em do}
delineate orbits. In this section, we briefly examine the consequences
for such techniques of making this faulty assumption.

Our experiment is based on the simulated tidal stream of
\figref{fig:nbody:isoch}. This particular example is for a stream at
apocentre: this is relevant, because many actual observed streams are
discovered close to apsis, for example the Orphan stream
\citep{orphan-discovery} and GD-1 \citep{gd1-discovery}.

We first
create two sets of pseudo-data: one corresponding to the progenitor
orbit, marked as a black line in \figref{fig:nbody:isoch}, and one
corresponding to the predicted stream track, marked as a blue line.
Each pseudo-data set contains approximately 40 phase-space
coordinates, sampled evenly from the length of the corresponding
track, as shown in \figref{fig:nbody:isoch}. 

We now wish to measure the quality with which an orbit for a given set
of isochrone-potential parameters can be made to fit the data. For the
purposes of this exercise, we have assumed the functional form of the
potential to be known. Further, we have granted ourselves pseudo-data
with full and accurate positional and velocity information.  Granting
ourselves an unrealistically complete pseudo-data set simplifies
considerably the matter of finding an orbit that is close to the best
fitting one.  In practice, one typically proceeds with one or more
phase-space coordinates unknown, or known with substantially less
precision than the others. In this case, our purpose is
solely to demonstrate the errors that can be made by naively assuming
that an observed stream can be fit with an orbit. Granting ourselves
an unrealistically complete set of pseudo-data does not diminish our
conclusions in this regard.

For each set of potential parameters, we choose an orbit as
follows. We first select a datum near the centroid of the stream, and
declare that our chosen orbit must pass directly through this
datum. Although it may be that some nearby orbit, one that does not
pass directly through this point, would make a better-fitting orbit,
any such orbit must pass very close to the selected datum, because it
is close to the centroid. Thus, such an orbit cannot be much
better-fitting than one that passes directly through the datum.
Having chosen a datum, for a given set of potential parameters, an
orbit is defined.

Having chosen our orbit, a goodness-of-fit
statistic $\chi^2$ is calculated as follows. For each datum in the stream, with
phase-space coordinate $\vect{w}_i$, a location along the orbit
$\vect{w}'_i$ is chosen that minimizes the square difference
\begin{equation}
(\vect{w}_i - \vect{w}'_i)^2.
\end{equation}
Having obtained the ${\vw'_i}$, the goodness-of-fit $\chi^2$ is defined by
\begin{equation}
\chi^2 = \sum_{i,j} {(w_{i,j} - w'_{i,j})^2 \over \sigma^2_j},
\label{eq:chisq}
\end{equation}
where $j$ are the phase-space coordinates, and $\sigma_j$ is the
rms of $(w_{i,j}+w'_{i,j})/2$ over $i$. This $\chi^2$ statistic
provides a dimensionless measure of the phase-space distance
between the best-fitting orbit in a given potential, and the pseudo-data.

If the pseudo-data set were a sample of a perfect orbit in some potential,
we expect the value of $\chi^2$ to be exactly zero, when the correct
potential parameters are used. As the potential parameters
are varied away from their true values, we expect the value of $\chi^2$
to rise, as the best-fitting orbit becomes a steadily worse
representation of the data. Hence, we expect minima in $\chi^2$ to be associated
with the potential parameters that are optimum, from the perspective
of fitting an orbit to the data. We seek such minima by plotting the value
of $\chi^2$ over a range of likely values for the potential parameters.

\subsection{Results}

\begin{figure}
  \centering{
    \includegraphics[width=\squarefigshrink\hsize]{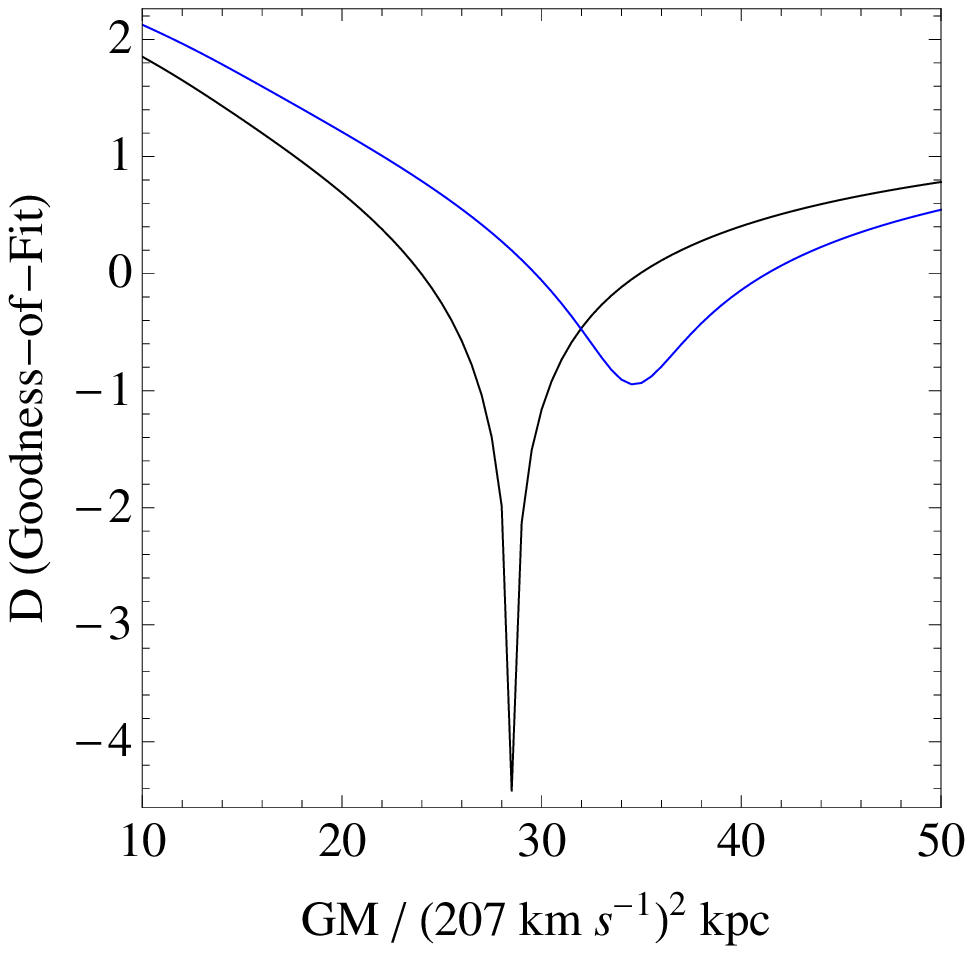}
    \includegraphics[width=\squarefigshrink\hsize]{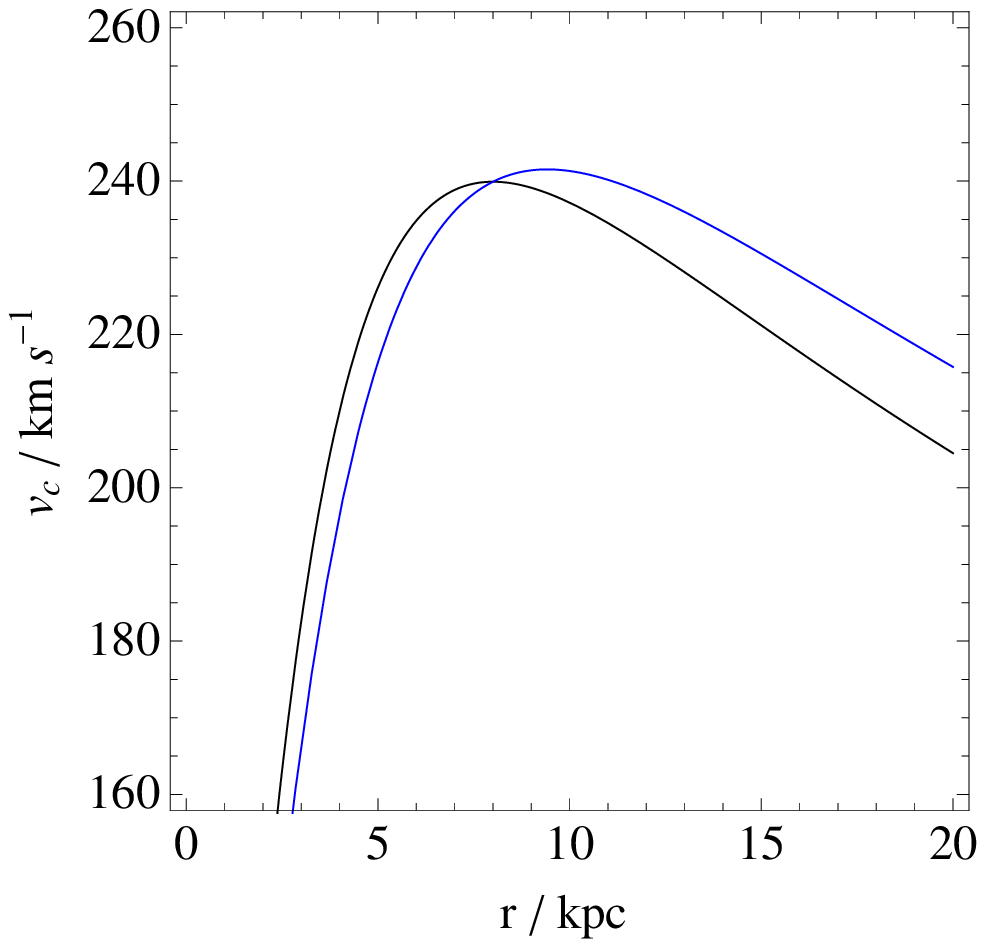}
  }
  \caption {Top panel: goodness-of-fit $\chi^2$ for each pseudo-data
    set, to the best-fitting orbit in the potential. The colours of
    the curves for each pseudo-data set identify it with the
    corresponding curve from \figref{fig:nbody:isoch} from which it
    was derived. Bottom panel: rotation curves for the optimum
    potential found for each pseudo-data set.  In all cases, we have
    required that the circular velocity $v_c = 240\kms$ at
    $\rsun=8\kpc$.  This reduces the potential to one of a single
    parameter; in this case, mass. The quality of the fit to the
    blue data set is significantly degraded, compared to that of
    the black. Utilising the high-curvature, blue
    data set as a proxy for an orbit causes us to overestimate the
    host mass by approximately 21\percent.}
\label{fig:orbit-fit}
\end{figure}

The upper panel of \figref{fig:orbit-fit} shows the goodness-of-fit
$\chi^2$, for the best-fitting orbit, in an isochrone potential with
mass parameter $GM$ as shown.  For each value for $GM$, the isochrone
parameter $b$ is chosen such that the potential reproduces the fiducial
rotational speed of $v_c = 240\kms$ at $\rsun=8\kpc$, since in
practical usage one would generally require any acceptable potential
to reproduce other observed features of the Milky Way galaxy, such as
the circular velocity at the Solar radius. This has the effect of
reducing the dimensionality of the search for the best-fitting
solution to a single parameter.

Along each of the black and blue curves in \figref{fig:orbit-fit},
$\chi^2$ has a minimum at
the potential parameters for which the best-fitting orbit
has been found. In the case of the black curve, which attempts
to fit to pseudo-data derived directly from the
progenitor orbit, the value of $GM$ is correctly identified with
high precision, thus validating the technique.

For the blue curve, which attempts to fit pseudo-data derived from the
stream track, the quality of the fit at optimum $GM$ is significantly
degraded when compared to the black curve. This indicates that,
although an optimal value for the mass parameter is being found, the
fit to the orbit is still not perfect there. Furthermore, the value of
$GM$ for which the optimal fit is found is $21\percent$ larger than
the true value.

The attempt to constrain the potential parameters by using misaligned
streams as proxies for orbits has led us to err.  The high-curvature
of the stream, relative to the orbits of the stars that comprise it,
has incorrectly led the fitting procedure to a significantly larger
enclosed mass, in order to provide the larger gravitational force
necessary to produce a more highly-curved orbit.  This can clearly be
seen from the lower panel of \figref{fig:orbit-fit}, which shows the
rotation curves for the optimal potentials for both sets of
pseudo-data.  In the case of the stream-derived pseudo-data, the
circular speed of the optimum potential is significantly higher than
the truth at those distances where our stream segment lies.

In conclusion, we find that there is a risk of substantial systematic errors
in parameter estimation being made, if one attempts to constrain the
potential using streams, and one assumes that streams perfectly delineate
orbits. In practice, one would hope that the significant degradation of the
minimum $\chi^2$ of the fit to the stream-derived data, when compared to
orbit-derived pseudo-data, would be noticeable. However, it is often the case
that random noise in the measurements are sufficient to reduce the quality of
fits such that such discrimination is not possible \citep{binney08,eb09a}.
If this is so, then even with the perfect error-free data used in this
section, the uncertainty on the deduced mass parameter could not be better
than $\pm 20\percent$. In practice, the uncertainty is likely to be much,
much worse. Only by fitting stream data to physically-motivated stream
tracks instead of orbits can we hope to overcome this limitation, and
obtain the tantalising diagnostic precision proffered by the analysis of
streams in such as \cite{binney08}.

\section{Non-spherical systems}
\label{sec:nonsph}
We have investigated the formation of streams in spherical potentials,
and demonstrated that they are not necessarily well modelled by
orbits, but that with some prior knowledge of the system, we can
accurately predict their tracks.  However, many real stellar systems
in the Universe are not spherical.  In particular, our own Galaxy,
whose potential we are interested in probing with streams, is
significantly flattened.

In the case of spherical systems, the stream can only be misaligned with the
progenitor orbit {\em within} the plane to which both stream and orbit are
confined.  Orbits in flattened potentials are generically three-dimensional
as a result of the instantaneous orbital plane precessing around the
potential's symmetry axis. Different parts of the stream have orbit planes
which precess at different rates, so the stream is not strictly confined to a
plane. Moreover, the plane that most nearly contains the stream generally
differs from the instantaneous orbit plane of the progenitor. These
complexities come on top of the misalignments within the best-fitting plane
that we have studied in spherical potentials.  They could well be the reason
that all attempts to find a single orbit that fits the Sagittarius dwarf
stream have had limited success \citep{fellhauer-sag,no-sagittarius}.
In this section, we investigate the formation of streams in a
flattened \stackel\ potential.

\subsection{\stackel\ potentials}
\label{sec:stackel}

Regular orbits in non-spherical potentials can be described by a system of
three actions, but the actions and angles can be computed by standard means
only if the potential is a \stackel\ potential (\S3.5.3 of BT08).  An
exhaustive treatment of these potentials 
is given in \cite{de-z-1} and \cite{de-z-2}, from which we adopt our
notation. The example we consider is an oblate axisymmetric potential, for
which the appropriate coordinate system is that of prolate spheroidal
coordinates, $(\lambda, \phi, \nu)$; $\lambda$ is constant on confocal
ellipsoidal surfaces of revolution, while $\nu$ is constant on the hyperbolic
surface of revolution that cut the ellipsoids orthogonally. Hence $\lambda$
is a radial coordinate and $\nu$ is analogous to latitude. The coordinate
system is specified by two negative quantities $(\alpha,\gamma)$.
The potential has the form
\begin{equation}
\Phi(\lambda, \nu) = -{(\lambda + \gamma)G(\lambda) - (\nu + \gamma)G(\nu)
\over \lambda - \nu},
\label{eq:stackelpotential}
\end{equation}
where de Zeeuw's function $G(\tau)$ is determined once the density
profile $\rho(z)$ along the $z$-axis has been chosen
\citep[see e.g.~equation (23),][]{de-z-2}. Thus, the model is
completely specified by $\rho(z)$ and the scaling
parameters $(\alpha, \gamma)$.
The actions for \stackel\ potentials of this type are $(J_\lambda,
L_z, J_\nu)$, corresponding to motion in
$(\lambda, \phi, \nu)$ respectively.

\subsection{Galaxy models with \stackel\ potentials}

\cite{de-z-2} shows that if one requires the density everywhere to be
non-negative, it is not possible to formulate a \stackel\ model in
which the density $\rho(r)$ falls off with distance from the $z$-axis more
rapidly than $r^{-4}$ as $r\rightarrow\infty$. This is because an
elementary density on the $z$-axis $\rho(z)=\delta(z - z_0)$ requires
an off-axis density term that falls as $r^{-4}$. This behaviour 
rules out many classes of galaxy models, including discs with
exponential density profiles.
However, we can construct
models in which the density falls more
slowly than $r^{-4}$ as $r\rightarrow\infty$. In particular,
models with asymptotically flat rotation curves,
i.e.~those in which $\rho(r) \sim r^{-2}$, are allowed.

In the models used in this section, we specify the $z$-axis density profile
\begin{equation}
\rho_z(z) = {-\gamma \rho_0 \over (z^2 - \gamma)}
= {-\gamma \rho_0 \over \tau},
\label{eq:rhozed}
\end{equation}
where we have made use of $z^2 = \tau + \gamma$. In this case, de Zeeuw's
function $G(\tau)$ can be written in closed form
\citep[equation (49) with $s=2$,][]{de-z-2}. Models specified by \eqref{eq:rhozed}
become spherical at large radii and have a rotation curve that
is asymptotically flat, with
\begin{equation}
\lim_{r\rightarrow\infty}v^2_c = -4\pi G \rho_0\gamma.
\end{equation}
In the core of these models, the surfaces of constant density are
approximately ellipsoidal, with axis ratio\footnote{Equation (46)
in \cite{de-z-2} presents a formula for this quantity, but it is
incorrect.}
\begin{equation}
{a_z^2 \over a_R^2} = {2 q^2 \over (1 - q^2)^2} \left(1 - q^2 + q^2 \log q^2\right),
\end{equation}
where the central potential axis ratio $q = \sqrt{\gamma/\alpha}$. The models
are completely specified by choosing a shape via $q$, a mass scale via $\rho_0$,
and a distance scale via $\gamma$.
The combination of an asymptotic logarithmic `halo' and a flattened
`disc' in these models allows them to make a fair representation of the
observed properties of disc galaxies, although the lack of freedom in the
models severely restricts the shape of the density distribution
that can be achieved.

The potential used in this section, SP1, was chosen to simulate a galaxy
with a massive disc. The axis ratio of the density distribution is fixed to
be 10 near the solar radius, $\rsun=8\kpc$, which is approximately the same
ratio as for the (exponential) thin disc profile of the Milky Way (see Table
2.3 of BT08). The specification is completed by requiring the rotation curve
to peak at $\rsun=8\kpc$ with a circular speed $v_c = 240\kms$. The resulting
parameters of the SP1 potential are: $\rho_0 = 3.61 \ttp 9 \msun\kpc^{-3}, \alpha =
-29.64 \kpc^2, \gamma = -8.89 \ttp {-3} \kpc^2.$

\begin{figure}
  \centering{
    \includegraphics[width=\squarefigshrink\hsize]{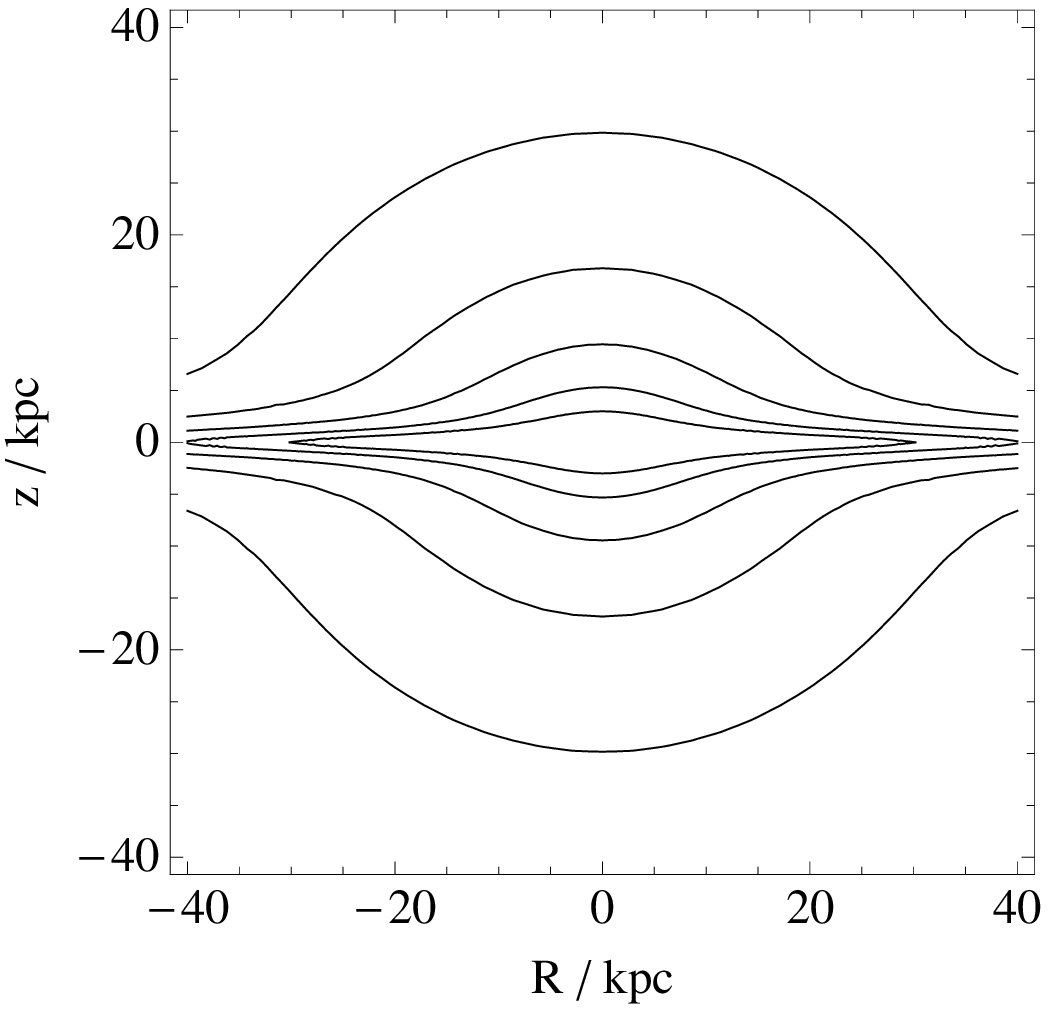}
    \includegraphics[width=\squarefigshrink\hsize]{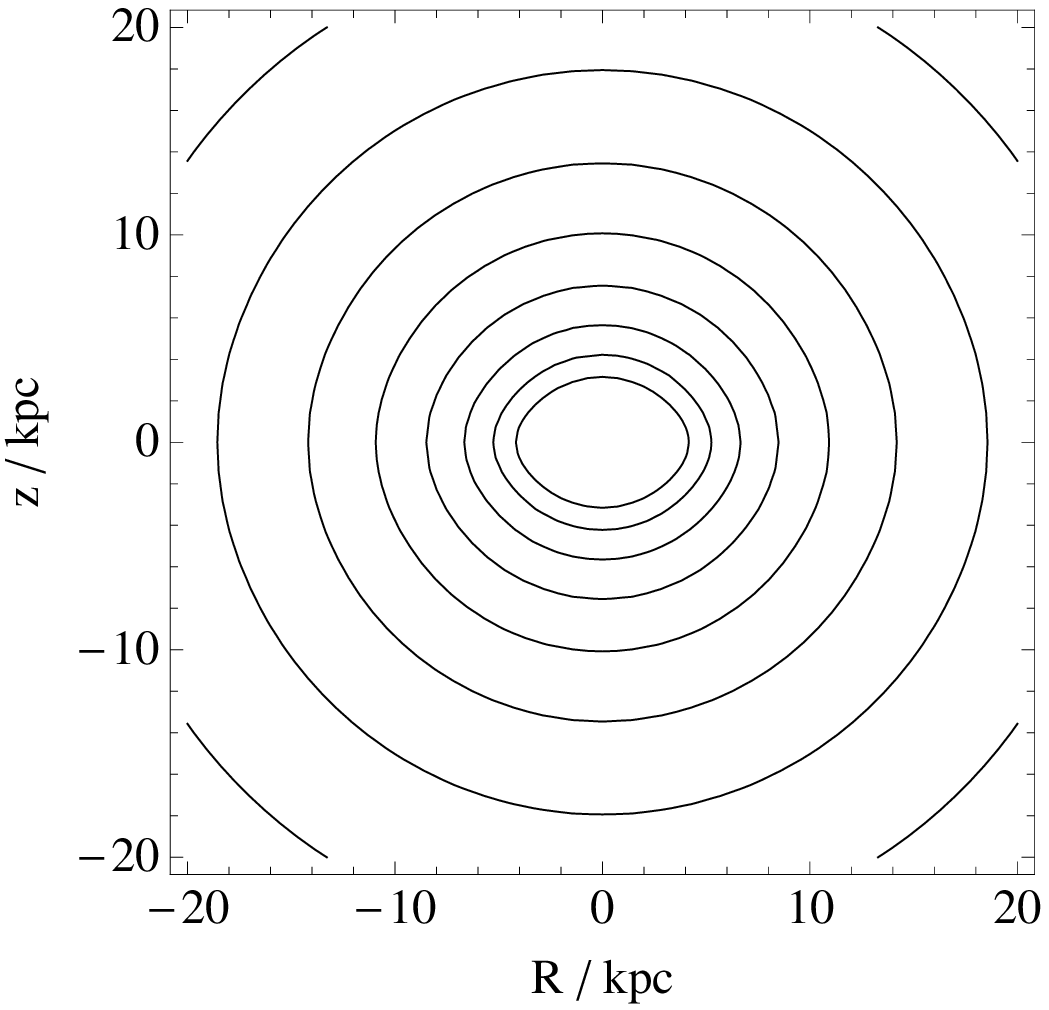}
  }
  \caption { The upper panel shows contours of $\log \rho/\rho_0$ for
    the \stackel\ model SP1, utilized in this section. The lower panel
    shows contours of the potential $\Phi$ for the same.  }
\label{fig:stackel-pots}
\end{figure}

We note that the asymptotic circular velocity in this model is
$v_c = 42\kms$, which can be regarded as the halo contribution to the
circular speed. The model is too centrally concentrated, and the halo
contribution is too weak, to realistically model the Milky
Way. However, it is highly flattened, and so makes an interesting
example in which to study stream geometry. Contours of constant density
and potential for this model are shown in \figref{fig:stackel-pots}.

\subsection{Stream misalignment in the \stackel\ potential SP1}
\label{sec:stackmisalignment}

\begin{figure*}
\centering{
    \includegraphics[width=\doubsquarefigshrink\hsize]{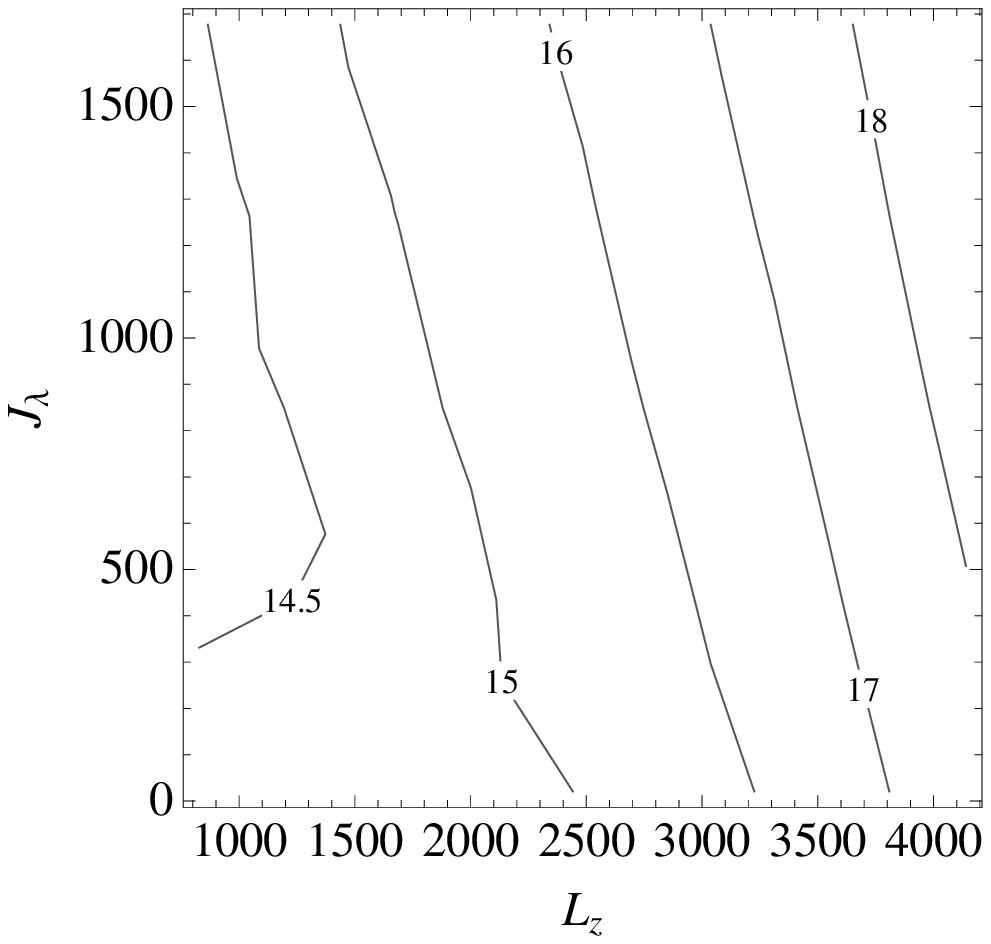}
    \includegraphics[width=\doubsquarefigshrink\hsize]{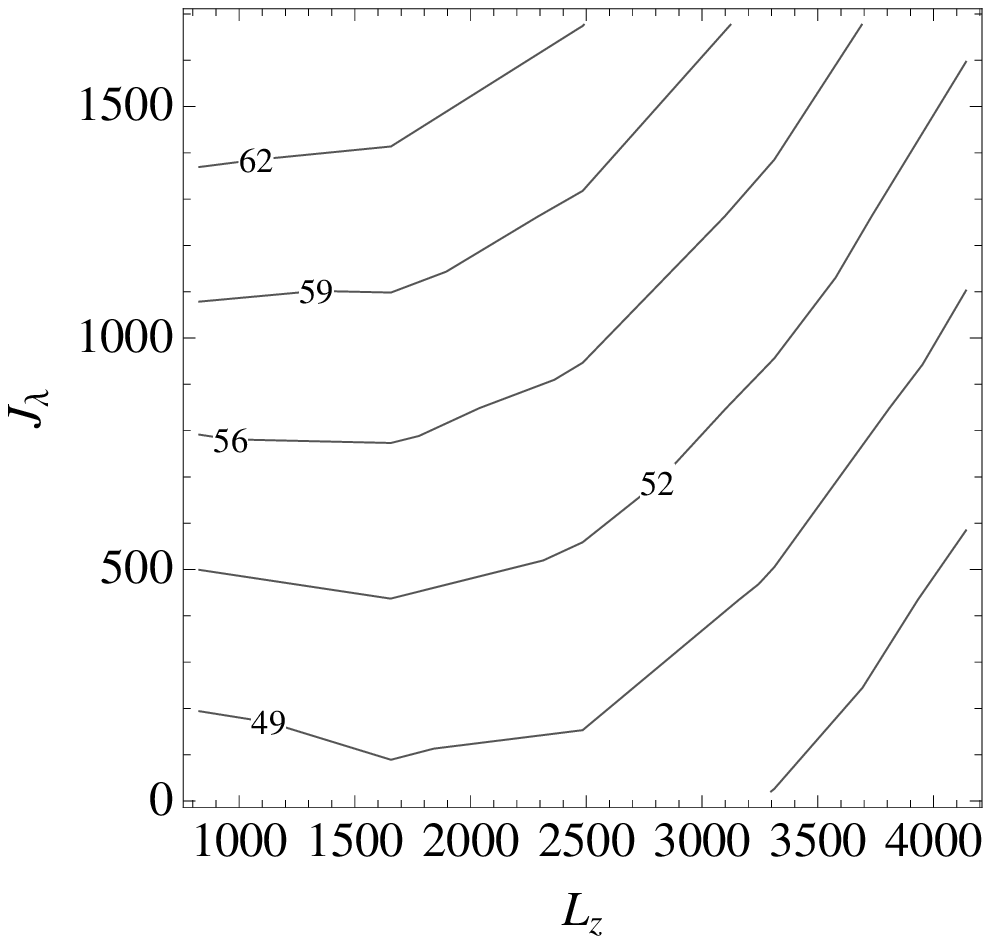}\\
    \includegraphics[width=\doubsquarefigshrink\hsize]{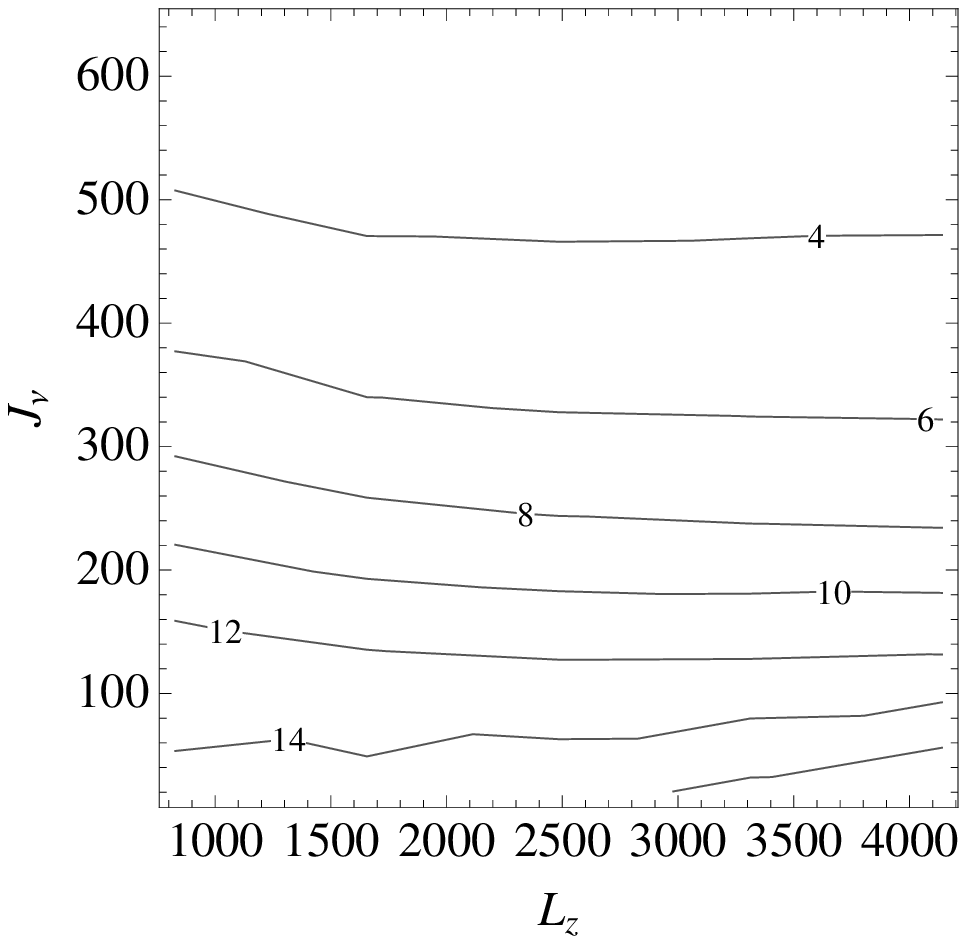}
    \includegraphics[width=\doubsquarefigshrink\hsize]{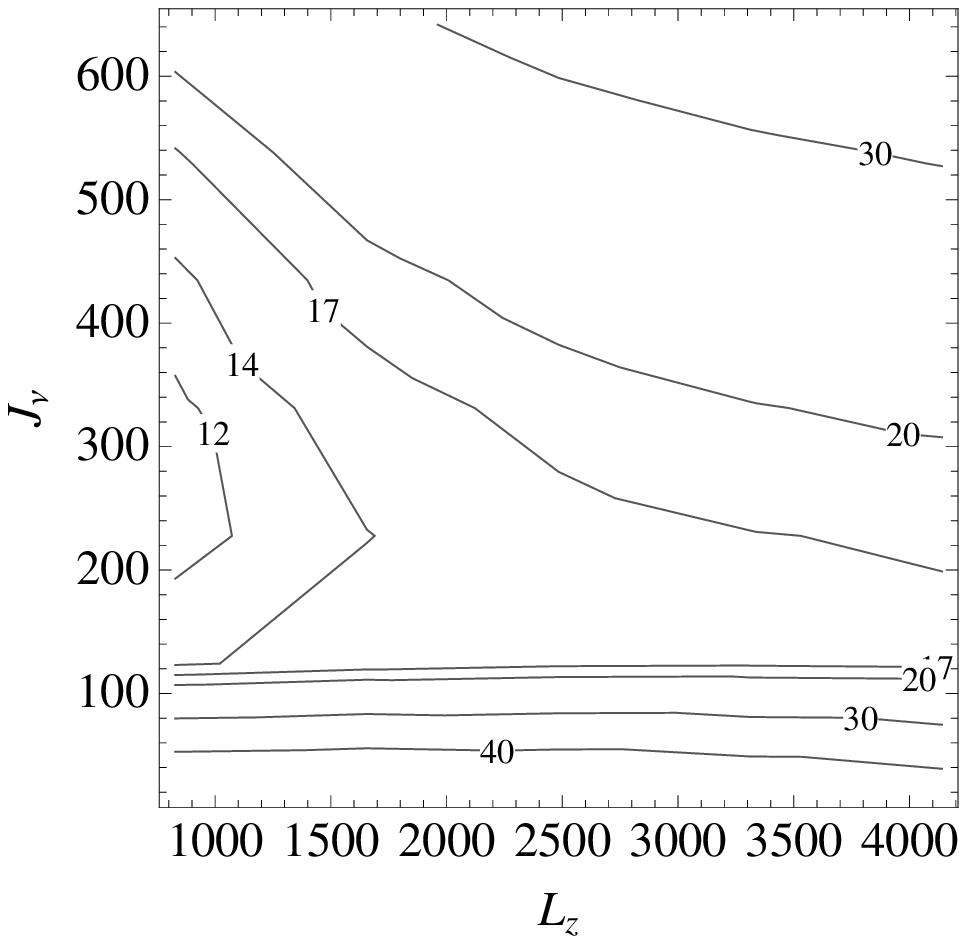}
  }
\caption
{Details of the stream geometry for the \stackel\ potential SP1.
Left panels: contours for the misalignment angle $\vartheta$, in degrees, between the principal
eigenvector of $\hessian$ and $\vO_0$, shown as a function of $\vJ$. Right panels:
contours for the eigenvalue ratio $\lambda_1/\lambda_2$.
The top panels show the plane in action-space with $J_\nu = 20.7\kpc\kms$, while
the bottom panels show the plane in action-space with $L_z=414\kpc\kms$.
The range of actions covers a variety of interesting orbits: details of 
orbits at the extremes of the range are given in \tabref{tab:sp1-orbit-extrema}
  }
  \label{fig:sp1-hessian}
\end{figure*}

We now consider the geometry of streams formed in the \stackel\
potential SP1. Although the actions $\vJ$ can be defined in
terms of an integral over a single coordinate, closed-form expressions for $\vJ$
do not exist. Instead, the integrals in the expressions
for $\vJ$ have to be evaluated numerically. Similarly, expressions for
both the frequencies $\vO$ and their derivatives $\nabla_\vJ \vO$ can
be derived, but not in closed form, and the integrals that they
contain must also be evaluated numerically.

The expressions for $\vJ$ and $\vO$ appear in \cite{de-z-1}. The
detailed expressions for $\nabla_\vJ \vO$ are algebraically rather
involved, and are presented in the appendix of \cite{thesis}. Here, we
simply note that having evaluated these quantities for a particular
orbit $\vJ_0$, the eigenvectors $\eigen_n$ and the eigenvalues $\lambda_n$ are
computed directly from the matrix $\hessian(\vJ_0) =
\left.\nabla_\vJ\Omega\right|_{\vJ_0}$ by standard methods.

\begin{table}
  \centering
  \caption
  {The coordinate extrema of selected orbits from \figref{fig:sp1-hessian},
    illustrating the variety of orbits covered by that figure.
    The actions are expressed in $\kpc\kms$, while the apses are in \kpc.}
  \begin{tabular}{llll|lll}
    \hline
    $J_\lambda$ & $L_z$ & $J_\nu$ & $R_{\rm p} $ & $R_\a$ & $|z|_{\text{max}}$ \\
\hline
    $20$ & $828$ & $20$ & $3.5$ & $5$ & $0.74$ \\
    $20$ & $4140$ & $20$ & $20$ & $26$ & $2.5$ \\
    $1680$ & $4140$ & $20$ & $14$ & $70$ & $7$ \\
    $1680$ & $828$ & $20$ & $1.75$ & $27$ & $3$ \\
    $414$ & $828$ & $20$ & $2$ & $10$ & $1.25$ \\
    $414$ & $4140$ & $20$ & $16$ & $38$ & $3.5$ \\
    $414$ & $4140$ & $640$ & $22$ & $56$ & $34$ \\
    $414$ & $828$ & $640$ & $3$ & $20$ & $17$ \\
\hline
  \end{tabular}
  \label{tab:sp1-orbit-extrema}
\end{table}

The left panels of \figref{fig:sp1-hessian} show contour plots of the
misalignment $\vartheta$ in angle-space between the principal
direction of $\hessian$ and the frequency vector $\vO_0$, where
$\vartheta$ is calculated from \eqref{eq:misangle}, as was the case
for systems of two actions. The right panels of the same figure show
contours of the ratio $\lambda_1/\lambda_2$.  The range of actions
shown in these plots covers a variety of interesting orbits; the apses
of the orbits at the extremes of the range are described in
\tabref{tab:sp1-orbit-extrema}.

As with the equivalent plots for the isochrone potential
(\figref{fig:isochrone-hessian}), we see that the principal direction of
$\hessian$ is never perfectly aligned with $\vO_0$.  In this very flattened
potential, streams with low $J_\nu$ have the greatest degree of misalignment,
at about $\near 15\deg$.  These orbits never go far from the disc.  The
misalignment diminishes with increasing $J_\nu$, falling to $\near 4\deg$ for
orbits with apses in $z$ of some tens of \kpc.  Hence, in this very flattened
potential, there is much more prospect for dramatic angle-space misalignment
than the small misalignments we encountered with the isochrone potential
(\figref{fig:isochrone-hessian}). \figref{fig:sp1-hessian} also shows that
the eigenvalue ratio $\lambda_1/\lambda_2 > 10$ everywhere for the SP1
potential; thus, we conclude that highly elongated streams will form on all
orbits which permit a cluster to be disrupted.

As with the spherical case, the precise behaviour of any given stream
depends on both the potential and the action-space distribution of its
stars.  To proceed further we must again consider a specific example,
by means of N-body simulation.

\subsection{A stream in the \stackel\ potential SP1}

\begin{figure}
  \centering{
    \includegraphics[width=\squarefigshrink\hsize]{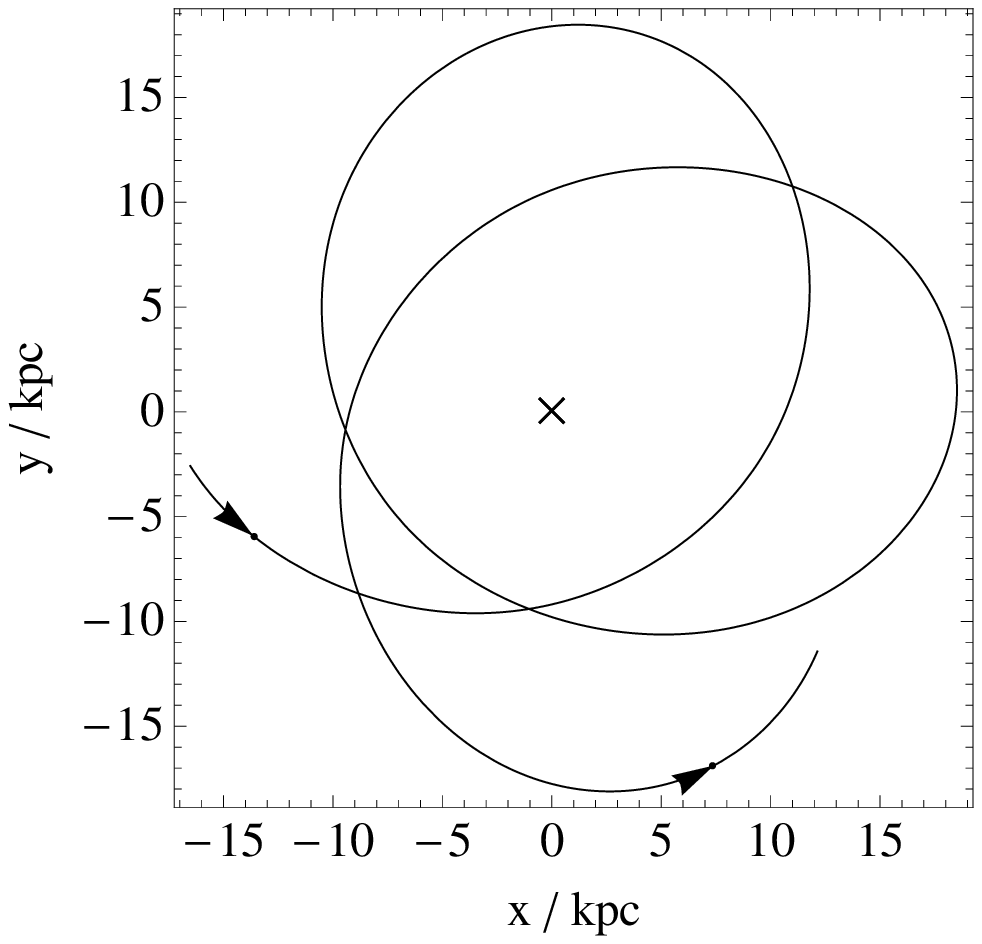}
    \includegraphics[width=\squarefigshrink\hsize]{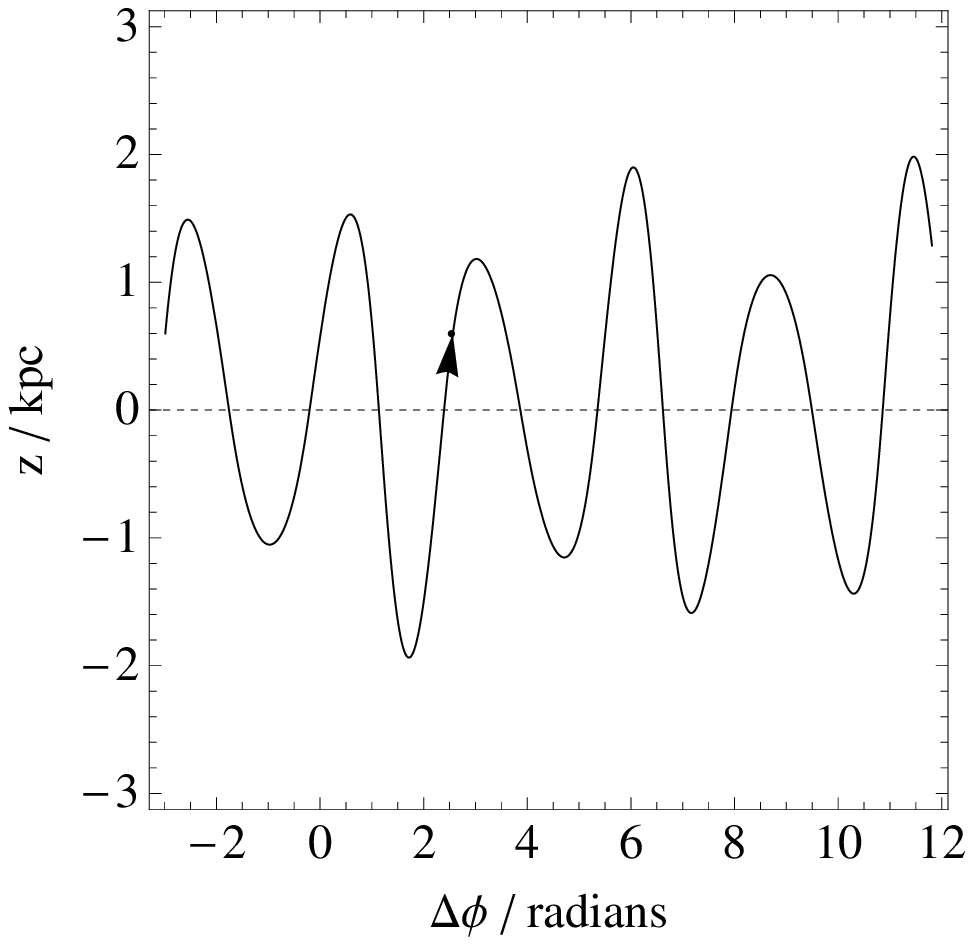}
  }
\caption
{
The real-space trajectory of the SO1 orbit
 in the \stackel\ potential SP1.
The upper panel shows a plan view of the galactic plane,
while the lower panel plots height above the plane, $z$, against
azimuthal coordinate, $\phi$.
}
\label{fig:stack-orbit-examples}
\end{figure}

\figref{fig:stack-orbit-examples} shows the real-space trajectory of
the orbit SO1, evaluated in the \stackel\ potential SP1. This orbit
has actions $(J_\lambda,L_z,J_\nu) = (252.3, 2618., 20.4) \kpc\kms$
and apses of approximately $R=(8,18)\kpc$ in the galactic plane, and
$z=(-2,2)$ above and below the plane. It is thus fairly representative
of an eccentric orbit that might be occupied by a globular cluster
orbiting close to a galactic disc.

The cluster model C5 (\tabref{tab:clusters}) is a King model specified for
the orbit SO1 according to the schema of \secref{sec:clusters}. The model has
the same mass and profile parameter as C1, and is very similar in all other
attributes, because the orbit SO1 is similar to the orbit I4 for which C1 was
specified.  A $10^4$ particle realization of the C5 was made by random
sampling of the King model distribution function.  This cluster was placed
close to apocentre on the orbit SO1 and evolved forward in time by the
\fvfps\ tree code, with time step $\d t=\tdyn/100$ and softening parameter
$\epsilon$ as specified in \tabref{tab:clusters}.  The total period of the
simulation was $2.15\Gyr$, or 7 complete radial oscillations.

\subsubsection{Action-space distribution}
\label{sec:stackeldistribution}

\begin{figure*}
  \centering{
    \includegraphics[width=\doubsquarefigshrink\hsize]{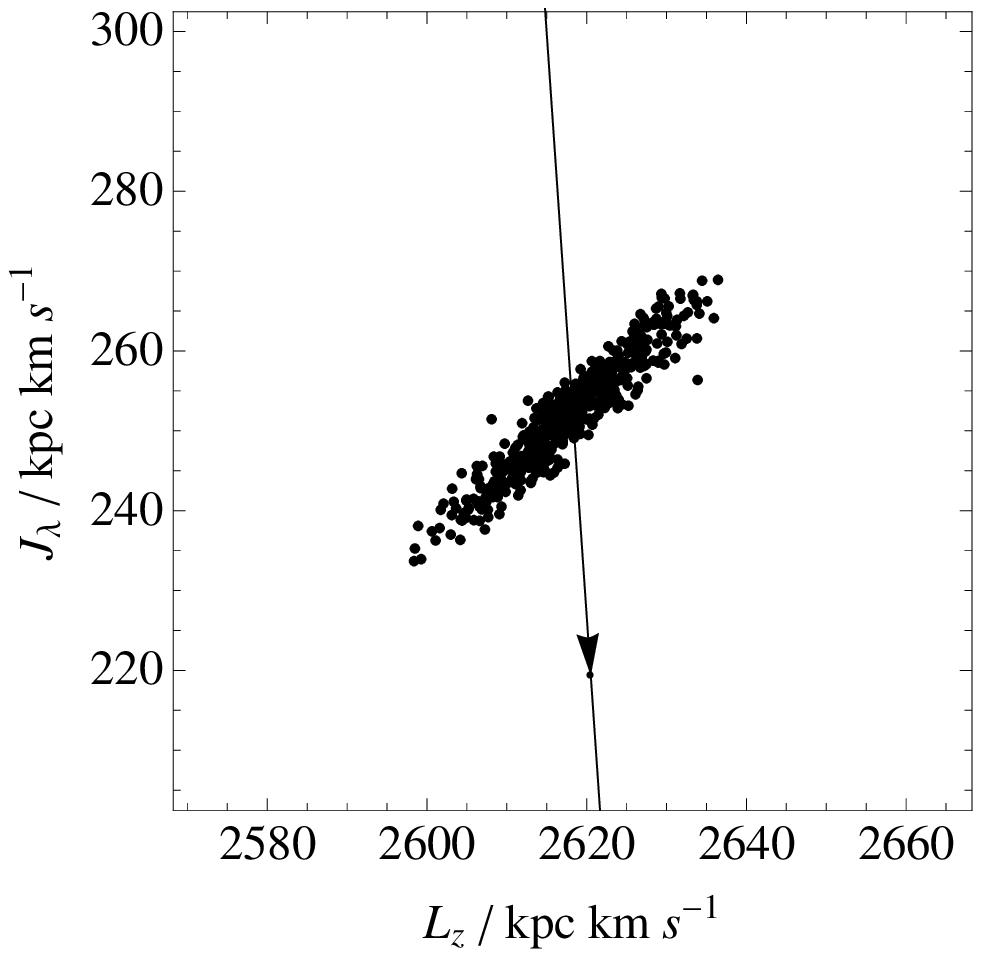}
    \includegraphics[width=\doubsquarefigshrink\hsize]{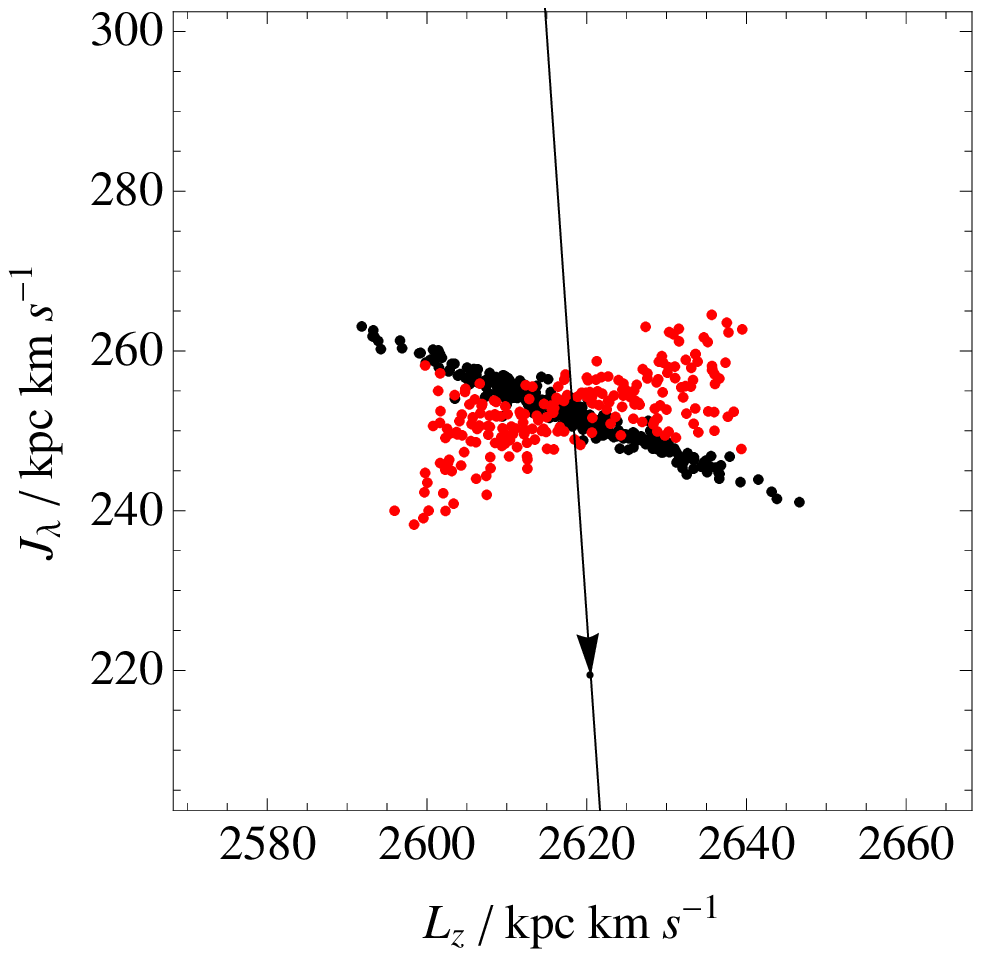}
    \includegraphics[width=\doubsquarefigshrink\hsize]{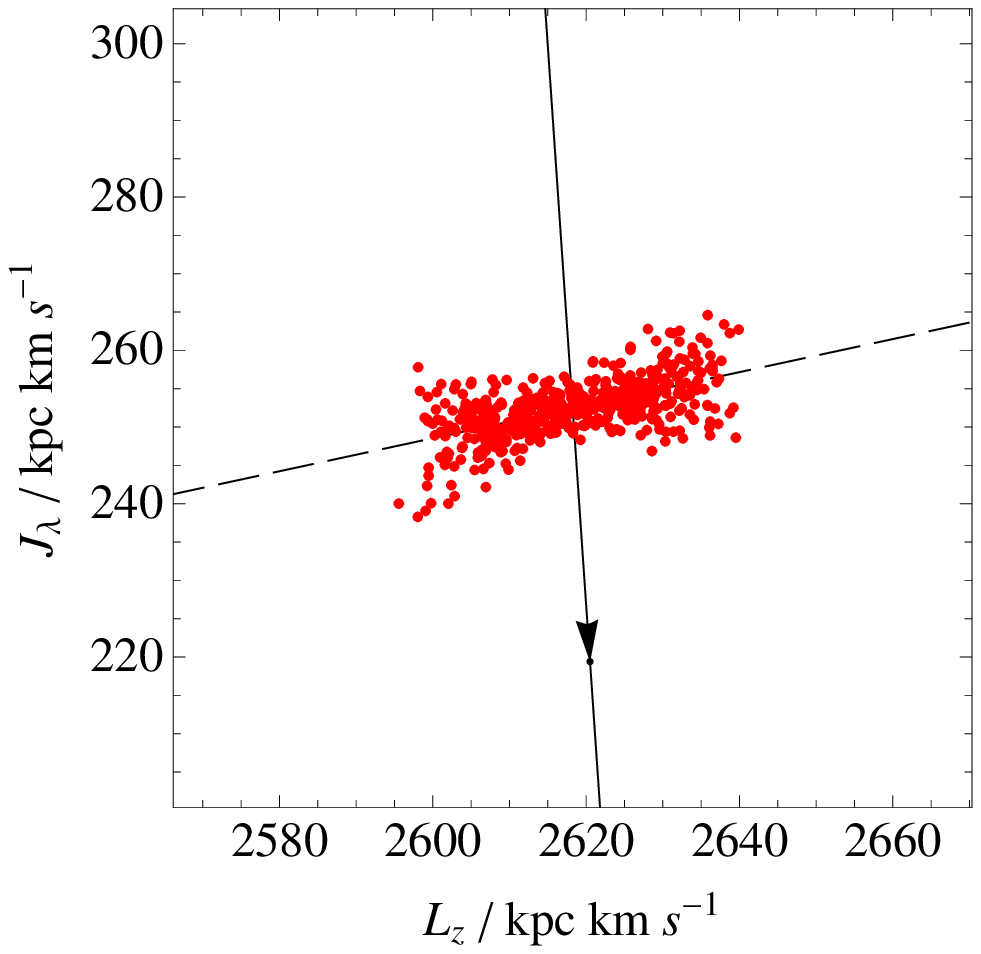}
  }
  \centering{
    \includegraphics[width=\doubsquarefigshrink\hsize]{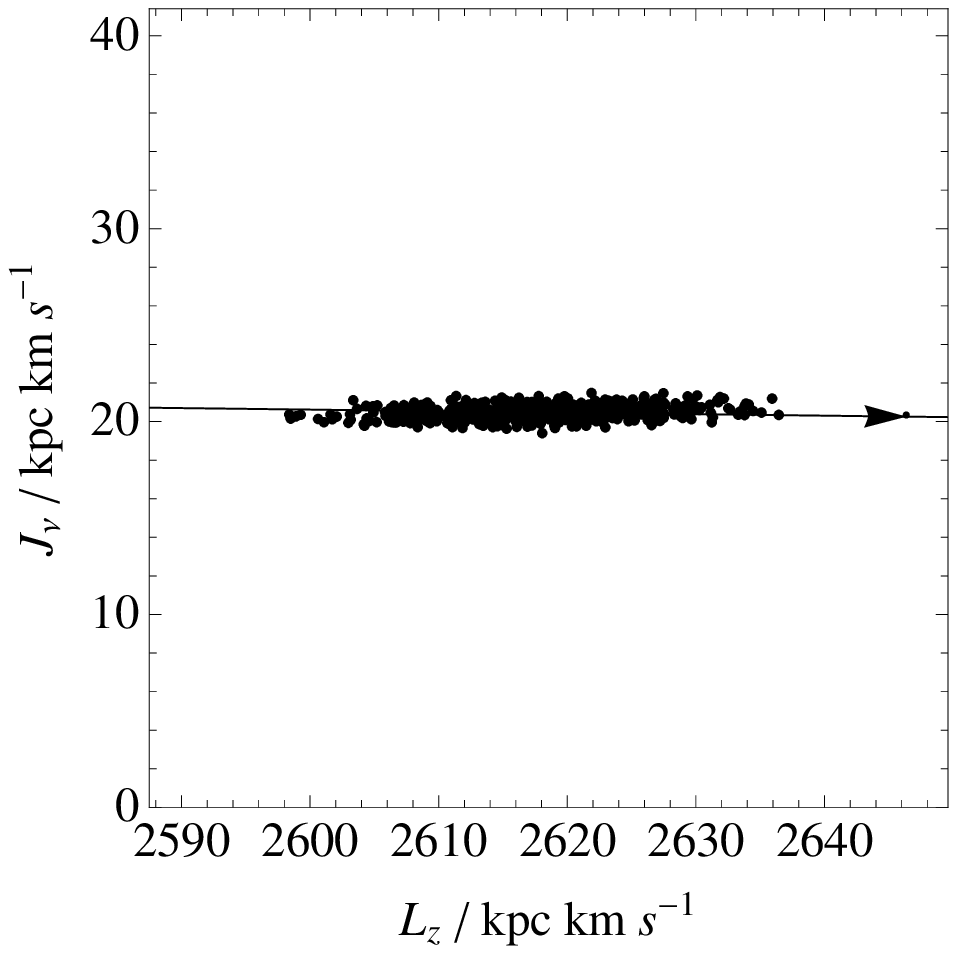}
    \includegraphics[width=\doubsquarefigshrink\hsize]{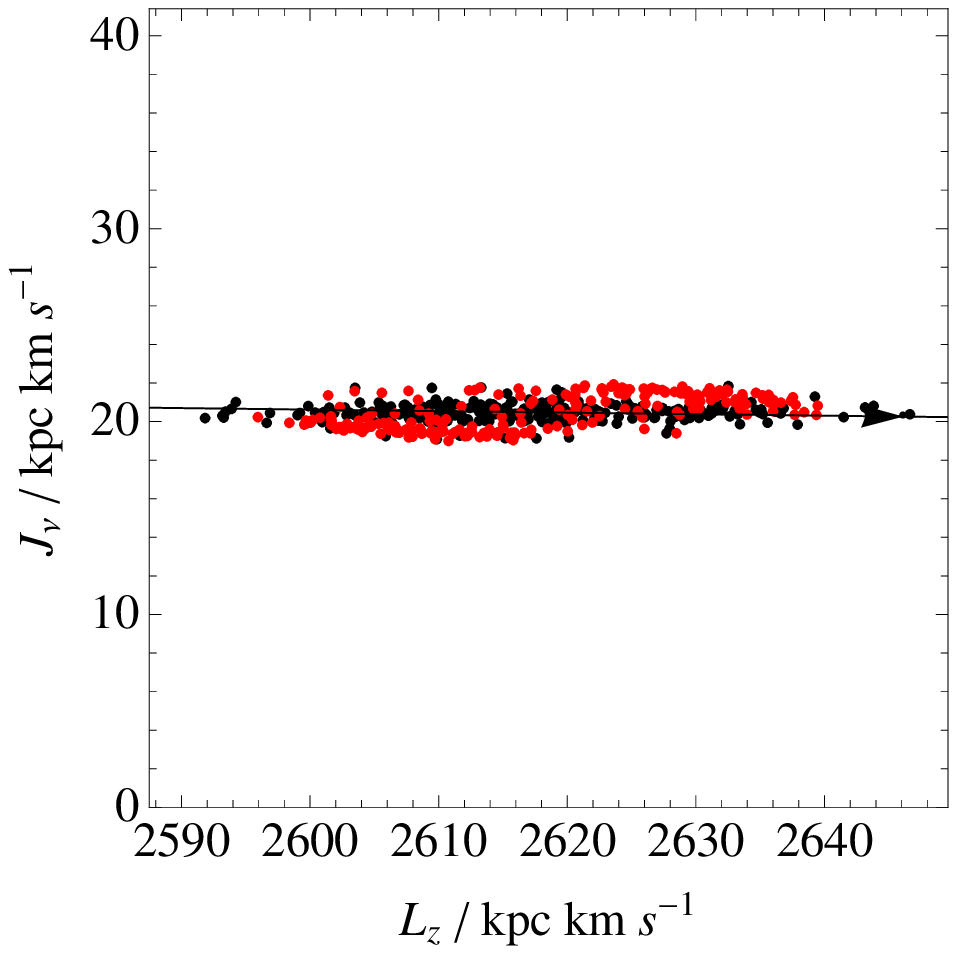}
    \includegraphics[width=\doubsquarefigshrink\hsize]{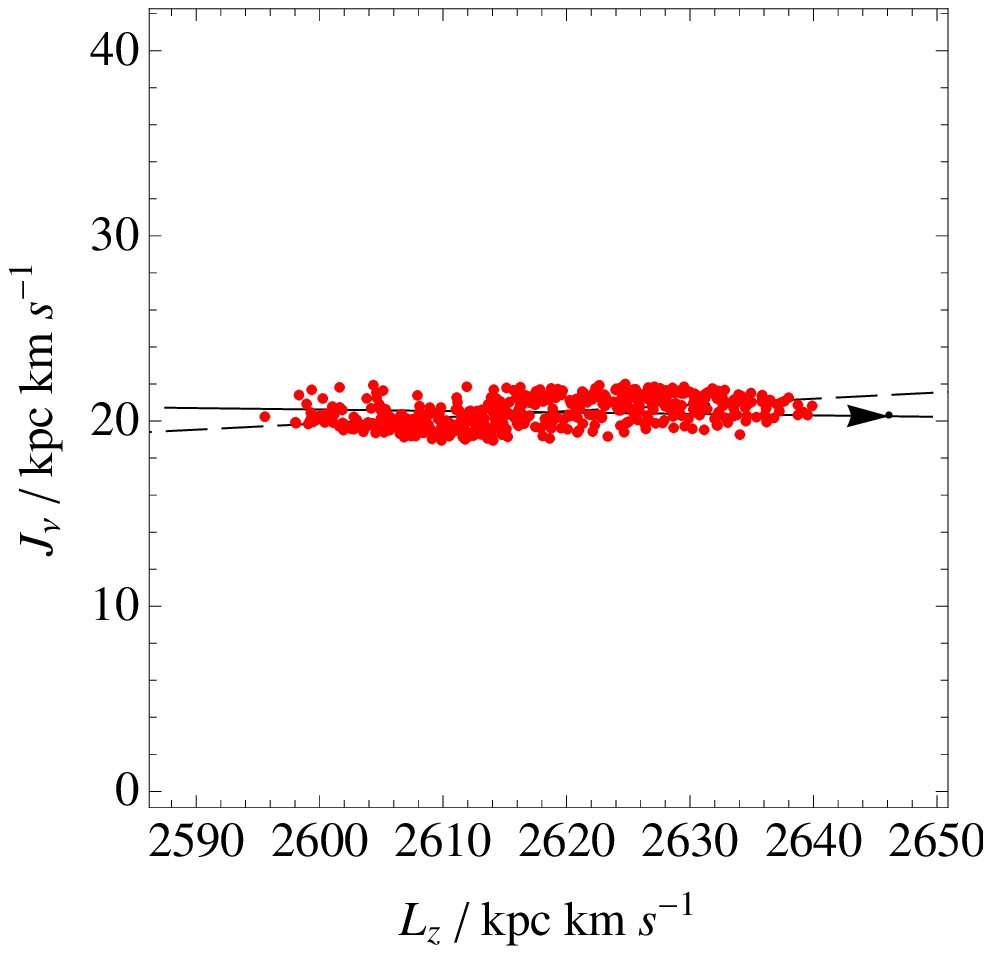}
  }
\caption{Action-space distribution for the simulated cluster
C5, on the orbit SO1 in the \stackel\ potential SP1,
at various points in time. The upper panels show the orthographic
projection of distribution onto the $(L_z,J_\lambda)$ plane, while
the lower panels show a similar projection onto the
$(L_z,J_\nu)$ plane. Distributions are shown at the following times:
left panels, the first pericentre passage;
middle panels, the subsequent apocentre passage;
right panels, the 7th apocentre passage.
Black particles are bound to the cluster, while
red particles orbit free in the host potential.
The dashed lines in the bottom panels are lines that have
been least-squares fitted to the particles.
}
\label{fig:stack-xt-jays}
\end{figure*}

\figref{fig:stack-xt-jays} shows the evolution in time of the action-space
distribution of this simulated cluster. Each column of panels
shows the distribution at a different point in time. The upper panel
in each row shows the orthographic
projection of the actions onto the $(L_z,J_\lambda)$ plane, while
the lower panel of each row shows a similar projection onto the
$(L_z,J_\nu)$ plane. In all panels, the appropriate projection
of the mapped frequency vector, $\hessian^{-1}\vO_0$, is shown
as an arrowed black line.

The left hand column shows the actions when the cluster is near to its first
pericentre passage. In the upper panel, the distribution is somewhat
flattened, and oriented with positive gradient in $\Delta J_\lambda/ \Delta
L_z$. This behaviour is analogous with that seen in the top-right panel of
\figref{fig:nbody-run1}: the motion of the cluster is predominantly in
$(\lambda, \phi)$, thus $(J_\lambda, L_z)$ are good proxies for the radial
action $J_r$ and angular momentum $L$ respectively.  $J_\lambda$ and $L_z$
are therefore highly correlated for a cluster near apsis on this orbit, in
analogy with the mechanism described by \eqref{eq:gradient} in
\secref{sec:disruption}.

Meanwhile, the distribution in the lower panel is both narrow and
oriented almost exactly along $\hat{L}_z$. We can understand the
shape as follows. For this orbit, which is confined to be close to the
plane, $J_\nu \sim J_z/2$, where the factor of 2 appears because
$J_\nu$ is defined on a path restricted to only one side of the
plane. $J_z$ can be estimated by close analogy with
\eqref{eq:deltajr}. Hence, the spread in $J_\nu$ for a cluster of
velocity dispersion $\sigma$ is approximately
\begin{equation}
\Delta J_\nu \simeq {1\over 2} \Delta J_z \sim {1 \over 2\pi} \delta p_z \Delta z
\simeq {1 \over 2\pi} \sigma \Delta z.
\end{equation}
By analogy with \eqref{eq:djr/dl} we find
\begin{equation}
{\Delta J_\nu \over \Delta L_z} \sim {\Delta z \over 2\pi R_{\rm p}},
\label{eq:djnu/dl}
\end{equation}
where $R_{\rm p}$ is the galactocentric pericentre radius in cylindrical
coordinates. Evaluating this expression for the orbit SO1 gives
$\Delta J_\nu/\Delta L_z \sim 0.04$, which we see from the lower-left
panel of \figref{fig:stack-xt-jays} is close to exact.

The flat orientation of the lower-left panel we explain by pointing out
that, as \figref{fig:stack-orbit-examples}
shows, the motion in $\nu$ in this example is almost decoupled from the
radial motion. This means that the $\nu$ coordinate need not be at
apsis when the cluster is at pericentre, and thus the arguments of
\secref{sec:disruption}, which force a correlation between
$J_\lambda \sim J_r$ and $L_z \sim L$ near pericentre, do not
apply. For an orbit in which $J_\nu$ is more strongly coupled to the
radial motion, we would expect to see the characteristic tilting of
the $(L_z, J_\nu)$ distribution near pericentre and apocentre, as a
correlation between $J_\nu$ and $L$ is forced.

The middle column of panels in \figref{fig:stack-xt-jays}
shows the action-space distribution at the subsequent apocentre passage.
Bound and unbound particles are shown in black and red, respectively.
The action-space structure of the unbound particles in the upper panel bears
striking similarity to that shown in \figref{fig:nbody-run1}, as might
be expected since $(J_\lambda, L_z)$ make good proxies for $(J_r,L)$.
The same physical principle for the disruption of the cluster applies
here as in the spherical case; that is, the particles will
escape the cluster through the Lagrange points $L_1$ and $L_2$ by first
travelling radially, so the action-space distribution will be
compressed in this direction. Thus, the range of $\Delta L_z$ and $\Delta J_\nu$
for the unbound stars is about the same as in the left hand panels, but
the range of $\Delta J_\lambda$ is markedly less.

The right hand panels of \figref{fig:stack-xt-jays} show the action-space
distribution at the 7th apocentre passage. The structure is
essentially the same as that of the middle panels, except that all
particles are now unbound.  Also plotted in each of the right hand panels
is a dashed line, which has been least-squares fitted to the unbound
particles.
We note that the image of the frequency vector and the dashed line are
highly misaligned in the upper-right plot;
because of this, we expect the stream to be significantly misaligned with $\vO_0$ in
angle-space.

In conclusion, we find that in very flattened potentials, disrupted
clusters form an action-space distribution that is wholly analogous with,
although necessarily more complicated than, that found for disrupted
clusters in spherical potentials. 

\subsubsection{The effects of disc shocks}

Unlike with a spherical potential, an axisymmetric potential allows
for tidal forces other than those felt during pericentre passage to
act upon an orbiting cluster.
In particular, the passage of a cluster through a massive galactic disc
will subject a cluster to a tidal force that is of comparable magnitude to
that felt when close to pericentre.

The tidal stress imposed on a cluster at pericentre has a tensile
component, which acts to strip stars from
the cluster. Conversely, the tidal stress imposed by a disc passage is entirely
compressive in nature. Hence, stars are not actively
stripped from a cluster during a disc passage. Instead, the action of
such `disc shocks' is to heat the cluster, perhaps repopulating the
outer edges of the cluster,  which were depopulated during
a previous pericentre encounter \citep[\S5.2a]{spitzer87}.

The net effect of disc shocks on the stripping process is a faster and
more complete disruption of the cluster than would take place for an
unshocked cluster exposed to equivalent pericentric tidal
stress. Since the vast majority of stars continue to be stripped at
pericentre even when the effect of disc shocks is significant, the
gross action-space distribution resulting from the stripping of a
shocked cluster will remain as previously described.  However, since the
disc shocks act to increase the velocity dispersion of the cluster
between stripping events, it is likely that the wings of the
resulting action-space distribution will be populated with more stars
than would otherwise have been the case.

\subsubsection{Predicting the stream track}

\begin{figure}
  \centering{
    \includegraphics[width=\squarefigshrink\hsize]{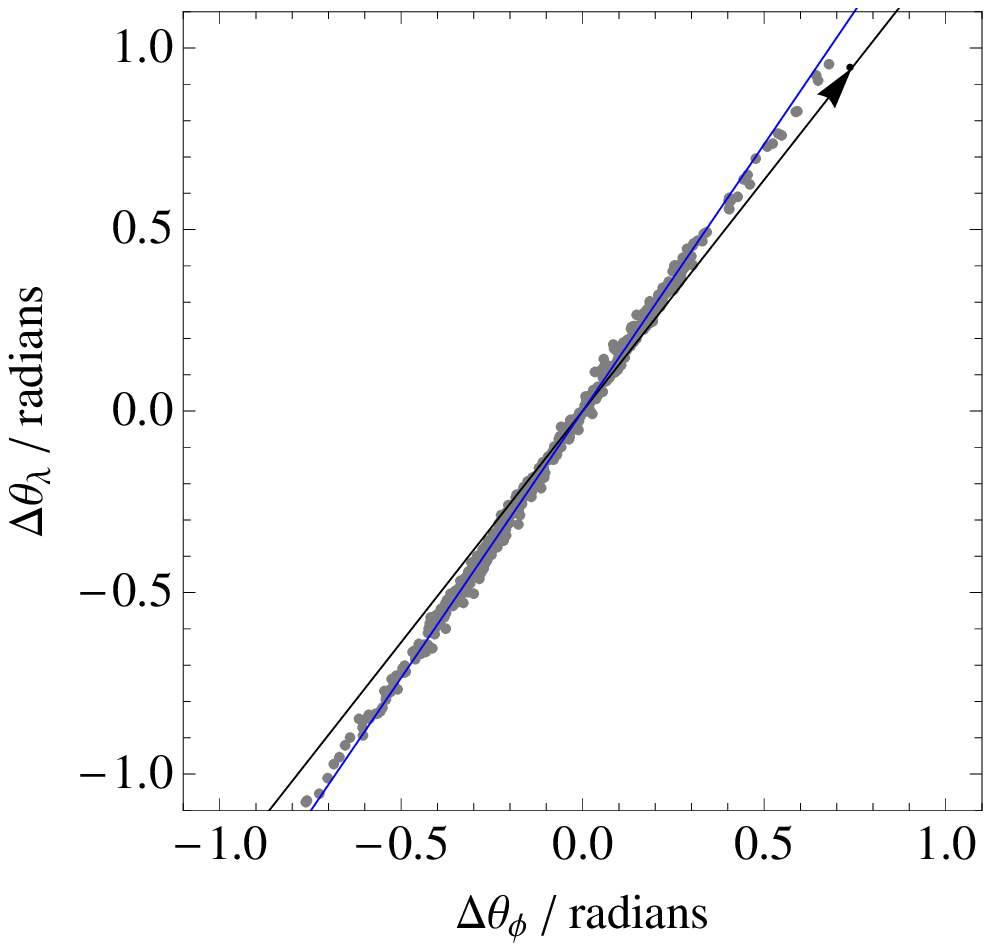}
    \includegraphics[width=\squarefigshrink\hsize]{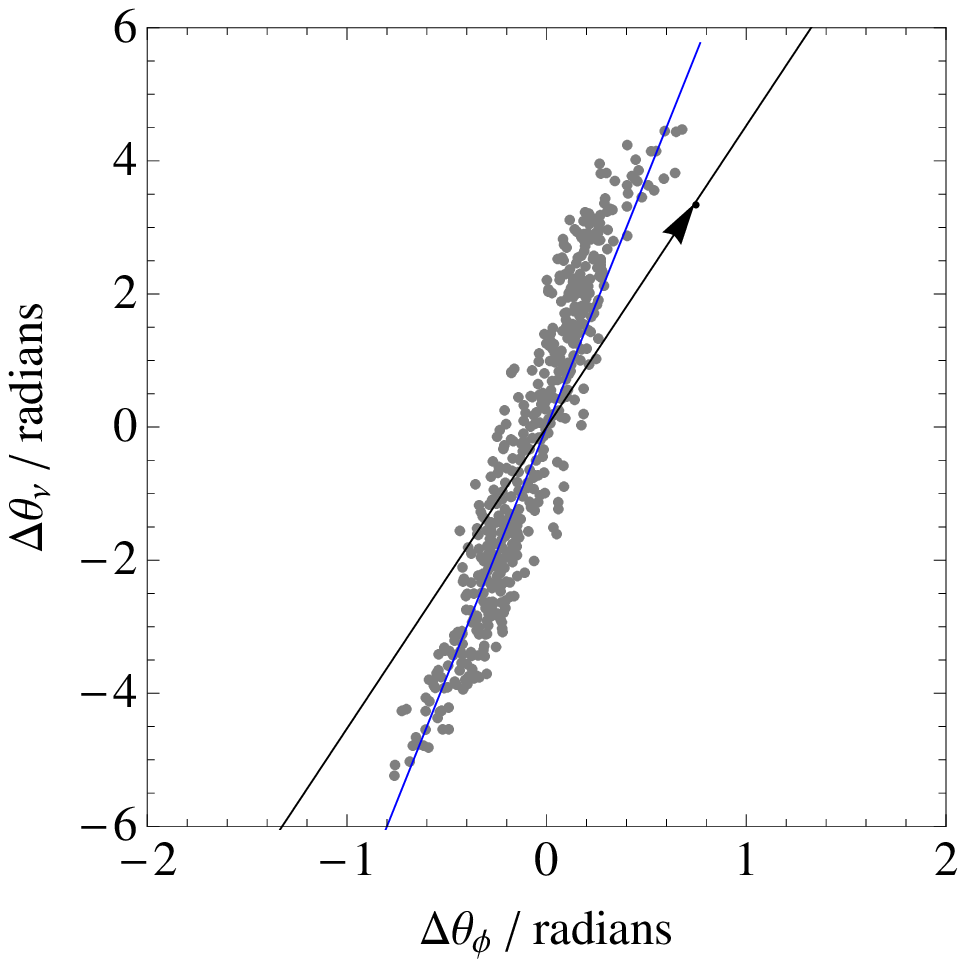}
  }
  \caption
{ Angles for the N-body cluster shown in the right hand panels
    of \figref{fig:stack-xt-jays}. The blue line shows the
    predicted stream, resulting from the mapping of the dashed
    line from the same plots. The blue
    line is clearly a much better representation of the stream than is
    $\vO_0$, represented by an arrowed black line.  }
\label{fig:stack-xt-angles}
\end{figure}

\figref{fig:stack-xt-angles} shows the angle-space configuration
for the simulated cluster near its 7th apocentre passage. The grey
particles are for angles that have been computed directly from the
output of the N-body simulation. The arrowed black line is $\vO_0$,
while the blue line is the map of the dashed-line from
\figref{fig:stack-xt-jays}. In both projections, the blue line is
a much superior match to the data than is the orbit.

\begin{figure}
  \centering{
    \includegraphics[width=\squarefigshrink\hsize]{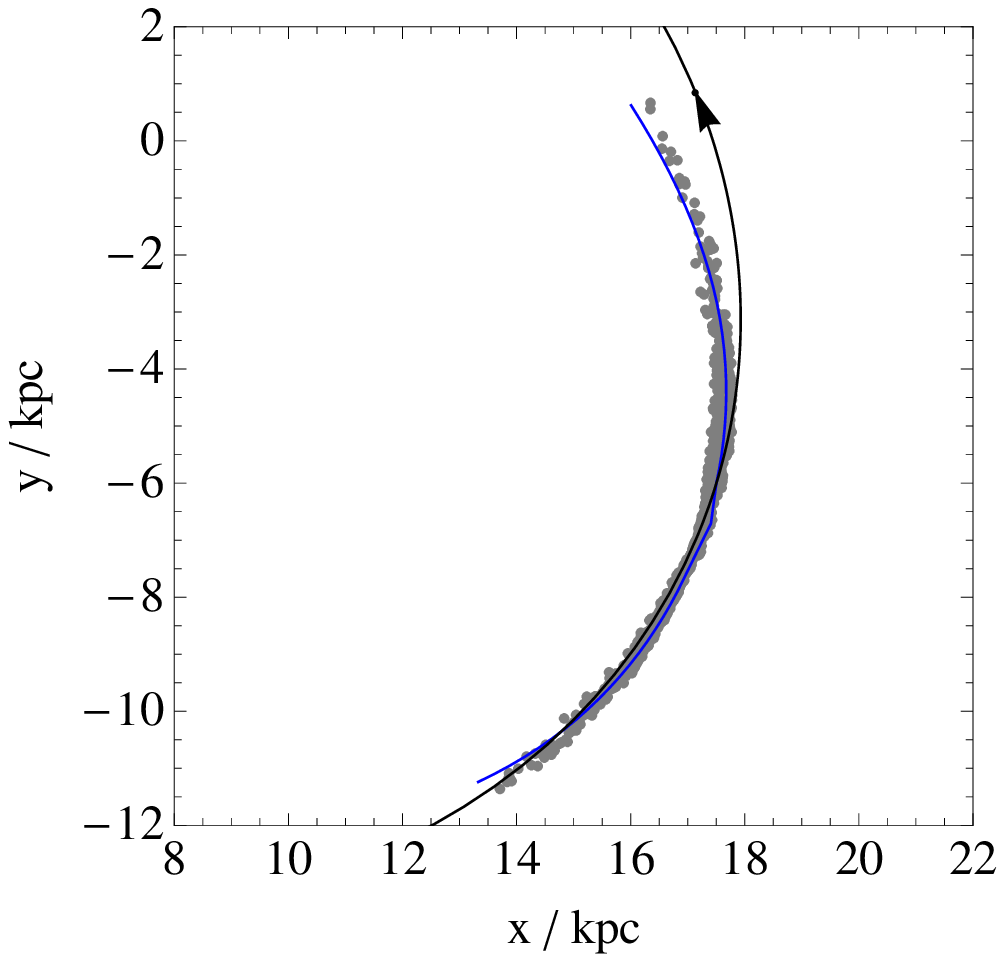}
    \includegraphics[width=\squarefigshrink\hsize]{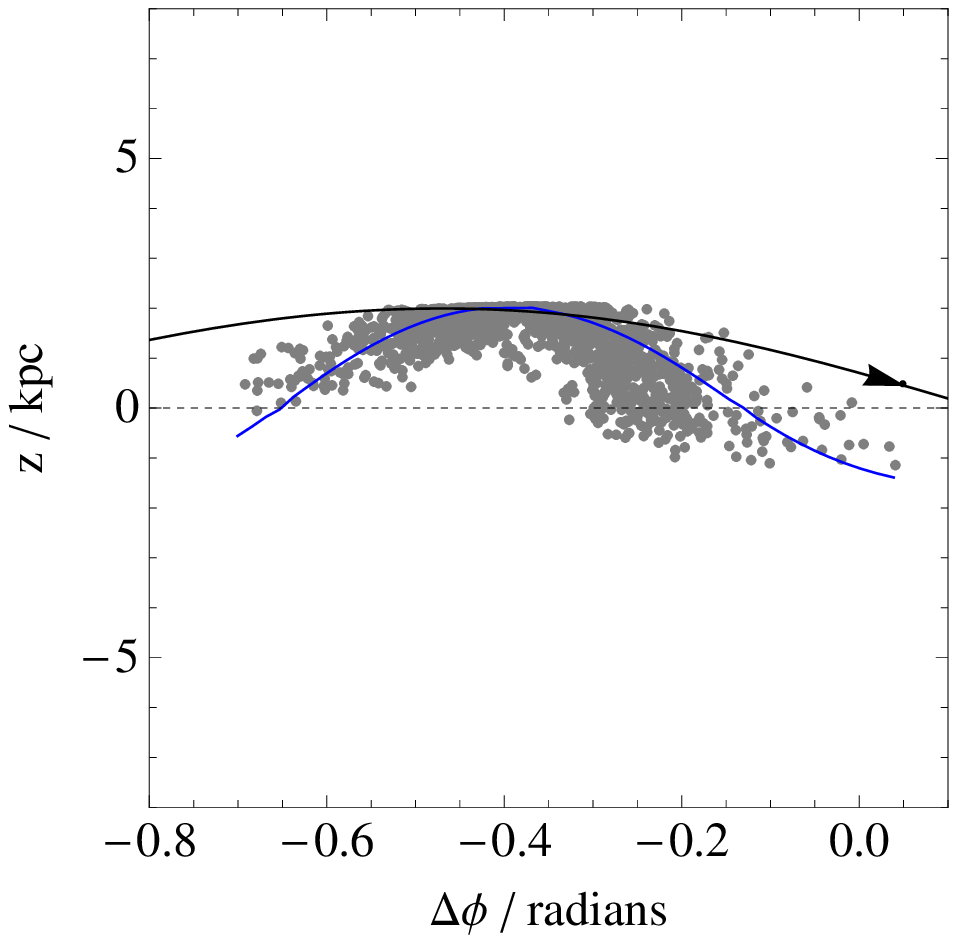}
  }
  \caption { Real-space representation of an N-body simulation of
    cluster model C5 evolved on the orbit SO1, in the SP1
    potential. The points are shown near the 7th apocentre passage
    following release. The black line shows the trajectory of an
    individual orbit; the blue line shows the predicted stream path.
  }
\label{fig:stack-xt-nbody}
\end{figure}

\figref{fig:stack-xt-nbody} shows equivalent plots to those of
\figref{fig:stack-xt-angles}, but in real space.  The misalignment
between the stream and the progenitor orbit in angle-space is seen to
map into a large misalignment in real space. Attempting to constrain
halo parameters by fitting orbits to the stream shown in
\figref{fig:stack-xt-nbody} would not produce sensible results.
Conversely, the map of the dashed-line model for the action-space
distribution, shown in \figref{fig:stack-xt-jays}, clearly provides an
excellent proxy for the track of the stream. We note that, as per
our expectation, there is significant distortion both in and out
of the effective orbital `plane'.

\section{Conclusions}
\label{sec:conclusions}

We have studied the mechanics of the disruption of low-mass clusters,
and the subsequent formation of tidal streams.  In particular, we
have examined the conditions required for the formation of tidal streams
from such clusters, and to what extent such streams delineate the
orbits of their stars. With regard to the latter, we are
motivated by techniques that promise to constrain the form and
parameters of the Galactic potential when observations of thin
streams are analyzed under the assumption that such streams precisely
delineate orbits \citep{binney08,eb09b,willett,newberg-orphan,koposov}.

We utilize action-angle variables extensively in our approach to the
problem. These coordinates allow a convenient and natural description
of the physical processes that occur in cluster disruption and stream
formation. The stream has a structure in action-space that is formed
at the point of disruption, and subsequently frozen for all time.  The
angle-space structure of the stream is the image of that action-space
distribution under a linear map, which is itself a function of the
host potential and the progenitor orbit only.  This angle-space image
elongates linearly with time, corresponding to the extension of the
stream, but its shape remains otherwise unchanged.  Disrupted clusters
elongate to form streams because this linear map preferentially
stretches the angle-space structure along a particular direction.  We
find this preferential stretch to occur for likely orbits in all the
realistic potentials that we consider: the only exception is the
spherical harmonic oscillator potential, in which clusters cannot be
stripped at all.

The real-space structure of a stream is a function of the
corresponding action-angle structure and the host potential. We have
investigated the conditions required for a stream to be perfectly
delineated in real space by its progenitor's orbit. One
necessary condition is that the angle-space stream be perfectly
delineated by the frequency vector.

We have shown that this will always be the case for a stream formed in
a Kepler potential.  Conversely, we have shown that this is generally
 not the case for a stream formed in an isochrone potential:
angle-space streams will generally be misaligned with
the frequency vector by a few degrees. Since the isochrone potential is a more
realistic representation of a galactic potential than is the Kepler
potential, we infer that streams formed in galactic potentials will
not generally be perfectly delineated by orbits.  Indeed, it would
appear that the alignment of streams with orbits is a property that
is peculiar to the Kepler potential. Since streams exist in the outer
parts of galaxies where the potential has a substantial monopole
component, we propose this as an explanation for the observation that
streams appear to be {\em approximately} delineated by the orbits of
their stars.

We have examined the real-space manifestation of angle-space
streams that are misaligned with their frequency vector.
We find that the real-space effect depends on the
orbital phase at which the stream is observed: at apsis,
the stream track will have a curvature that differs
from that of the progenitor orbit, while away from apsis
the stream and its orbits will be spatially misaligned.

We have also examined the real-space effects of a finite
action-space distribution for a stream. We conclude that, with a
cluster described by a large, but realistic, velocity dispersion,
knowledge of the angle-space structure and the progenitor orbit is no
longer sufficient to accurately predict the real-space track of a
stream. We describe a first-order correction to the actions of the
stream stars that is deduced from observables, plus an estimate of the
time elapsed since the initial cluster disruption event. This
correction suffices to allow accurate predictions of real-space tracks
to be made for streams formed from larger clusters.

We have confirmed the validity of our analyses with the use of N-body
simulations of the disruption of clusters in the isochrone potential.  These
show that disrupted clusters indeed form streams that are poorly represented by the
trajectory of the progenitor orbit.

We have used these simulations to inform an explanation of the
action-space structure of disrupting clusters. We find that the
action-space distribution of a disrupted cluster takes a
characteristic shape, the details of which depend on the
cluster model, the cluster orbit and the host potential.  We are able
to explain all features of this distribution in terms of the basic
physical processes that apply to clusters undergoing tidal
disruption.
We further show that by utilising a simple, straight-line model of
the action-space distribution, we are able to predict the real-space
stream track of a stream with very high accuracy.

We have briefly examined the consequences of our findings for techniques that
attempt to constrain the Galactic potential under the assumption that streams
can be fit with orbits. We utilize such an algorithm to attempt to constrain
the mass of an isochrone potential, subject to the rotation speed at the
Solar circle being equal to a fiducial value, and subject to input
pseudo-data being fit with an orbit.  We validate our optimisation procedure
using input pseudo-data for a track that perfectly delineates an orbit.
When the
input pseudo-data are for a simulated stream in which the track shows an
increased curvature with respect to orbits in the correct potential,
the algorithm fails to identify the correct parameters: in our example, the
response of the algorithm was to report an optimum mass parameter
$21\percent$ greater than the true value, albeit with a substantially
decreased quality-of-fit compared to attempts to fit to pseudo-data
representative of a true orbit.  We conclude that large systematic errors can
be made when attempting to optimize potential parameters under the assumption
that streams act as proxies for orbits.

We have examined stream formation in a flattened, axisymmetric \stackel\
model, which provides a fair approximation to the potential near a massive
galactic disc.  We find that in this potential the principal direction of the
linear map is misaligned with the frequency vector by $\near 10\deg$. Hence
in highly flattened potentials like this, streams will not be well
represented by orbits.

We performed an N-body simulation of a disrupting cluster on an
approximately planar orbit in the flattened potential.  We found that
the angle-space misalignment between the stream and the frequency
vector is indeed large, and that the resulting real-space track is
very poorly represented by the progenitor orbit. However, we again
find that a simple straight-line model of the action-space
distribution predicts the corresponding real-space stream track with
superb accuracy.  We also find that in our \stackel\ example, the
action-space distribution resulting from the disruption of clusters is
directly comparable to that found in the isochrone potential,
confirming the generality of our earlier conclusions.

The consequences of our findings for potential-optimization techniques
are significant: large systematic errors are possible if orbits
are fitted to streams. Before attempting such activity,
it is now necessary to check whether the segment of the
stream in question could be legitimately represented by an orbit. Similarly,
is also important for this check to be performed retrospectively for those
streams---e.g. Orphan stream analysis of \cite{newberg-orphan} and the
GD-1 analysis of \cite{koposov}---that have already been used to
constrain the parameters of the Galactic potential, since the validity
of those analyses is now in question.

Such a check could always be performed using N-body simulation, where
it is believed that the progenitor model, the orbit and the applicable
potential are well known. However, the application of the techniques
in this paper would constitute a more general approach, in which the
implications of an incorrect model or orbit can be readily addressed.

The techniques presented in this paper depend on the ability  to
compute action-angle variables from conventional phase-space
coordinates, and vice versa. In this paper,
we have restricted our consideration to models for which
global transformation relations can be written down. The only such
models are those with potentials of the \stackel\ type, which includes
all spherical potentials as a limiting case \citep{de-z-1}. However,
it is difficult to construct a realistic flattened galaxy model that
is satisfactory in all respects whilst restricted to using only
\stackel\ potentials. In general, one would like to work with more
sophisticated models for which no general transformation between
action-angle variables and phase-space coordinates can be made.

Fortunately, the ``torus machine'' \citep{mcmillan-torus} enables the
actions, the frequencies and their derivatives to be accurately and quickly
computed for regular orbits in realistic Galactic potentials. By utilising
the torus machine, the techniques explored in this paper could be extended to
work with such models.  This approach could be used to investigate the impact
on streams of orbital resonances and chaos: with the torus machine one can
construct approximate angle-action variables even for potentials that support
chaos \citep{KaasalainenB94}.  Resonances and chaos play rather a small role
in the dynamics of axisymmetric galactic potentials, but one might expect the
sensitivity of chaotic orbits to their initial conditions to lead to a
measurably more rapid spread of tidal streams in chaotic rather than
integrable regions of phase space.

If a check---however performed---confirms that the stream is well
modelled by an orbit, then diagnosing the potential by the fitting of
orbits to streams is appropriate. However, if the stream is not well
modelled by an orbit, as will generally be the case, the technique of
fitting orbits should not be used.

In such circumstances, one might resort to N-body shooting methods to
compute stream tracks to feed to an optimization algorithm
\citep{johnston-nbody-fitting}. However, the computational expense
of such a technique would be a major burden, and would severely
limit the quality of the optimization.

The results of this paper present a possible alternative.  We have
found that with simple models of the action-space distribution of a
disrupted cluster, such as can be readily obtained from a single
N-body simulation, we can reliably and accurately predict the track of
a stream, even when it diverges significantly from the trajectory of
the progenitor orbit. In principle, we can compute these tracks with
no more computational effort than it takes to integrate an
orbit. Existing techniques for constraining potential parameters by
fitting orbits could be readily adapted to fit stream tracks instead.

To achieve this goal, the key hurdle is the development of a
predictive theory for the action-space structure of a disrupted
cluster, valid for any problem parameters of our choice.
A possible approach would be to use N-body simulations to obtain the
action-space distribution for a small number of cluster models on a
set of possible orbits, for a range of potentials.  The resulting
distributions could be used as a basis set, which would be
interpolated and distorted to provide an estimate for the action-space
structure for any given cluster on any chosen orbit.  The required
distortions for changes to cluster model and orbit parameters have
already been touched upon in this paper, although in order to be of
practical use, a complete quantitative theory of these distortions
will be required.

In the future, the techniques presented here may well be applicable to
the Sagittarius dwarf stream. In this paper we have assumed that our
low-mass clusters do not affect the host potential.
This may not be true in the case of a heavy Sagittarius
progenitor.  Further study of the effect of a live host potential
on the mechanics of stream formation will be required for the
techniques to be reliably applicable to the Sagittarius dwarf stream.
Also, the action-space structures discussed will be, in part,
composed of dark matter in the case of streams formed from dwarf spheroidal
galaxies. Further investigation of the stripping of satellites using
models including both dark particles and stars will be required
to determine which parts of the stream will remain visible, and
which will be composed solely of dark matter.

\section*{Acknowledgments}

We would like to thank Ben Burnett and the other members Oxford
dynamics group for helpful discussions. AE would like to acknowledge
the support of both PPARC/STFC and the Rudolf Peierls Centre for
Theoretical Physics at Oxford during the preparation of this work.
Parts of this paper were derived from work previously presented in the
thesis of \citet{thesis}.

\bibliographystyle{mn2e}
\bibliography{streamdirs}

\bsp

\label{lastpage}

\end{document}